\documentclass[preprint, 12pt,nofootinbib,aps]{revtex4}

\usepackage{amsfonts}
\usepackage{amsmath}
\usepackage{amssymb}
\usepackage{mathptmx}
\usepackage[dvips]{color}
\usepackage{epsfig}
\usepackage{graphicx}
\usepackage{cancel}
\usepackage{afterpage}

\newcommand{\qslash}{~\rlap/\!\!q}

\newcommand{\rmRe}{{\rm Re}}
\newcommand{\rmIm}{{\rm Im}}

\newcommand{\hc}{\textrm{h.c.}}

\newcommand{\Br}{\textrm{BR}}
\newcommand{\fb}{\textrm{fb}}

\newcommand{\GeV}{{\rm GeV}}

\newcommand{\TeV}{{\rm TeV}}

\newcommand{\calA}{{\cal A}}

\newcommand{\calL}{{\cal L}}
\newcommand{\calO}{{\cal O}}
\newcommand{\calU}{{\cal U}}

\newcommand{\calW}{{\cal W}}

\newcommand{\calZ}{{\cal Z}}

\newcommand{\lam}{\lambda}

\newcommand{\xvec}{{\bf x}}

\newcommand{\zvec}{{\bf z}}
\begin{document}

\title{Phenomenology in minimal cascade seesaw for neutrino mass}
\author{Ran Ding~$^{a}$}
\email{dingran@mail.nankai.edu.cn}
\author{Zhi-Long Han~$^{a}$}
\email{hanzhilong@mail.nankai.edu.cn}
\author{Yi Liao~$^{a,c}$}
\email{liaoy@nankai.edu.cn}
\author{Hong-Jun Liu~$^{a}$}
\email{liuhj@mail.nankai.edu.cn}
\author{Ji-Yuan Liu~$^{b}$}
\email{liujy@tjut.edu.cn}

\affiliation{
$^a$~School of Physics, Nankai University, Tianjin 300071,
China
\\
$^b$~College of Science, Tianjin University of Technology, Tianjin
300384, China
\\
$^c$ Kavli Institute for Theoretical Physics China, CAS, Beijing
100190, China}

\begin{abstract}

We make a comprehensive analysis on the phenomenology in the minimal version of cascade
seesaw for tiny neutrino mass. The seesaw induces at tree level a neutrino mass operator
at dimension nine, by introducing a quadruple scalar $\Phi$ of hypercharge unity and a
quintuple fermion $\Sigma$ of hypercharge zero.
We work in a framework that handles the complicated Yukawa couplings in a nice way
without losing generality. All mixing matrices are essentially expressed in terms of the
vacuum expectation value of the quadruple scalar $v_\Phi$, a free complex parameter $t$,
and known neutrino parameters. We show that the low-energy lepton flavor violating
transitions of the charged leptons set strong constraints on the free parameters. The
constraints have a significant impact on collider physics, and are incorporated in our
signal analysis at the LHC. We investigate the signatures of new particles by surveying
potentially important channels. We find that the $4j2\ell^\pm$ signal is most important
for the detection of the scalars and the $2\ell^{\pm}2\ell^{\mp}2j$,
$3\ell^{\pm}\ell^{\mp}2j$ and $3\ell^{\pm}2\ell^{\mp}+\cancel{E_T}$ signals are quite promising for the fermions.

\end{abstract}

\maketitle

\section{Introduction}
\label{sec:intro}

The origin of tiny yet nonvanishing neutrino mass has remained mysterious since its discovery
in oscillation experiments.
Although such a tiny mass can be incorporated by a trivial extension of the standard model (SM) with right-handed neutrinos, it has to appeal to unnaturally small
Yukawa couplings. In this circumstance, it is more useful
to regard SM as an effective field theory in which the neutrino mass appears as a low energy remnant of some high scale physics. Such low energy effects can be systematically organized by higher dimensional operators in terms of the SM fields. Indeed, it has been known for long that such an operator, that is relevant to neutrino mass, first appears at dimension five and has the unique form,
$\calO_5=\Big(\overline{F^C_{L}}\epsilon\phi\Big)\Big(\phi^T\epsilon F_{L}\Big)$,
the so-called Weinberg operator \cite{Weinberg:1979sa}.
Here $F_L$ and $\phi$ are respectively the SM left-handed lepton doublet and Higgs doublet,
and $\epsilon$ is the antisymmetric matrix in the weak isospin space. The operator is
suppressed by an effective coupling $\lambda/\Lambda$, where $\Lambda$ is a heavy mass
scale and $\lambda$ a product of fundamental couplings of some new physics.

What high scale new physics would be responsible for the operator $\calO_5$,
and is it accessible in current experiments? A nice analysis shows \cite{Ma:1998dn} that,
if the operator is a tree
level effect of some high-scale fundamental physics, there are three and only three ways to
realize it.
It is amusing that they correspond exactly to the three types of conventional seesaws that were
suggested previously from different points of view \cite{type1,type2,Foot:1988aq}.
While completely equivalent as far as the neutrino mass at low energies is concerned, these seesaws
are indeed vastly different at high energies. The issue becomes whether they are discernable
in the current or near-future experiments, in particular at the Large Hadron Collider (LHC).

The LHC physics of the three seesaws has been explored in this spirit. The type I seesaw introduces singlet neutrinos whose impact on SM physics enters mainly through their mixing
with the SM neutrinos (see, e.g., Ref.
\cite{Han:2006ip,Atre:2009rg,del Aguila:2007em,Kersten:2007vk,Dev:2013wba}).
This is generically very hard to detect since an appreciable mixing clashes apparently with the desire of tiny neutrino mass and not too heavy new particles.
The seesaw has thus been studied in an effective sense, namely by
decoupling the correlation between the heavy mass and mixing parameters that would appear in a genuine type I seesaw.  In type II seesaw one assumes a scalar triplet carrying the same hypercharge as the SM Higgs. The tininess of neutrino mass is
then shared by the vacuum expectation value (VEV) of the triplet and its Yukawa coupling to leptons.
The phenomenology of type II seesaw is rich, and has been extensively investigated
in the literature
\cite{Han:2007bk,Perez:2008ha,Garayoa:2007fw,Akeroyd:2007zv}.
The type III seesaw attributes the tiny neutrino mass to the mixing with heavy triplet fermions of zero hypercharge. This seesaw is potentially rich in phenomenology but more involved than
type II, and has been studied in
\cite{Franceschini:2008pz,Li:2009mw,Biggio:2011ja}.
A comparative study has been made on all three seesaws in Ref. \cite{delAguila:2008cj}.
Both CMS and ATLAS groups at LHC have set constraints on those seesaws based on
various simplifying assumptions
\cite{Chatrchyan:2012fla,ATLAS:2012yoa,Chatrchyan:2012ya,ATLAS:2012hi,CMS:2012ra,ATLAS:2013hma},
which will be briefly discussed in section \ref{sec:production}.
Due to the residual tension in the conventional seesaws between heavy mass and small couplings,
there are attempts that combine them in one way or another, or include additional structures;
see for example, Ref. \cite{Mohapatra:1986aw} for the inverse seesaw as an
alternative realization of the type I seesaw and Refs. \cite{Malinsky:2005bi,Dev:2012sg} for
its variants. For a partial list of more references, see,
\cite{Keung:1983uu,Arhrib:2009mz,Bajc:2007zf,Han:2012vk,BarShalom:2008gt,Perez:2009mu,
delAguila:2013mia,delAguila:2008cj,Dev:2013oxa}.

Because of the tension mentioned above, it is natural to try to go beyond the conventional seesaws.
One approach is to diminish the effect of the operator $\calO_5$ by associating it with
additional factors of couplings and loop factors. This is achieved usually by employing heavy particles that carry new exact or softly broken discrete symmetries,
so that the operator can only be induced at the loop level
\cite{Zee:1980ai,Krauss:2002px,Ma:2006km}.
This helps to alleviate the tendency to accommodate tiny neutrino
mass by inaccessibly heavy particles and minute couplings to SM particles. An interesting
example in this context is the so-called color-octet model \cite{FileviezPerez:2009ud}
in which neutrinos gain mass by interacting with new colored particles that could be detectable
\cite{FileviezPerez:2010ch} at LHC.

A second approach to relax the tension is to raise the dimension of leading operators that are
responsible for the neutrino mass, so that they are naturally suppressed by more factors of a
high scale. We recall that regarding SM as an effective field theory the neutrino mass
operators are unique at each dimension and have the simple form,
$\calO_{5+2n}=\calO_5(\phi^\dagger\phi)^n$, where $n$ is a positive integer \cite{Liao:2010ku}.
For this to work at the tree level, one has to appeal to heavy particles that constitute a higher
dimensional irreducible representation of the SM gauge group
\cite{Babu:2001ex,Babu:2009aq,Liao:2010cc}.
It has been shown in Ref. \cite{Liao:2010cc} that, by choosing the representations judiciously, a higher-dimensional operator can be induced in a systematical and
economical manner: the seesaw operates step by step through a cascading process, with each step
offering certain amount of suppression.
Such models typically contain multiply charged particles that have characteristic decay modes
into like-sign multiple leptons or $W^\pm$ bosons, which could be utilized to remove the SM
backgrounds. The purpose of this work is to explore the feasibility of detecting new particles
in the minimal version of the cascade seesaw models.

In the next section we outline the basic idea of the cascade seesaw and describe in detail
its minimal version whose phenomenology will be investigated in the remainder of the paper.
In section \ref{sec:LFV}, we work out the lepton flavor violating (LFV) transitions at
low energies that are induced in the model. They turn out to set stringent constraints on
the model parameters, and make our analysis of LHC physics more realistic. The decays of new particles,
their production and detection channels at LHC are studied in section \ref{sec:production}.
A brief summary of our main results is recapitulated in the last section. Some details of the
model, decays, and loop functions are reserved to several appendices.

\section{The model}
\label{sec:model}

There are too many possible ways to introduce new fields in order to induce a higher dimensional
neutrino mass operator at the tree level.
It was proposed in Ref. \cite{Liao:2010cc} to use as our criteria the following points.
First, for a given set of fields, we assume that the lowest dimension operator $\calO_{5+2n}$
dominates the neutrino mass; namely, we do not consider accidental cancellations in the couplings
associated with the mass operators. Second, for a given mass operator, we employ as few new fields
as possible to realize it. And finally, we do not impose any symmetry other than the SM gauge
symmetries.
After a careful analysis, the consequences turn out to be very simple \cite{Liao:2010cc}.
Both new scalars $\Phi$ and fermions $\Sigma$ are necessary to go beyond the three conventional
seesaws. And the possibilities are classified according to whether the SM Higgs $\phi$ couples to
the new fermions $\Sigma$ or not. If it does, the option is unique -- we need a fermion of weak
isospin and hypercharge $(I,Y)=(1,2)$ and a scalar of $(I,Y)=(3/2,3)$.
This is the model composed earlier in Ref. \cite{Babu:2009aq} which yields the operator $\calO_7$
for the neutrino mass.
If $\phi$ does not couple to $\Sigma$, the result is a class of models \cite{Liao:2010cc}.
We need one fermion $\Sigma$ with $(I,Y)=(n+1,0)$ with integer $n\ge 1$ and a sequence of scalars
$\Phi^{(m+1/2)}$ of $(I,Y)=(m+1/2,1)$ with $m=1,~2,~\dots,n$. The SM gauge symmetries dictate that
only the scalar of the highest isospin, $\Phi^{(n+1/2)}$, can have a Yukawa coupling to the fermions $\Sigma$
and $F_L$, and that only the scalar of the lowest isospin, $\Phi^{(3/2)}$, can develop a naturally
small VEV from interactions with the SM Higgs $\phi$.
The VEV is then transmitted by a cascading procedure from $\Phi^{(3/2)}$ up to $\Phi^{(n+1/2)}$
through scalar interactions, at each step earning an additional suppression from heavy scalar masses.
The end result is a neutrino mass operator $\calO_{5+4n}$ that is multiplied by the square of
the VEV and Yukawa coupling of $\Phi^{(n+1/2)}$.

In this work, we focus on the minimal version of the cascade seesaw; i.e., we introduce one scalar
$\Phi$ with $(I,Y)=(3/2,1)$ and one fermion $\Sigma$ with $(I,Y)=(2,0)$, whose members are
\begin{eqnarray}
\Phi=(\Phi_{+2},\Phi_{+1},\Phi_0,\Phi_{-1}),~
\Sigma=(\Sigma_{+2},\Sigma_{+1},\Sigma_0,\Sigma_{-1},\Sigma_{-2}),
\label{fields}
\end{eqnarray}
where the subscripts refer to the electric charge. The SM Higgs doublet and lepton fields are,
\begin{eqnarray}
\phi=(\phi_+,\phi_0),~F_L=(n_L,f_L),~f_R,
\end{eqnarray}
where the subscripts $L,~R$ denote the chirality.
We describe some details of the model in the remainder of this section.

\subsection{Scalars}

The complete scalar potential is
\begin{eqnarray}
V&=&-\mu_\phi^2\phi^\dagger\phi+\lam_\phi(\phi^\dagger\phi)^2+\mu_\Phi^2\Phi^\dagger\Phi%
\nonumber
\\
&&
-\lambda_1(\Phi\tilde\Phi)_0(\phi\tilde\phi)_0%
-\lambda_2\big((\Phi\tilde\Phi)_1(\phi\tilde\phi)_1\big)_0%
+\lambda_3\big((\Phi\Phi)_1(\tilde\Phi\tilde\Phi)_1\big)_0%
+\lambda_4\big((\Phi\Phi)_3(\tilde\Phi\tilde\Phi)_3\big)_0%
\nonumber
\\
&&
-\big[\kappa_1(\Phi\tilde\phi\phi\tilde\phi)_0+\textrm{h.c.}\big]%
-\big[\kappa_2\big((\Phi\Phi)_1(\tilde\phi\tilde\phi)_1\big)_0
+\textrm{h.c.}\big]%
-\big[\kappa_3\big((\Phi\Phi)_1(\tilde\Phi\tilde\phi)_1\big)_0
+\textrm{h.c.}\big],
\end{eqnarray}
where
$\tilde\Phi=(\Phi_{-1}^*,-\Phi_0^*,\Phi_{+1}^*,-\Phi_{+2}^*)$ and
$\tilde\phi=(\phi_0^*,-\phi_+^*)$ transfer under weak isospin precisely as $\Phi$ and $\phi$
respectively. The subscript to a pair of parentheses refers to the isospin of the normalized
product inside which is obtained by Clebsch-Gordan coefficients. The couplings
$\kappa_{1,2,3}$ are generally complex and the other parameters are real.
Together with the Yukawa couplings to be discussed in the next subsection, the $\kappa_{1,3}$
terms violate the lepton number by one unit and the $\kappa_2$ term by two units, thus it looks
plausible to assume $\kappa_2\sim\kappa_{1,3}^2$.
For simplicity, we will assume when diagonalizing the scalar masses that
$\kappa_1\approx\kappa_3\approx\kappa$ and $\kappa_2\approx\kappa^2$ with $\kappa$ being real.
We assume $\mu^2_\phi>0$ and $\mu_\Phi^2>0$ so that $\Phi$ can only develop a naturally small
VEV out of that of $\phi$. For small $\kappa$'s and perturbative $\lam$'s, they are found to be
\begin{eqnarray}
  v_\phi\approx\sqrt{\frac{\mu^2_\phi}{2\lambda_\phi}},~
  v_\Phi\approx\frac{\kappa_1 v_\phi}{2\sqrt{3}r_\Phi},
  \label{vPhi}
\end{eqnarray}
where $r_\Phi=\mu_\Phi^2/v_\phi^2+\lambda_1/(2\sqrt{2})+\lambda_2/(2\sqrt{30})$.
Inspection of the second derivatives of $V$ confirms that this is indeed the correct vacuum.

The neutral and singly-charged members of $\phi$ and $\Phi$ mix respectively due to the small
$\kappa$ terms. Denoting $X_0=(\rmRe X_0+i\rmIm X_0)/\sqrt{2}$ for
$X=\Phi,~\phi$, the mass matrices for $(\rmRe\Phi_0,\rmRe\phi_0)$ and
$(\rmIm\Phi_0,\rmIm\phi_0)$ are diagonalized approximately by the mixing angles respectively,
\begin{eqnarray}
  \theta_R&\approx&\frac{\sqrt{2}\lambda_1+2\lambda_2/\sqrt{30}
  -6r_\Phi}{4\sqrt{3}r_\Phi(r_\Phi-4\lambda_\phi)}\kappa,~
  \theta_I~\approx~-\frac{1}{2\sqrt{3}r_\Phi}\kappa.
\end{eqnarray}
The physical states are the CP-even $H_0,~h$ and the CP-odd $A_0$ respectively. Their masses
are only modified by $O(\kappa^2)$ terms, which are safely ignored. The other state from the imaginary
fields is the would-be Goldstone field
$G^0=\rmIm\phi_0+\kappa/(2\sqrt{3}r_\Phi)\rmIm\Phi_0$.
For the singly charged fields, $(\Phi_{+1},\Phi_{-1}^*,\phi_+)$, noting that there is no mixing
between $\Phi_{+1}$ and $\Phi_{-1}^*$, the $\Phi_{+1}-\phi_+$ and $\Phi_{-1}^*-\phi_+$ mixing is
diagonalized by the angle $\omega$ and $\varpi$ respectively,
\begin{eqnarray}
  \omega\approx-\frac{1}{\sqrt{3}r_\Phi}\kappa,~
  \varpi~\approx~\frac{1}{2r_\Phi}\kappa.
\end{eqnarray}
Two of the eigenstates have approximately a mass of $\Phi_{+1}$ and $\Phi_{-1}^*$, and the third
one is the would-be Goldstone
$G^+=\phi_+ +\kappa/(\sqrt{3}r_\Phi)\Phi_{+1}-\kappa/(2r_\Phi)\Phi_{-1}^*$.

The gauge covariant derivative is standard,
\begin{eqnarray}
D=\partial-i\frac{g_2}{\sqrt{2}}(I_+W^+ +I_-W^-)-i\frac{g_2}{c_W}(I_3-Qs_W^2)Z-ieQA,
\end{eqnarray}
with the usual notations, $s_W=\sin\theta_W,~c_W=\cos\theta_W$. For a field of weak isospin
$3/2$ and unity hypercharge like $\Phi$, one has
\begin{eqnarray}
I_+^\dag=I_-=\begin{pmatrix}
    0&&&\\ \sqrt{3}&0&&\\ &2&0&\\&&\sqrt{3}&0
  \end{pmatrix},~
I_3=\text{diag}\left(\frac{3}{2},\frac{1}{2},\frac{-1}{2},\frac{-3}{2}\right),~
Q=\text{diag}\left(2,1,0,-1\right).
\end{eqnarray}
The masses of the $W$ and $Z$ bosons are modified by $O(\kappa^2)$ terms
\begin{eqnarray}
  m_W\approx\frac{g_2v_\phi}{\sqrt{2}}\left(1+\frac{7}{24r_\Phi^2}\kappa^2\right),~
  m_Z\approx\frac{g_2v_\phi}{\sqrt{2}}\left(1+\frac{1}{24r_\Phi^2}\kappa^2\right),
\end{eqnarray}
resulting in a negligible deviation from unity in the $\rho$ parameter,
$\rho-1\approx\kappa^2/(4r_\Phi^2)$.
The gauge couplings of $\Phi$ are recorded in Appendix A.

\subsection{Fermions}

We employ here a vector-like fermion $\Sigma$ with both left-handed and right-handed chiralities.
Denoting its Dirac-barred field by $\tilde\Sigma$ that transfers under isospin as $\Sigma$ itself and
has the components in the order of descendent $I_3$,
\begin{eqnarray*}
\overline{\Sigma_{-2}},~-\overline{\Sigma_{-1}},~\overline{\Sigma_{0}},~
-\overline{\Sigma_{+1}},~\overline{\Sigma_{+2}},
\end{eqnarray*}
the Yukawa couplings involving $\Phi$ are
\begin{eqnarray}
  -\calL_\Phi^\text{Yuk}&=&2\sqrt{5}\left[x_i(\overline{F^C_{Li}}\Phi\Sigma)_0
  +z_i(\tilde\Sigma\Phi F_{Li})_0+\text{h.c.}\right],
\end{eqnarray}
where the sum over generation $i=1,~2,~3$ is understood. Redefining the fermion fields,
\begin{eqnarray}
&&  \Sigma^0_{1L}=\frac{1}{\sqrt{2}}\big(\Sigma_{0L}+\Sigma_{0R}^C\big),
  ~\Sigma^0_{2L}=\frac{i}{\sqrt{2}}\big(\Sigma_{0L}-\Sigma_{0R}^C\big),
  \nonumber\\
&&  \Sigma^{-}_{1}=\frac{1}{\sqrt{2}}\big(\Sigma_{-1}-\Sigma_{+1}^C\big),
~\Sigma^{-}_{2}=\frac{i}{\sqrt{2}}\big(\Sigma_{-1}+\Sigma_{+1}^C\big),
  \nonumber\\
&&  \Sigma^{--}_{1}=\frac{1}{\sqrt{2}}\big(\Sigma_{-2}+\Sigma_{+2}^C\big),
~\Sigma^{--}_{2}=\frac{i}{\sqrt{2}}\big(\Sigma_{-2}-\Sigma_{+2}^C\big),
  \label{lepton.redefine}
\end{eqnarray}
the Yukawa couplings can be rewritten as
\begin{eqnarray}
  -\calL_\Phi^\text{Yuk}&=&
  \sum_{m=-2}^{+2}Y_{ix}^m\left[\sqrt{2+m}\Phi_m\overline{\Sigma_x^m}P_Ln_i
  +\sqrt{2-m}\Phi_{m+1}\overline{\Sigma_x^m}P_Lf_i\right]+\text{h.c.}~.
  \label{Yukawa}
\end{eqnarray}
where the sum over $x=1,~2$ is also implied and the new $3\times 2$ Yukawa coupling matrices
$Y^m$ are
\begin{eqnarray}
  &&Y^{+2}=Y^{-2}=-Y^{+1}=Y^{-1}=Y^{0}=\frac{1}{\sqrt{2}}[(x+z),i(z-x)].
\end{eqnarray}

Including the SM Yukawa couplings
\begin{eqnarray}
  -\calL_\phi^\text{Yuk}=(y_\phi)_{ij}\overline{F_{Li}}\phi f_{Rj}+\text{h.c.},
\end{eqnarray}
and a bare mass for $\Sigma$,
the fermion mass terms read
\begin{eqnarray}
  -\calL_m=\frac{1}{2}\overline{N_L}M_NN_R+\overline{E_L}M_EE_R
  +\overline{D_L}M_DD_R+\text{h.c.},
\end{eqnarray}
where the mass matrices are
\begin{eqnarray}
  &&M_N=\begin{pmatrix}
    0_3&(x+z)^*v_\Phi&i(x-z)^*v_\Phi\\
    (x+z)^\dag v_\Phi&M_\Sigma&0\\
    i(x-z)^\dag v_\Phi&0&M_\Sigma
  \end{pmatrix},
\nonumber\\
  &&M_E=\begin{pmatrix}
    y_\phi v_\phi&\sqrt{3/2}(x+z)^*v_\Phi&i\sqrt{3/2}(x-z)^*v_\Phi\\
    0&M_\Sigma&0\\
    0&0&M_\Sigma
  \end{pmatrix},
\nonumber\\
  &&M_D=M_\Sigma \mathbf{1}_2,
\end{eqnarray}
in the basis
\begin{eqnarray}
  N_L=\begin{pmatrix}
    n_L\\\Sigma^0_{1L}\\
    \Sigma^0_{2L}
  \end{pmatrix},~
  N_R=N_L^C,~
  E=\begin{pmatrix}
    f\\\Sigma^{-}_{1}\\\Sigma^{-}_{2}
  \end{pmatrix},~
  D=\begin{pmatrix}
    \Sigma_1^{--}\\\Sigma_2^{--}
  \end{pmatrix}.
  \label{lepton.fields}
\end{eqnarray}

\subsubsection{Diagonalization of fermion mass matrices}

As we will see later, the mixing between heavy and light charged particles
is tiny, with negligible corrections to their mass eigenvalues. We first diagonalize the
submatrix for light charged leptons by bi-unitary transformations,
$F_L\to\calU_LF_L,~f_R\to\calU_Rf_R$, so that
\begin{eqnarray}
  \calU_L^\dag v_\phi y_\phi\calU_R=\text{diag}(m_e,m_\mu,m_\tau).
\end{eqnarray}
The transformations will change nothing else in the Lagrangian but the
Yukawa couplings,
\begin{eqnarray}
  \calU_L^\dag(x+z)^*=\frac{2\sqrt{3}r_\Phi}{v_\phi} \mathbf{u}_1,~~~~
  \calU_L^\dag i(x-z)^*=\frac{2\sqrt{3}r_\Phi}{v_\phi} \mathbf{u}_2,
  \label{coupling.trans}
\end{eqnarray}
where $\mathbf{u}_{1,2}$ are two three-component column vectors which are
determined by $x,~z$, and $\calU_L$. With the help of eq. (\ref{vPhi}),
the mass matrices become
\begin{eqnarray}
  M_N=\begin{pmatrix}
    0_3&\kappa U\\
    \kappa U^T&M_\Sigma\mathbf{1}_2
  \end{pmatrix},~~~~
  M_E=\begin{pmatrix}
    \text{diag}(m_e,m_\mu,m_\tau)
    &\sqrt{3/2}\kappa U\\
    0&M_\Sigma\mathbf{1}_2
  \end{pmatrix},
\end{eqnarray}
where $U=(\mathbf{u}_1,\mathbf{u}_2)$.

To diagonalize $M_N$, we first deal with the light-heavy mixing,
\begin{eqnarray}
  (U_N^\text{l-h})^\dag M_N (U_N^\text{l-h})^\ast=
  \begin{pmatrix}
    M^v_\text{light}&\\&M_\Sigma 1_2
    \end{pmatrix},
\end{eqnarray}
by
\begin{eqnarray}
  U_N^\text{l-h}=
  \begin{pmatrix}
  \mathbf{1}_3-\kappa^2\frac{UU^\dag}{2M_\Sigma^2}&\kappa U/M_\Sigma\\
  -\kappa U^\dag/M_\Sigma&\mathbf{1}_2-\kappa^2\frac{U^\dag U}{2M_\Sigma^2}
  \end{pmatrix},
\end{eqnarray}
so that
\begin{eqnarray}
  M^v_\text{light} = -ZZ^T,
\end{eqnarray}
where $Z=(\mathbf{z}_1,\mathbf{z}_2)$ is a $3\times2$
matrix in terms of the column vectors
$\mathbf{z}_{1,2}=\kappa \mathbf{u}_{1,2}/\sqrt{M_\Sigma}$.
The matrix $M_\text{light}^\nu$ can be diagonalized by the PMNS matrix,
\begin{eqnarray}
  U_\text{PMNS}^\dag M_\text{light}^\nu U_\text{PMNS}^\ast
  =\text{diag}(m_{\nu_1},m_{\nu_2},m_{\nu_3}),
\end{eqnarray}
with tiny corrections to the definition of $U_\text{PMNS}$ from the heavy-light mixing of
singly charged fermions. Now we employ an algebraic trick in Ref. \cite{Liao:2009fm} to solve
$Z$ in terms of $U_\text{PMNS}$, $m_{\nu_i}$ and free physical parameters.
Writing $U_\text{PMNS}=(\xvec_1,\xvec_2,\xvec_3)$ and noting that one of the light neutrinos
is massless in the considered model \cite{Liao:2010cc},
we can parameterize $Z$ for normal hierarchy (NH) and inverted hierarchy (IH) of neutrino mass,
\begin{eqnarray}
\text{NH:}&&m_{\nu_1}=0,~m_{\nu_2}=\lambda_-,~m_{\nu_3}=\lambda_+,
\nonumber\\
&&\zvec_1=c_-\xvec_2+c_+\xvec_3,~\zvec_2=d_-\xvec_2+d_+\xvec_3,
\nonumber\\
\text{IH:}&&m_{\nu_3}=0,~m_{\nu_1}=\lambda_-,~m_{\nu_2}=\lambda_+,
\nonumber\\
&&\zvec_1=c_-\xvec_1+c_+\xvec_2, ~\zvec_2=d_-\xvec_1+d_+\xvec_2,
\label{matrix.Z}%
\end{eqnarray}
where $\lambda_+>\lambda_->0$ are the two non-zero mass eigenvalues.
The coefficients $c_\pm$, $d_\pm$
can be expressed in terms of the eigenvalues $\lam_\pm$ plus a free
complex parameter $t$. For both hierarchies, we have
\begin{eqnarray}
c_-=i\sqrt{\lambda_-}\frac{2t}{1+t^2},&&
d_-=i\sqrt{\lambda_-}\frac{1-t^2}{1+t^2},
\nonumber\\
c_+=i\sqrt{\lambda_+}\frac{1-t^2}{1+t^2},&&
d_+=-i\sqrt{\lambda_+}\frac{2t}{1+t^2}.
\label{c.and.d}%
\end{eqnarray}
The preceding matrices can now be determined in terms of $Z$. For instance,
\begin{eqnarray}
 U=\frac{\sqrt{M_\Sigma}}{\kappa}Z,
\end{eqnarray}
and the complete transformation matrix for neutral fermions,
$\nu_L=U_N^\dag N_L,~\nu_R=\nu_L^C$, reads
\begin{eqnarray}
  U_N=\begin{pmatrix}
    \left(\mathbf{1}_3-\frac{ZZ^\dag}{2M_\Sigma}\right)U_\text{PMNS}&
    \frac{Z}{\sqrt{M_\Sigma}}\\
    -\frac{Z^\dag U_\text{PMNS}}{\sqrt{M_\Sigma}}&
    \mathbf{1}_2-\frac{1}{2M_\Sigma}Z^\dag Z
  \end{pmatrix}.
\end{eqnarray}

The mass matrix $M_E$ for singly charged fermions is diagonalized
by bi-unitary transformation, $\ell_L=U_L^\dag E_L,~\ell_R=U_R^\dag E_R$, to
$M_\ell=\text{diag}(m_e,m_\mu,m_\tau,m_{\Sigma^-_1},m_{\Sigma^-_2})$.
$U_L$ diagonalizes $M_EM_E^\dag$, and is found in the limit of $M_\Sigma\gg m_\tau$, to be
\begin{eqnarray}
  U_L=\begin{pmatrix}
  \mathbf{1}_3-\frac{3}{4M_\Sigma}ZZ^\dag&\sqrt{\frac{3}{2M_\Sigma}}Z\\
  -\sqrt{\frac{3}{2M_\Sigma}}Z^\dag&\mathbf{1}_2-\frac{3}{4M_\Sigma}Z^\dag Z
  \end{pmatrix}.
  \label{UL}
\end{eqnarray}
Similarly, $U_R$ diagonalzies $M_E^\dag M_E$, and is found to be
\begin{eqnarray}
  U_R=\begin{pmatrix}
  \mathbf{1}_3-\frac{3}{4}\eta\eta^\dag
  &\sqrt{\frac{3}{2}}\eta
  \\
  -\sqrt{\frac{3}{2}}\eta^\dag &\mathbf{1}_2-\frac{3}{4}\eta^\dag\eta
  \end{pmatrix},
  \label{UR}
\end{eqnarray}
with $\eta=M_\Sigma^{-3/2}\text{diag}(m_e,m_\mu,m_\tau)Z$. The mixing is suppressed by
an additional factor of $m_{e,\mu,\tau}/M_\Sigma$ compared to $U_L$.

The above mixing matrices will enter into gauge interactions of the fermions as well as
Yukawa couplings. In the basis of eq. (\ref{lepton.fields}),
the weak and electromagnetic currents are
\begin{eqnarray}
J^{+\mu}_W&=&\overline{N}\gamma^\mu(w_LP_L+w_RP_R)E
+2\overline{E}\gamma^\mu w_DD,\nonumber\\
J^\mu_Z&=&\overline{N}\gamma^\mu z_L^NP_LN
+\overline{E}\gamma^\mu(z_L^EP_L+z_R^EP_R)E
-2c_W^2\overline{D}\gamma^\mu D,\nonumber\\
J^\mu_\text{em}&=&-\overline{E}\gamma^\mu E-2\overline{D}\gamma^\mu D,
\end{eqnarray}
where
\begin{eqnarray}
&&w_L=\left(\begin{array}{cc}\mathbf{1}_3&\\
&\sqrt{6}\mathbf{1}_2\end{array}\right),~
w_R=\left(\begin{array}{ccc}0_3&\\
&\sqrt{6}\mathbf{1}_2\end{array}\right),~
w_D=\begin{pmatrix}0_{3\times2}\\\mathbf{1}_2\end{pmatrix},\nonumber\\
&&z_L^N=\left(\begin{array}{cc}\frac{1}{2}\mathbf{1}_3&\\
&0_2\end{array}\right),~z_L^E=\left(\begin{array}{cc}\left(-\frac{1}{2}+s_W^2\right)\mathbf{1}_3&\\
&-c_W^2\mathbf{1}_2\end{array}\right),~~~
z_R^E=\left(\begin{array}{cc}s_W^2\mathbf{1}_3&\\
&-c_W^2\mathbf{1}_2\end{array}\right).
\end{eqnarray}
In terms of the mass eigenstates, they become
\begin{eqnarray}
J^{+\mu}_W&=&\overline{\nu}\gamma^\mu(\calW_LP_L+\calW_RP_R)\ell
+2\overline{\ell}\gamma^\mu(\calW_L^DP_L+\calW_R^DP_R)D,\nonumber\\
J^\mu_Z&=&\overline{\nu}\gamma^\mu\calZ_L^\nu P_L\nu
+\overline{\ell}\gamma^\mu(\calZ_L^\ell P_L+\calZ_R^\ell P_R)\ell
-2c_W^2\overline{D}\gamma^\mu D,\nonumber\\
J^\mu_\text{em}&=&-\overline{\ell}\gamma^\mu
\ell-2\overline{D}\gamma^\mu D,
\label{eq_currents}
\end{eqnarray}
where the matrices $\calW_L$ etc are given in Appendix A.
As can be seen from there, the flavor changing neutral currents (FCNC) of the light
charged leptons are suppressed by a factor of light neutrino mass over $M_\Sigma$ for
the left-handed chirality and by even an additional factor of light charged lepton mass
over $M_\Sigma$ for the right-handed chirality.

\section{Lepton flavor violating transitions}
\label{sec:LFV}

We will study the LHC production and detection of new particles in section
\ref{sec:production}. To make this realistic, we have to consider the constraints
that are already available on the new interactions. As we will show in this section,
the precise measurements in LFV transitions indeed set strong bounds on the relevant
couplings.

Since the deviation from SM gauge interactions is significantly suppressed by
a small mass ratio $\sqrt{m_\nu/M_\Sigma}$ or even more, we focus on
the new Yukawa couplings. With the fields redefined
in eq. (\ref{lepton.redefine}) and the transformations in eq. (\ref{coupling.trans}),
the relevant Yukawa couplings are written as
\begin{eqnarray}
  -\calL_\Phi^\text{Yuk}&\supset&
  \sum_{m=-2}^{+1}\sqrt{2-m}Y^m_{ix}~\Phi_{m+1}\overline{\Sigma^m_x}P_L\ell_i+\text{h.c.},
\end{eqnarray}
where now
\begin{eqnarray}
  Y^{-2}=Y^{-1}=Y^{0}=-Y^{+1}=\frac{\sqrt{M_\Sigma}}{\sqrt{2}v_\Phi}Z^\ast,
\end{eqnarray}
and further mixing of the SM charged leptons $\ell_i$ can be safely ignored as we discussed
in the last section.

\subsection{Radiative transitions and electromagnetic dipole moments}

\begin{figure}[ht]
\centering
\resizebox{.9\textwidth}{!}{%
\includegraphics{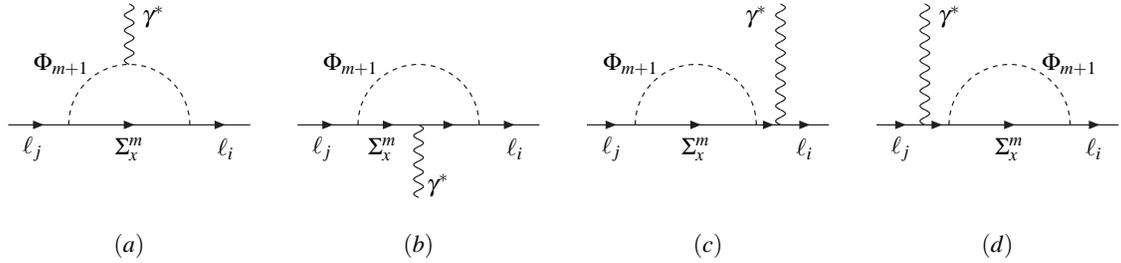}}
\caption{Diagrams for radiative transitions.} %
\label{radiative.diag}
\end{figure}

The diagrams contributing to radiative transitions are shown in
Fig. \ref{radiative.diag}. Ignoring the mass splitting among $\Sigma$'s (and $\Phi$'s) of
various charges and working in the small mass limit of light charged leptons,
a calculation similar to that in Ref. \cite{Liao:2009fm} yields the amplitude for
the process $\ell_j(p)\to\ell_i(p-q)\gamma^\ast(q)$,
\begin{eqnarray}
  \calA^m_\mu(\ell_j\to\ell_i\gamma^\ast)&=&
  \frac{-e(ZZ^\dag)_{ij}}{2(4\pi)^2v_\Phi^2M_\Sigma}
  \bar{u}_i(p-q)\Big[F_m(r)(P_Rm_j+P_Lm_i)i\sigma_{\mu\nu}q^\nu\nonumber\\
  &&\hspace{4em}+G_m(r)P_R(q^2\gamma_\mu-\qslash q_\mu)\Big]u_j(p),
\end{eqnarray}
where $m=-2,\dots,+1$ refers to the charge of the virtual $\Sigma^m$ in the loop, and
the functions are
\begin{eqnarray}
  F_m(r)&=&r(2-m)[(m+1)F^a(r)-mF^b(r)],
  \nonumber\\
  G_m(r)&=&r(2-m)[(m+1)G^a(r)-mG^b(r)],
\end{eqnarray}
with $r=M_\Sigma^2/M_\Phi^2$ and the functions $F^{a,b}$ and $G^{a,b}$ are given in
Appendix B.

For on-shell transitions, only the dipole term survives and yields the branching ratio
\begin{eqnarray}
  \Br(\ell_j\to\ell_i\gamma)=
  \Br(\ell_j\to\ell_i\bar{\nu}_i\nu_j)\times
  \frac{3\alpha\left|(ZZ^\dag)_{ij}\right|^2}{64\pi G_F^2 v_\Phi^4M_\Sigma^2}
  \left[\sum_{m=-2}^1F_m(r)\right]^2,
\label{mu2egamma}%
\end{eqnarray}
and the contribution to the anomalous magnetic moment of $\ell_i$ is obtained as a by-product,
\begin{eqnarray}
a(\ell_i)=\frac{m_i^2(ZZ^\dag)_{ii}}{(4\pi)^2v_\Phi^2M_\Sigma}\sum_{m=-2}^1 F_m(r).
\end{eqnarray}

\subsection{Purely leptonic decays}

With three generations of SM leptons, there are three possible types of purely leptonic
transitions:
\begin{eqnarray}
(1)&&\ell_l(p)\to\ell_i(k_1)\ell_i(k_2)
\bar\ell_k(k_3),~\ell_i\ne\ell_k,
\nonumber\\
(2)&&\ell_l(p)\to\ell_i(k_1)\ell_j(k_2)
\bar\ell_j(k_3),~\ell_i\ne\ell_j,
\nonumber\\
(3)&&\ell_l(p)\to\ell_i(k_1)\ell_i(k_2)
\bar\ell_i(k_3).
\end{eqnarray}
As we discussed at the end of sec \ref{sec:model}, FCNC at tree level contributes little to the transitions because of strong suppression in both chiralities.
We thus focus on the loop contributions due to new Yukawa couplings. The calculation is
again similar to that in Ref. \cite{Liao:2009fm} for the color-octet model.

\begin{figure}[ht]
\centering
\resizebox{.25\textwidth}{!}{%
\includegraphics{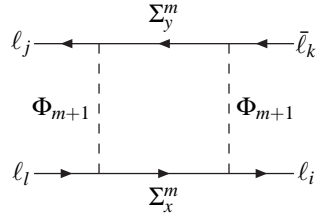}}
\caption{Additional box diagram for purely leptonic transitions.} %
\label{box.diag}
\end{figure}

The type-(1) decay is contributed only by the box diagram in Fig. \ref{box.diag},
\begin{eqnarray}
\calA^m(1)=\frac{-(ZZ^\dag)_{il}(ZZ^\dag)_{ik}}{2(4\pi)^2v_\Phi^4}
  \left[\bar u(k_1)\gamma_\mu P_Lu(p)\right]
  \left[\bar u(k_2)\gamma^\mu P_Lv(k_3)\right]H_m(r),
\end{eqnarray}
where a factor of 2 has been attached since the two minus signs from identical fermions in the
final state and from the Fierz identity cancel each other, and $H_m(r)=(2-m)^2H(r)$ with $H(r)$  given in Appendix B. A summation over the fermion charge $m$ is always implied in
the amplitude.

For type-(2) decay, both radiative and box diagrams contribute,
\begin{eqnarray}
\calA^m(2)=\calA^m_\triangle+\calA^m_\Box,
\end{eqnarray}
where
\begin{eqnarray}
\calA^m_\triangle&=&%
\frac{\alpha(ZZ^\dag)_{il}}{8\pi v_\Phi^2M_\Sigma}%
\bar u(k_2)\gamma^\mu v(k_3)
\nonumber
\\
&&\times\bar u(k_1)%
\Big[F_m(r)(P_Rm_l+P_Lm_i)i\sigma_{\mu\nu}(k_2+k_3)^\nu
s_{23}^{-1}+G_m(r)P_R\gamma_\mu\Big]u(p),\nonumber\\
\calA^m_\Box&=&%
\frac{-1}{4(4\pi)^2v_\Phi^4}\Big[(ZZ^\dag)_{il}(ZZ^\dag)_{jj}
+(ZZ^\dag)_{jl}(ZZ^\dag)_{ij}\Big]
\nonumber
\\
&&\times\bar u(k_2)\gamma^\mu P_Lv(k_3)~\bar u(k_1)\gamma_\mu P_Lu(p)H_m(r),
\label{eq_decay2}
\end{eqnarray}
with $s_{ij}=(k_i+k_j)^2$.
Finally, the type-(3) decay arises as a special case of type-(2), with all leptons in the
final state of the same flavor,
\begin{eqnarray}
\calA^m(3)=\left.\calA^m(2)\right|_{j=i}-(k_1\leftrightarrow k_2).
\label{eq_A3}
\end{eqnarray}

The branching ratios for all three types of decays are worked out using the phase space
integrals computed in Ref. \cite{Liao:2009fm} to be,
\begin{eqnarray}
\Br(1)&=&\Br(\ell_l\to\ell_i\nu_l\bar\nu_i)
\frac{\left|(ZZ^\dag)_{il}(ZZ^\dag)_{ik}\right|^2}
{2^{14}\pi^4v_\Phi^8G_F^2}\left[\sum_{m=-2}^1H_m(r)\right]^2,
\\
\Br(2)&=&\Br(\ell_l\to\ell_i\nu_l\bar\nu_i)
\frac{1}{2^{13}\pi^4v_\Phi^4G_F^2}\bigg\{
\left|\sum_m(B^m+T_1^m)\right|^2+\left|\sum_mT_1^m\right|^2
\nonumber\\
&&-4\textrm{Re}\left(\sum_mB^mT_2^{m*}\right)
-8\textrm{Re}\left(\sum_mT^m_1T_2^{m*}\right)+\Big[-\frac{14}{3}
+8\ln\frac{m_l^2}{4m_j^2}\Big]\left|\sum_mT^m_2\right|^2
\bigg\},
\\
\Br(3)&=&\Br(\ell_l\to\ell_i\nu_l\bar\nu_i)
\frac{1}{2^{13}\pi^4v_\Phi^4G_F^2}\bigg\{
2\left|\sum_m(B^m+T_1^m)\right|^2+\left|\sum_mT_1^m\right|^2
\nonumber\\
&&-8\textrm{Re}\left(\sum_mB^mT_2^{m*}\right)
-12\textrm{Re}\left(\sum_mT^m_1T_2^{m*}\right)+\Big[-\frac{8}{3}
+8\ln\frac{m_l^2}{4m_i^2}\Big]\left|\sum_mT^m_2\right|^2
\bigg\},
\label{mu23e}
\end{eqnarray}
where for type-(2) decay,
\begin{eqnarray}
B^m&=&-\frac{1}{2v_\Phi^2}\Big[(ZZ^\dag)_{il}(ZZ^\dag)_{jj}
+(ZZ^\dag)_{jl}(ZZ^\dag)_{ij}\Big]H_m(r),
\nonumber\\
T_1^m&=&\frac{e^2(ZZ^\dag)_{il}}{M_\Sigma}G_m(r),
\nonumber\\
T_2^m&=&\frac{e^2(ZZ^\dag)_{il}}{M_\Sigma}F_m(r),
\label{eq_BT}
\end{eqnarray}
and for type-(3) decay $j=i$ is set in the above functions.

We note incidentally that the amplitude for the type-(1) decay also implies the effective
interaction for muonium-anti-muonium oscillation:
\begin{eqnarray}
\calL_\textrm{eff}=-G_{M\bar M}\bar\mu\gamma^\alpha P_Le
~\bar\mu\gamma_\alpha P_Le,
\end{eqnarray}
where the effective Fermi constant is,
\begin{eqnarray}
\frac{G_{M\bar M}}{\sqrt{2}}=\frac{\left[(ZZ^\dag)_{\mu e}\right]^2}
{4(4\pi)^2v_\Phi^4}\sum_mH_m(r).
\end{eqnarray}

\subsection{$\mu-e$ conversion in nuclei}

Our result in the previous subsections can also be applied to the $\mu-e$ conversion
in nuclei. The effective Lagrangian can be written as
\begin{eqnarray}
  -\calL_{\mu e}=\frac{4}{\sqrt{2}}
  \left(m_\mu A_R\bar e P_R\sigma^{\mu\nu}\mu F_{\mu\nu}+\hc\right)
  +\frac{1}{\sqrt{2}}\sum_{q=u,d,s}
  \left(g_{LV(q)} \bar e \gamma^\mu P_L\mu \bar q\gamma_\mu q+\hc\right),
\end{eqnarray}
where $F_{\mu\nu}$ is the electromagnetic field strength and the effective couplings are
\begin{eqnarray}
  A_R&=&\frac{\sqrt{2}e(ZZ^\dag)_{e\mu}}{16(4\pi)^2v_\Phi^2M_\Sigma}
  \sum_mF_m(r),
  \nonumber\\
  g_{LV(u)}&=&\frac{2}{3}\frac{\alpha(ZZ^\dag)_{e\mu}}{\sqrt{2}
  (4\pi)v_\Phi^2M_\Sigma}\sum_mG_m(r),
  \nonumber\\
  g_{LV(d,s)}&=&-\frac{1}{3}\frac{\alpha(ZZ^\dag)_{e\mu}}{\sqrt{2}(4\pi)v_\Phi^2M_\Sigma}
  \sum_mG_m(r).
\end{eqnarray}
Then the $\mu-e$ conversion branching ratio is given by \cite{Kitano:2002mt}
\begin{eqnarray}
  \Br(\mu^- N\to e^-N)=\frac{2|A_R D+\tilde g^{(p)}_{LV}V^{(p)}
  +\tilde g^{(n)}_{LV}V^{(n)}|^2}{\omega_\text{capt}},
\label{mueconv}%
\end{eqnarray}
where
\begin{eqnarray}
  \tilde g^{(p)}_{LV}&=&2g_{LV(u)}+g_{LV(d)}=
  \frac{\alpha(ZZ^\dag)_{e\mu}}{\sqrt{2}(4\pi)v_\Phi^2M_\Sigma}\sum_mG_m(r),\nonumber\\
  \tilde g^{(n)}_{LV}&=&g_{LV(u)}+2g_{LV(d)}=0,
\end{eqnarray}
and $D,~V^{(p)}$ and $V^{(n)}$ are overlap integrals which are numerically evaluated
together with the corresponding ordinary muon capture rate $\omega_\text{capt}$ \cite{Kitano:2002mt}.

\begin{figure}[ht]
\centering
\resizebox{.9\textwidth}{!}{%
\includegraphics{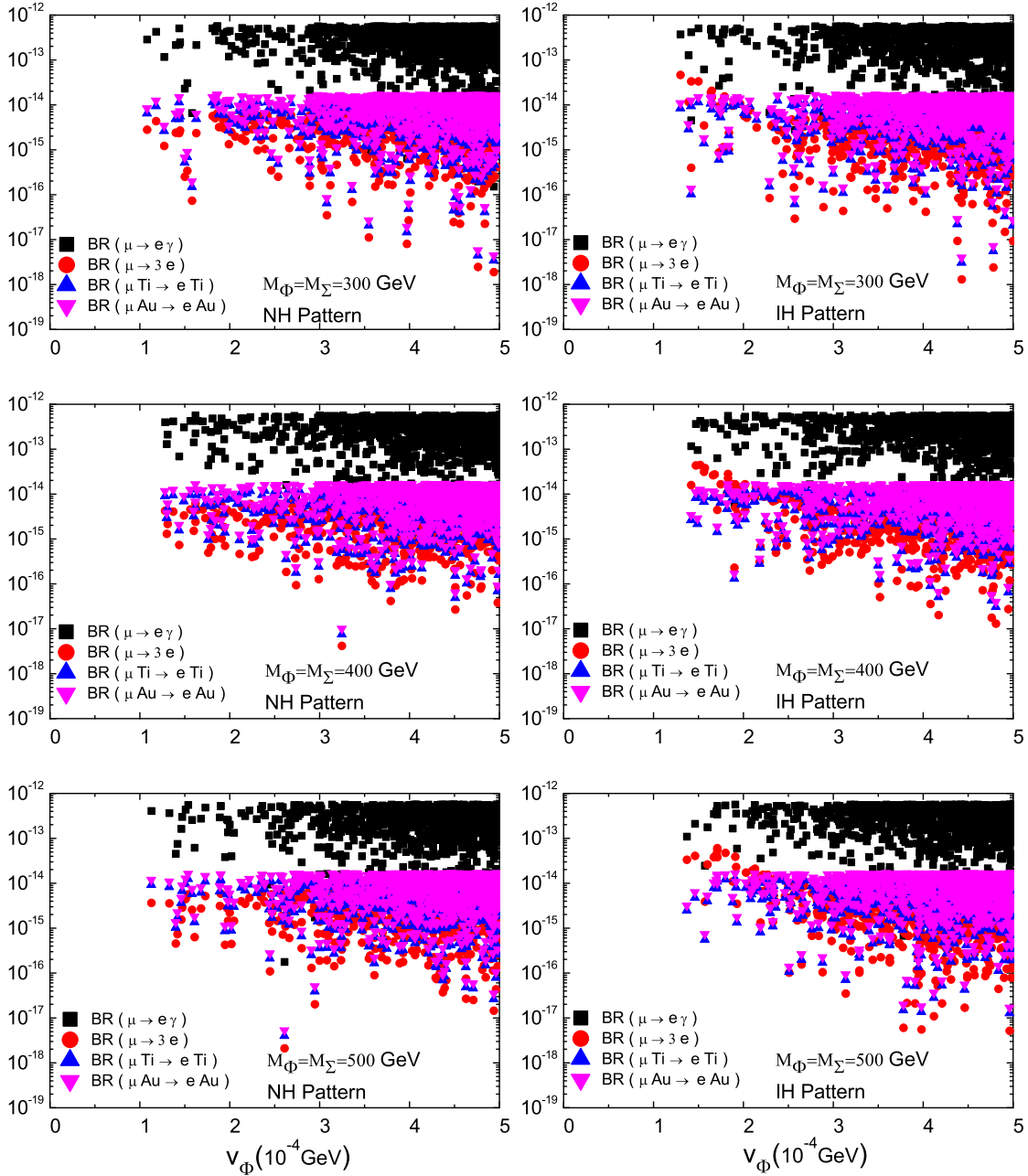}}
\caption{Points allowed by current bounds on LFV transitions are shown for $M_\Phi=M_\Sigma=300,~400,~500~\GeV$ and for NH and IH patterns in the small VEV range
$v_\Phi\lesssim 5\times 10^{-4}~\GeV$.} %
\label{LFV.Scatter}
\end{figure}

The most stringent upper bound on the radiative LFV decays comes from
$\Br(\mu\to e\gamma)<5.7\times10^{-13}$ ($90\%$ C.L.) \cite{Adam:2013mnn}.
Concerning the pure leptonic decays, the record is still held by the old result
$\Br(\mu\to 3e)<1.0\times10^{-12}$ ($90\%$ C.L.) \cite{Bellgardt:1987du}.
The current bound on $\mu-e$ conversion in nuclei also looks competitive,
$\Br(\mu^-\text{Ti}\to e^-\text{Ti})<4.3\times10^{-12}$ ($90\%$ C.L.) and
$\Br(\mu^-\text{Au}\to e^-\text{Au})<7\times10^{-13}$ ($90\%$ C.L.)
\cite{Beringer:1900zz}, while the current bound on the muonium-antimuonium
oscillation, $G_{M\bar M}\le 3.0\times 10^{-3}~G_F$ ($90\%$ C.L.)
\cite{Willmann:1998gd} is still too poor to be useful.
Using the analytical results in this section, we have made a
numerical scanning on the parameter regions that respect the above experimental
bounds and are potentially accessible at LHC.

The most relevant parameters for our later analysis of the LHC phenomenology are the masses
$M_{\Sigma,\Phi}$ of the new particles and the VEV $v_\Phi$. As usual, a larger mass tends
to suppress the LFV transitions. We scan parameters for relatively light, degenerate new particles that would be accessible at LHC, i.e., $M_\Sigma=M_\Phi=300,~400,~500~\GeV$.
Similarly, since a larger $v_\Phi$ implies smaller Yukawa couplings and is thus safe for LFV
transitions, we scan it in the range of smaller values, $v_\Phi\lesssim 5\times 10^{-4}~\GeV$. This is also the range in which the like-sign dilepton signals are important
at LHC. We allow the neutrino mass squared differences and mixing angles to assume
a random value in their $3\sigma$ ranges, set the unknown CP phases to zero and the parameter $t$ to take a value randomly. Then, the matrix $Z$ is fully determined by eqs. (\ref{matrix.Z}) and (\ref{c.and.d}) for both hierarchy patterns, and the branching ratios for the decays $\mu\to e\gamma,~3e$ and the $\mu-e$ conversion in nuclei can be evaluated with the help of eqs. (\ref{mu2egamma}), (\ref{mu23e}), and (\ref{mueconv}).
The scanning results passing all constraints are shown in Fig. \ref{LFV.Scatter}.

As one can see from Fig. \ref{LFV.Scatter}, the current bound on BR$(\mu\to e\gamma)$ sets the most stringent constraint. Respecting it implies that all other LFV transitions are well below their current bounds. The allowed lower bound on the VEV is about $v_\Phi\sim O(10^{-4})~\GeV$. Generally speaking, the bound should increase with heavy masses, though this is not very obvious in Fig. \ref{LFV.Scatter}. This is partly because the three values chosen for the masses are too close
and partly because of fluctuations in sampling the parameters with a limited number of points.
This does not affect our interest in a small $v_\Phi$ at LHC when
the like-sign dilepton signals are relevant. We will thus assume $v_\Phi=10^{-4}~\GeV$ for the purpose of illustration.
Since the light neutrino masses (and thus elements in $Z$) in the IH case are larger than in the NH case, the lower bound on $v_\Phi$ in the IH case is a bit larger. Finally, we note in passing that this lower bound on the scalar VEV from LFV transitions is much stronger than in the type II seesaw, see for example, Ref. \cite{Fukuyama:2009xk}. The reason for this
difference is clear.
Our neutrino mass arises from a dimension-nine operator, resulting in $m_\nu\sim (v_\Phi^2/M_\Sigma)YY^\dagger$ \cite{Liao:2010cc},
while it is from a dimension-five operator in the latter
case with $m_\nu\sim v_\Delta Y$, where $v_\Delta$ is the triplet VEV. For given $m_\nu$
and assuming a similar mass for all heavy particles, the LFV upper bounds on $Y$ translate
roughly into a relation between the lower bounds on the VEV's, $v_\Phi\sim v_\Delta(M_\Sigma/m_\nu)^{1/2}$.

\section{Collider phenomenology of cascade seesaw model}
\label{sec:production}

We explore in this section the collider signatures of the minimal version of the cascade seesaw detailed in the last section. Our analysis procedure is as follows.
We implement the model in the {\em Mathematica} package {\tt FeynRules1.7} \cite{Christensen:2008py}, whose output {\tt UFO} model file is taken by {\tt Madgraph5} \cite{Alwall:2011uj} to generate the parton level events for the relevant physical processes. Those events then pass through {\tt Pythia6} \cite{Sjostrand:2006za} to include the initial- and final-state radiation, fragmentation, and hadronization. We use {\tt PGS} for the detector simulation and {\tt MadAnalysis5} \cite{Conte:2012fm} for the analysis. In our simulation, we employ the {\tt CTEQ6L1} parton distribution function (PDF) \cite{Nadolsky:2008zw}.
Concerning the physical parameters, we recall that, using our
parametrization, all mixing matrices are expressed in terms of the two free parameters, the quadruplet VEV $v_\Phi$ and the complex parameter $t$. Together with the masses of the new particles, $M_\Phi$
and $M_\Sigma$, all production rates and decay widths are fixed. And to simplify the matter, we
assume that the scalars (fermions) of various charges are degenerate.
However, it is straightforward to include the non-degenerate case in our code and
we will leave this general case for another work.
The constraints from low energy processes are respected in our analysis of collider phenomenology, which allows us to set comprehensive bounds on the model in the future.
For the purpose of illustration, we often work with the benchmark parameter points, $M_\Phi=M_\Sigma=300~\GeV$, $t=1+i$, and $v_\Phi=10^{-4}$ or $10^{-2}~\GeV$.
Our numerical results are not particularly sensitive to the $t$ parameter except at the
singular points $t=\pm i$ where our parametrization (\ref{c.and.d}) is not appropriate.
For instance, LFV transitions are not affected by its magnitude being larger or small
than unity because the transformation $t\to -t^{-1}$ only flips the global sign of the matrix
$Z$ \cite{Liao:2009fm}.

The new physical particles in our model are,
the scalar quadruplet which includes as its members the neutral CP-even (-odd) $H_0$ ($A_0$),
the singly charged $\Phi_{-1}/\Phi_{-1}^{*},~\Phi_{+1}/\Phi_{+1}^{*}$, and the doubly charged $\Phi_{+2}/\Phi_{+2}^{*}$, and the fermion quintuplet which includes the neutral $\Sigma^{0}$, the singly charged $\Sigma^{\pm}$, and the doubly charged $\Sigma^{\pm\pm}$ particles.
The dominant production of these particles at hadron colliders proceeds via the Drell-Yan process through an s-channel exchange of a photon and $Z$ boson for the pair production,
\begin{eqnarray}
pp &\to& \gamma^{*}/Z^{*} \to\Phi_{+2}^{*}\Phi_{+2}/\Phi_{+1}^{*}\Phi_{+1}/\Phi_{-1}^{*}\Phi_{-1}
/A_0H_0,
\nonumber\\
 &\to& \gamma^{*}/Z^{*} \to \Sigma^{++}\Sigma^{--}/\Sigma^{+}\Sigma^{-},
\end{eqnarray}
or of a $W$ boson for the associated production,
\begin{eqnarray}
pp &\to& W^{*} \to \Phi_{+1}^{*}\Phi_{+2}/A_0\Phi_{+1}/A_0\Phi_{-1}^{*}/
H_0\Phi_{+1}/H_0\Phi_{-1}^{*},
\nonumber\\
&\to& W^{*} \to \Sigma^{++}\Sigma^{-}/\Sigma^{+}\Sigma^{0},
\end{eqnarray}
plus their charge conjugates.
The subdominant channels involving $h$, $A_0$, $H_0$, $\Phi_{-1}^{*}$, and $\Phi_{+1}$ exchanges
and the vector boson fusion process with two extra jets \cite{delAguila:2013mia}
have much smaller cross sections and can be neglected.
\footnote{
In this paper, we only consider the tree-level contributions. The QCD correction to doubly
charged scalar pair production was computed in \cite{Muhlleitner:2003me}, with a $K$-factor of
about $1.25$, while the contribution from real photon annihilation tends to increase the production
by $10\%$, resulting in an overall $K$-factor of 1.35 \cite{Han:2007bk}. The associated production of scalars in principle gives a similar $K$-factor $\simeq$ 1.25 \cite{Perez:2008ha}. However, to our knowledge, the similar study is missing for heavy fermions.}

\begin{figure}
\begin{center}
\includegraphics[width=0.45\linewidth]{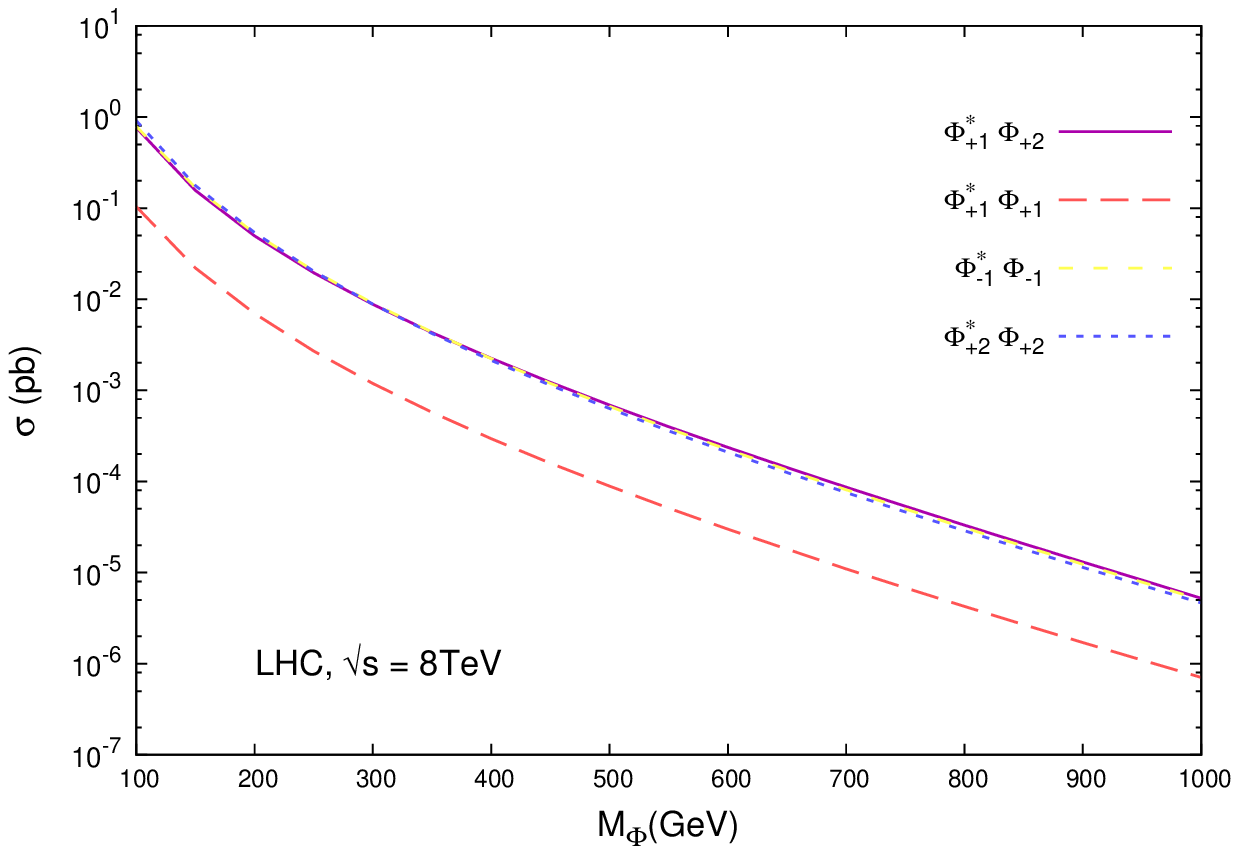}
\includegraphics[width=0.45\linewidth]{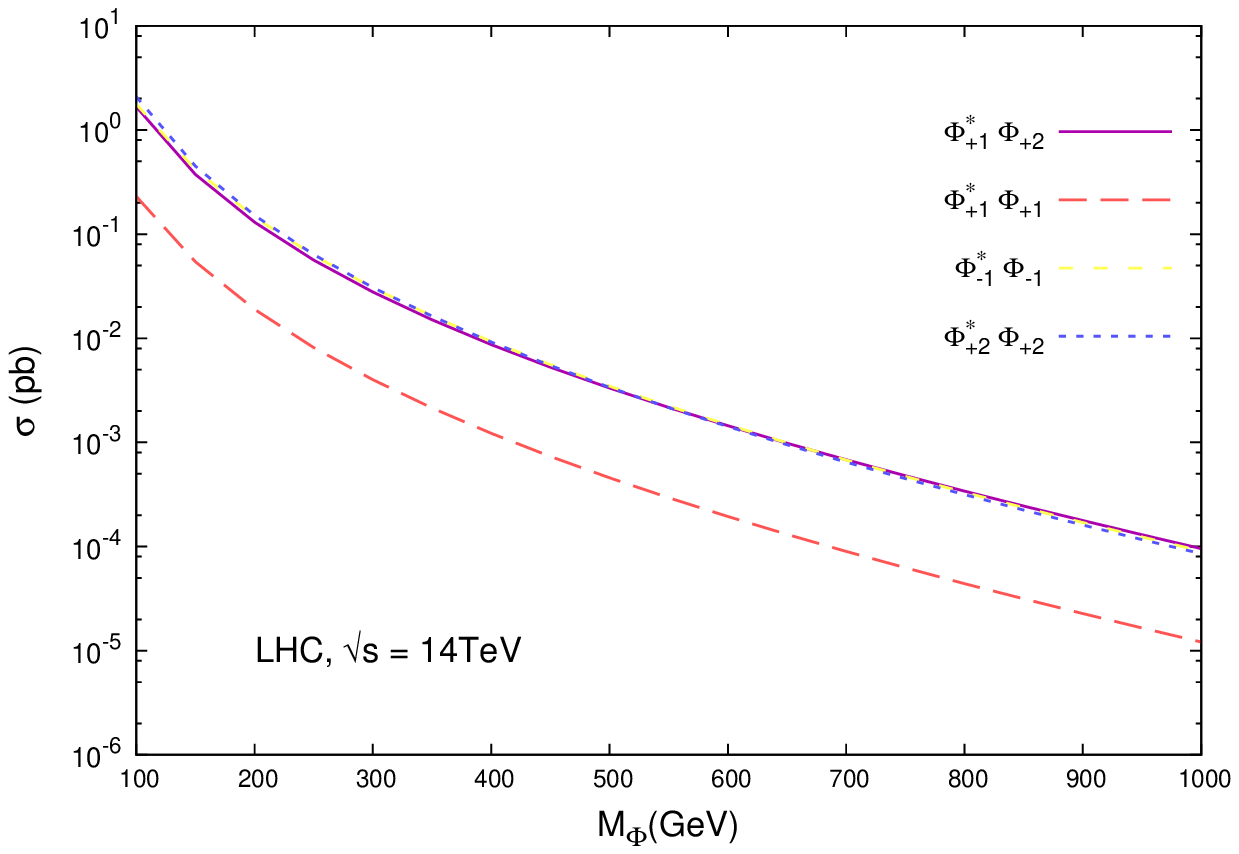}
\includegraphics[width=0.45\linewidth]{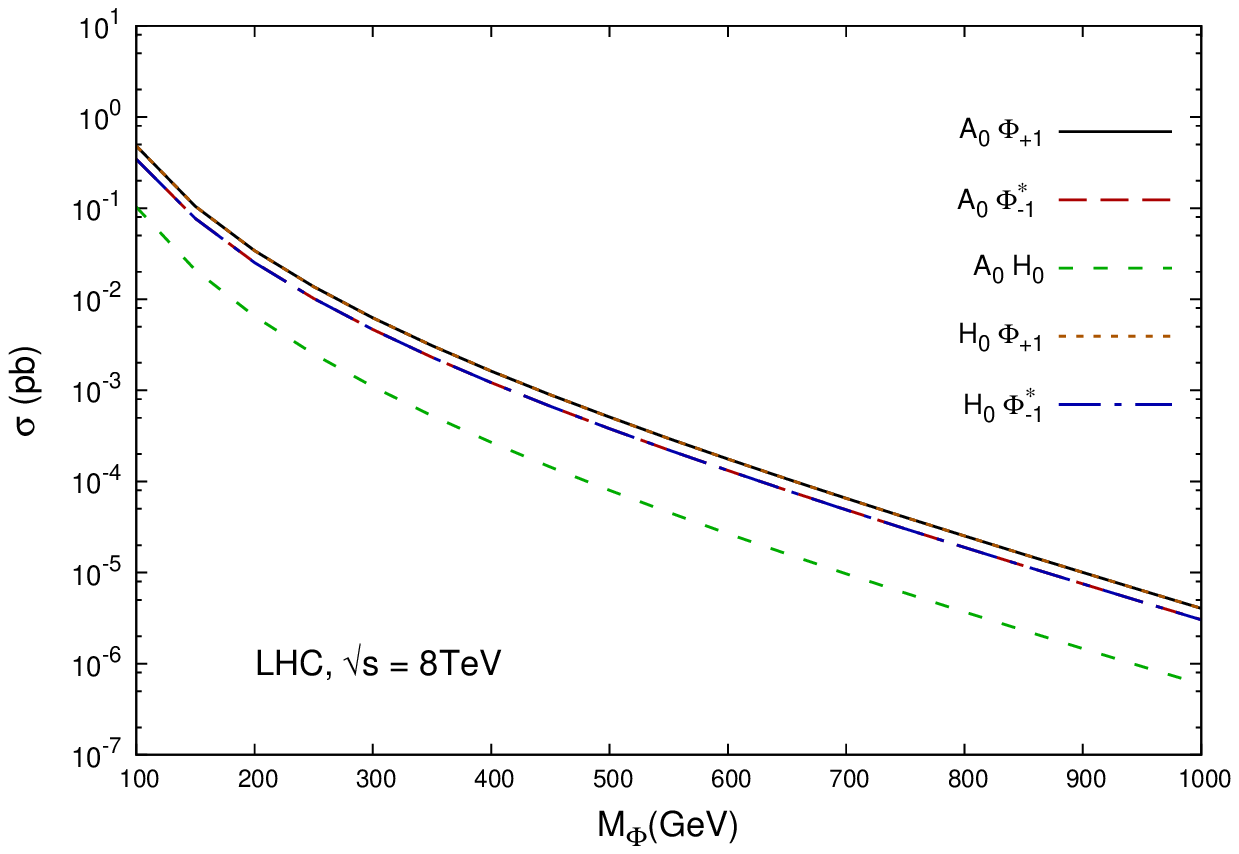}
\includegraphics[width=0.45\linewidth]{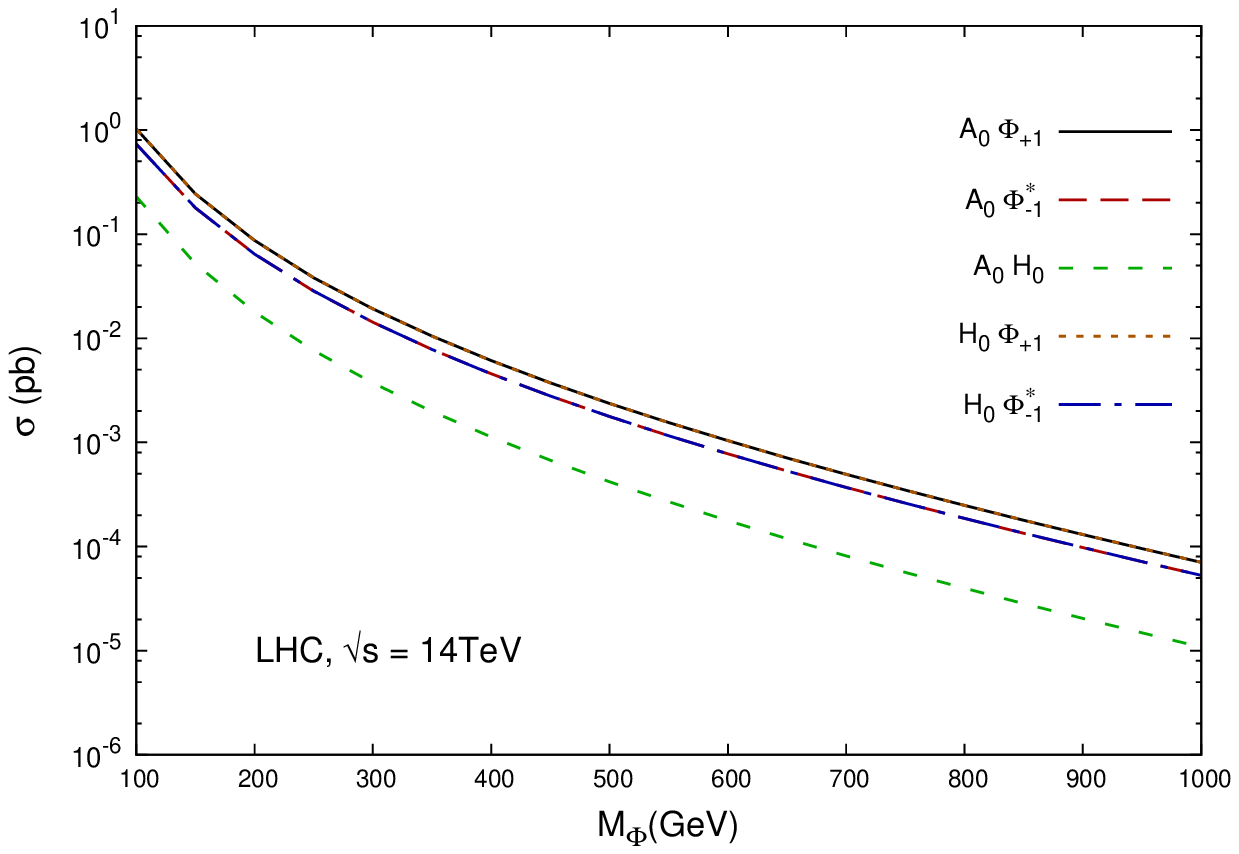}
\includegraphics[width=0.45\linewidth]{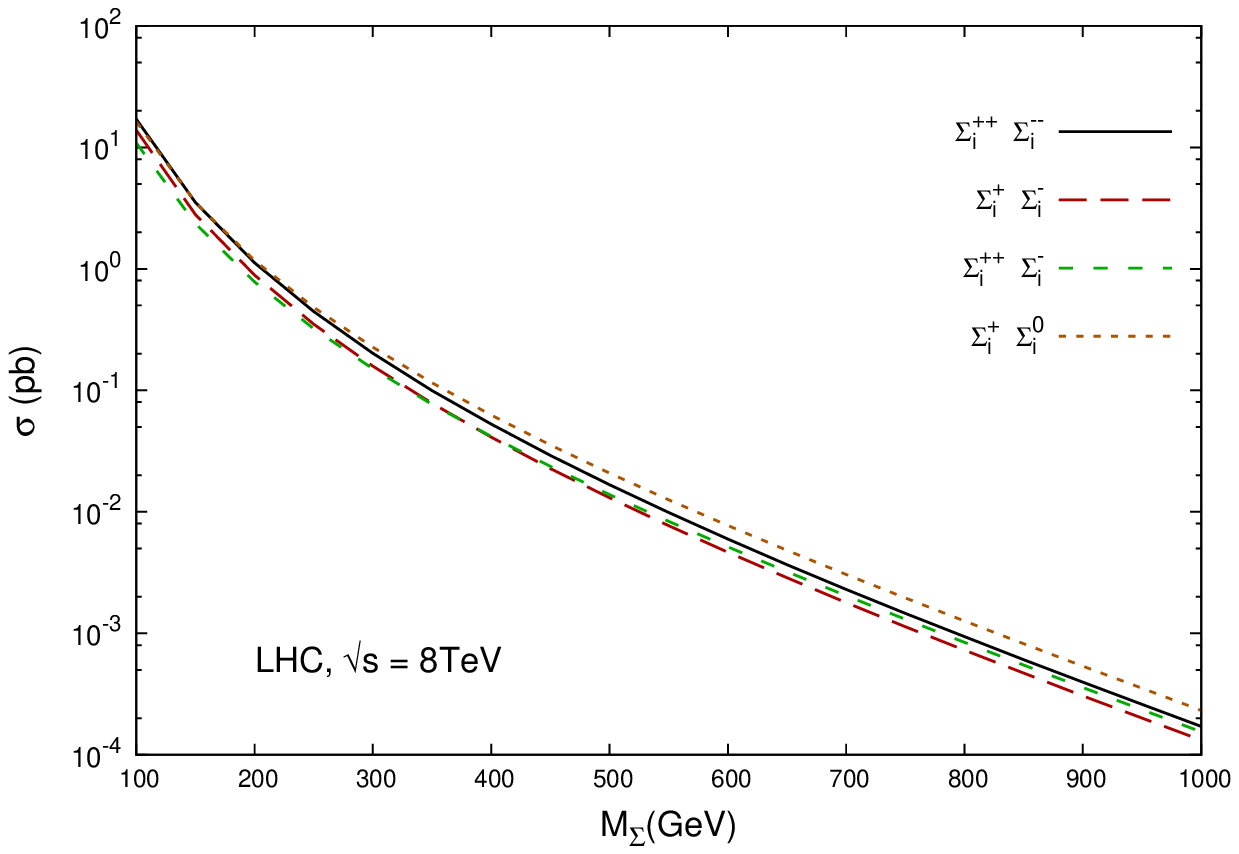}
\includegraphics[width=0.45\linewidth]{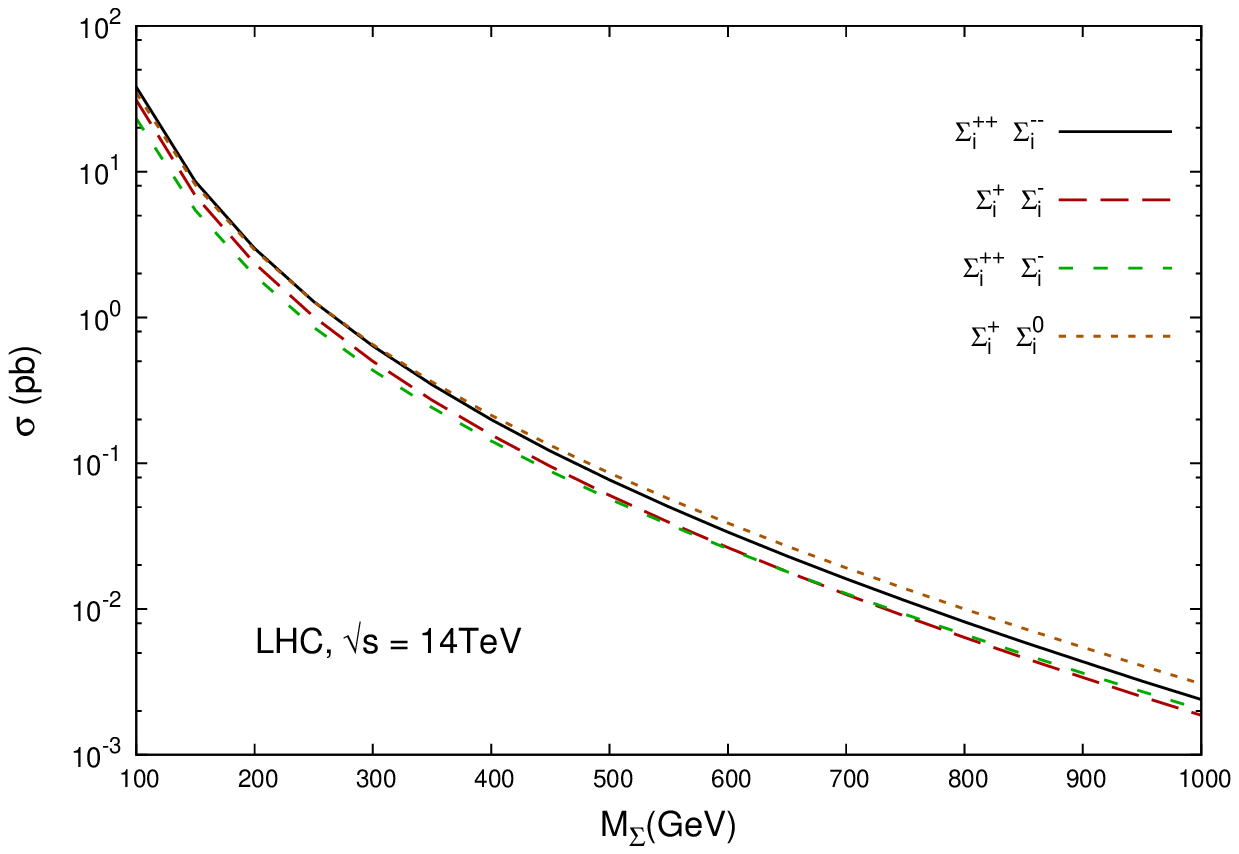}
\end{center}
\caption{Total cross section for various channels at LHC $8~\TeV$ ($14~\TeV$) is shown as a function of the mass $M_\Phi$ or $M_\Sigma$ on the left (right) panel.
\label{fig:cross}}
\end{figure}

In Fig.~\ref{fig:cross}, the total cross section for various channels at LHC is plotted as a function of the masses $M_{\Phi,\Sigma}$. These channels have sizable rates, as they do not suffer from small
mixing suppression. For instance, at LHC 14 TeV, the cross section is larger than $0.01~\fb$ ($1~\fb$) in each $\Phi$ ($\Sigma$) production channel up to a heavy mass of order $1~\TeV$.
Nevertheless, to see whether it is really feasible to observe those new particles, we have to examine their decay properties and employ them to devise appropriate kinematical cuts to suppress the SM background.

\subsection{Decay properties of new particles}
\label{subsec:decay}
In this subsection we study the decays of new particles in the minimal cascade seesaw model. All relevant decay widths are listed in Appendix C. As we stated earlier we assume for simplicity a degenerate spectrum for both the scalar quadruplet and the fermion quintuplet. Then, all new particles decay directly into the SM particles. For the four free parameters $M_{\Phi},~M_\Sigma,~v_\Phi$ and $t$, we evaluate at the benchmark points unless otherwise stated. In particular, $t=1+i$ is always assumed.

\begin{figure}
\begin{center}
\includegraphics[width=0.45\linewidth]{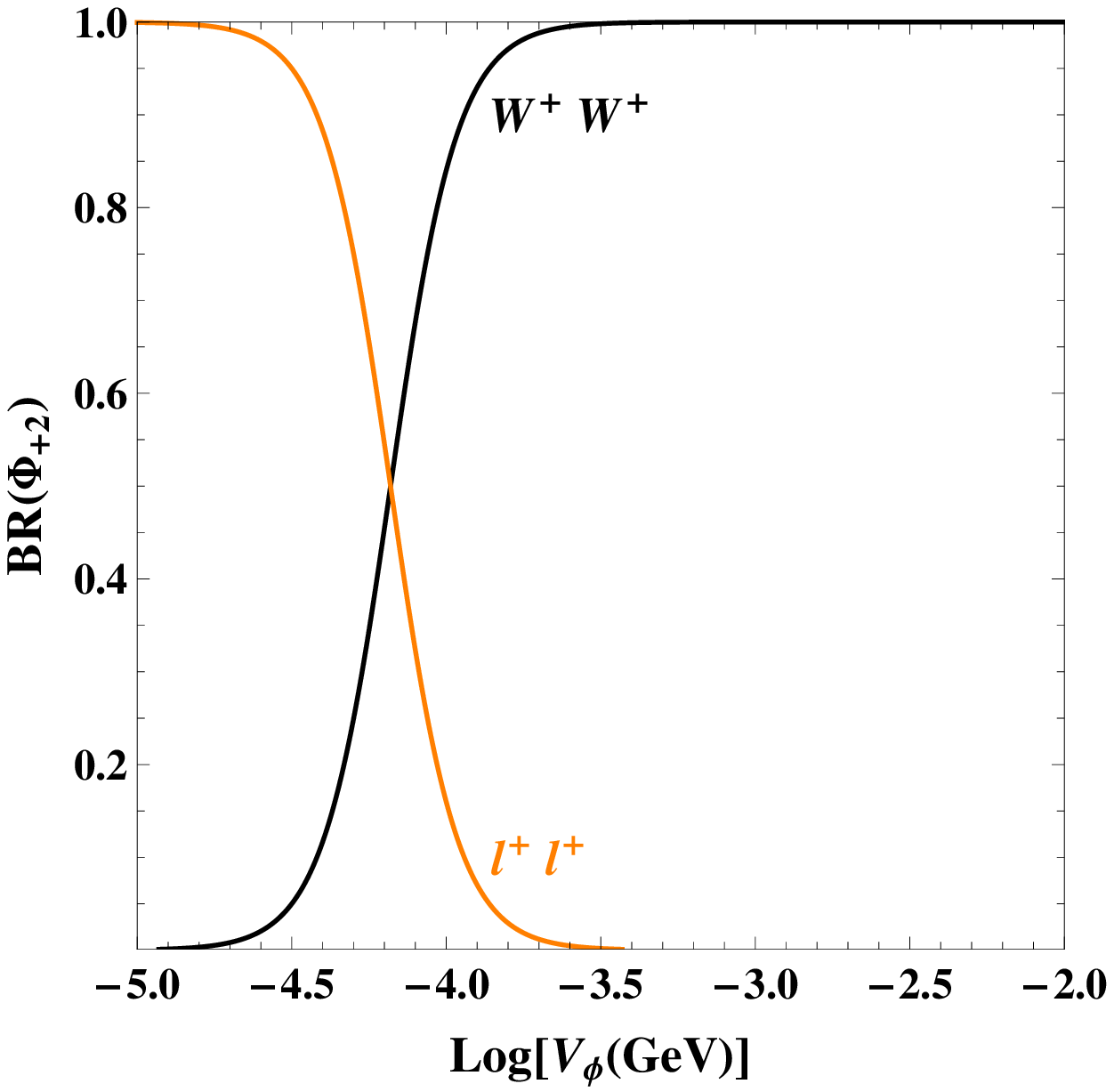}
\includegraphics[width=0.45\linewidth]{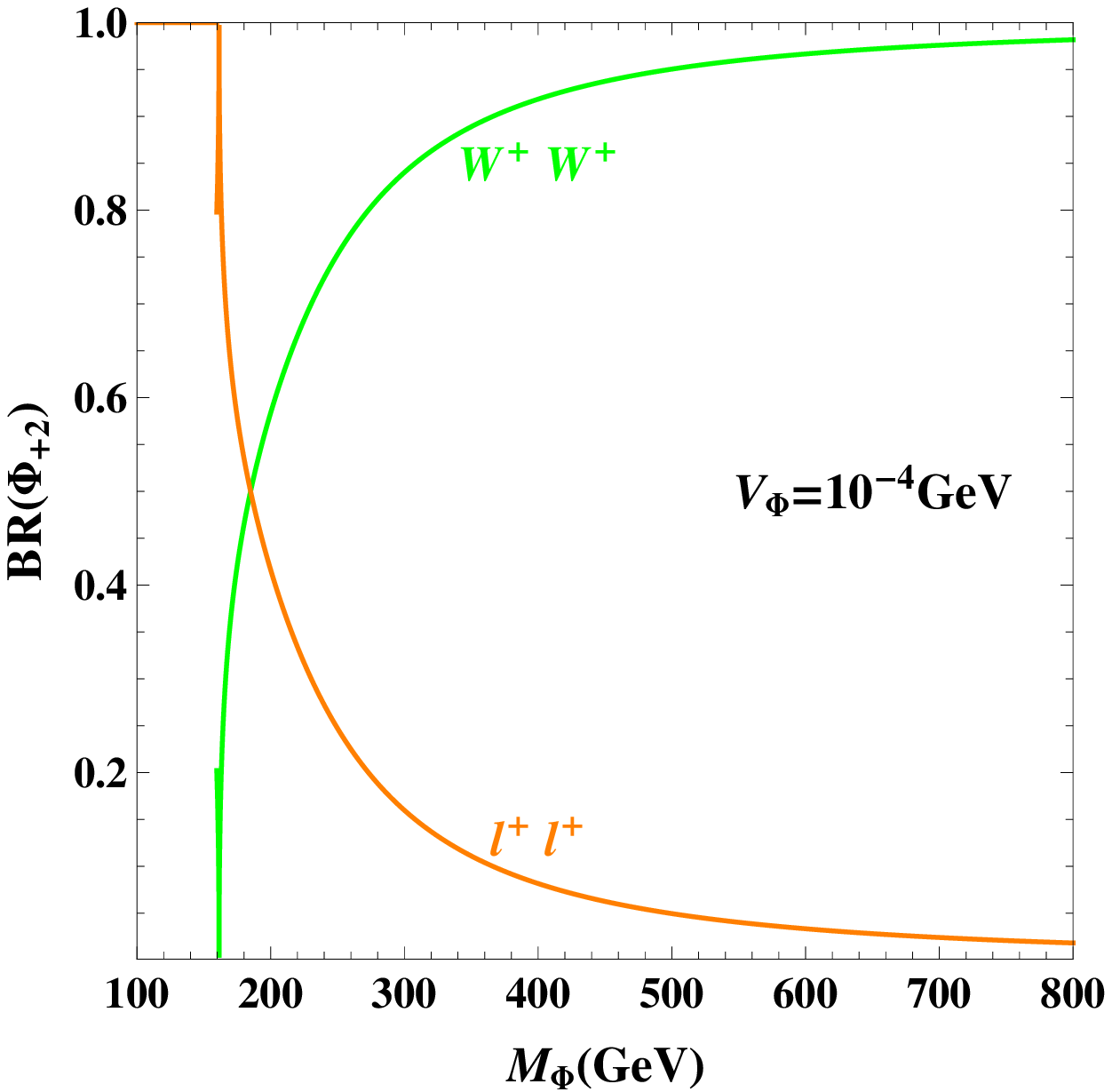}
\end{center}
\caption{Branching ratios of $\Phi_{+2}$ as a function of $v_{\Phi}$ for $M_{\Phi}=300$ GeV (left panel) and of $M_{\Phi}$ for $v_{\Phi}=10^{-4}$ GeV (right).
\label{fig:decayphi1}}
\end{figure}

\subsubsection{Doubly charged scalar $\Phi_{+2}$ decays}
\label{subsubsec:phidecay1}

There are two decay modes for the doubly charged scalars, the lepton number violating (LNV) like-sign dilepton decays $\Phi_{+2}\to \ell_i^+ \ell_j^+$ ($\ell=e,~\mu,~\tau$) and the like-sign di-$W$ decay
$\Phi_{+2}\to W^+W^+$. The amplitude for the former is proportional to the Yukawa coupling matrix for neutrinos and inversely proportional to $v_\Phi$ while the amplitude for the latter is proportional to $v_\Phi$. The ratio between the two decay widths is given by
\begin{eqnarray}
\frac{\Gamma(\Phi_{+2}\to \ell_i^+ \ell_j^+)}{\Gamma(\Phi_{+2}\to W^+ W^+)}\simeq \frac{|(ZZ^{\dag})_{ij}|^2 v_\phi^4}{M_{\Phi}^2 v_{\Phi}^4}\sim \left(\frac{m_{\nu}}{M_{\Phi}}\right)^2\left(\frac{v_\phi}{v_{\Phi}}\right)^4.
\label{eq:decayratio}
\end{eqnarray}

The branching ratios are presented in Fig. \ref{fig:decayphi1}. In the left panel, $\Br(\Phi_{+2})$ is plotted as a function of $v_\Phi$ at $M_\Phi=300~\GeV$, while in the right panel it is plotted as a function of $M_\Phi$ at $v_\Phi=10^{-4}~\GeV$. From Fig. \ref{fig:decayphi1} and eq. (\ref{eq:decayratio}), one finds that the two decay modes are comparable at, for instance, $v_\Phi\sim 10^{-4}$ GeV and $M_\Phi\sim 200~\GeV$. For a given $M_\Phi$, the di-$W$ decay dominates at a larger
$v_\Phi$ while the dilepton decay dominates at a smaller $v_\Phi$.
As we discussed in section \ref{sec:LFV}, $v_\Phi\approx 10^{-4}~\GeV$ is almost the lower bound
allowed by the LFV transitions and thus the like-sign dilepton decay is suppressed in the
majority of the parameter space.
This sets a stringent constraint on the LHC search for doubly charged scalars in the dilepton channel.
We will discuss this issue in more detail in subsection \ref{subsec:signal}.

\begin{figure}
\begin{center}
\includegraphics[width=0.45\linewidth]{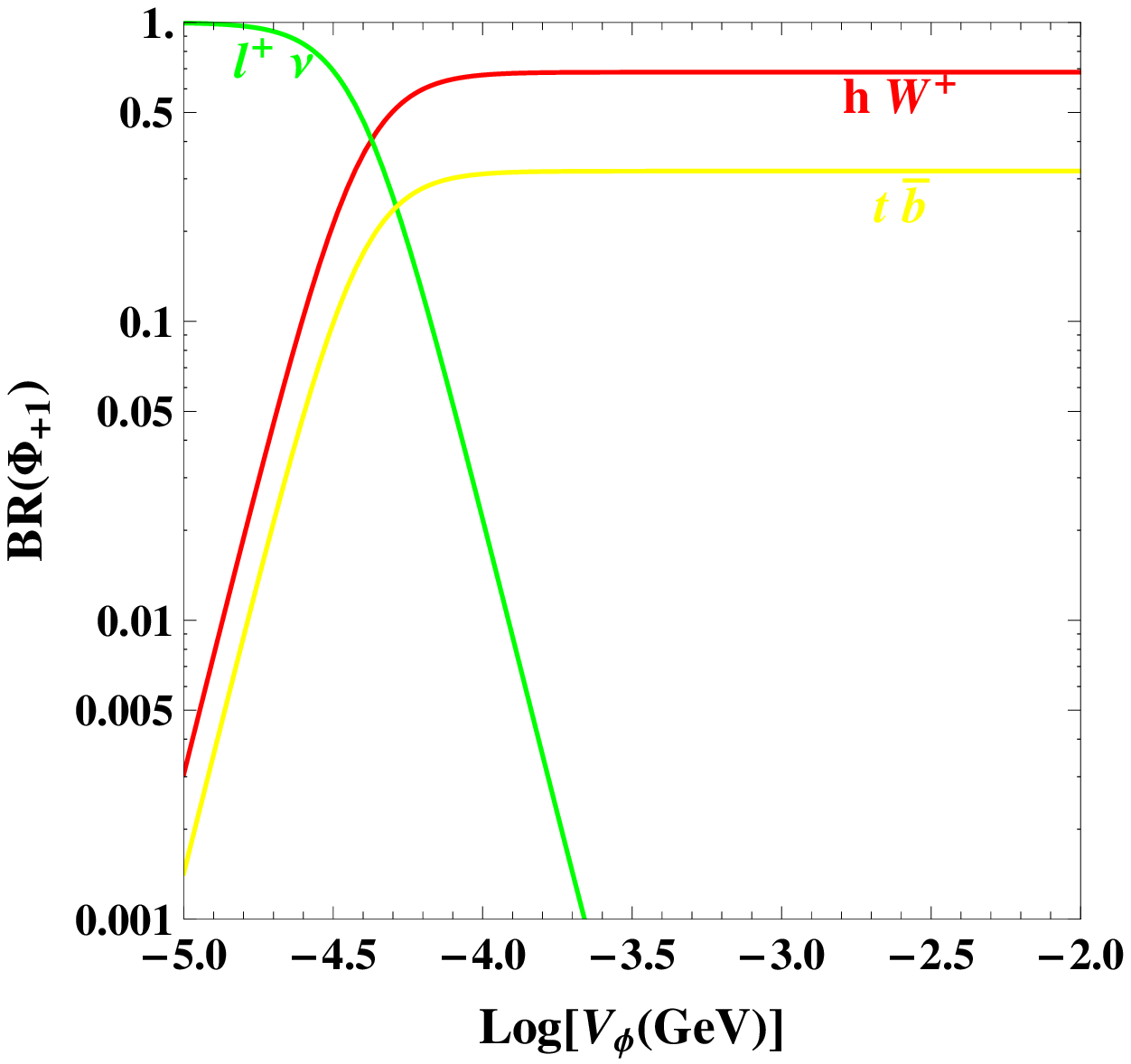}
\includegraphics[width=0.45\linewidth]{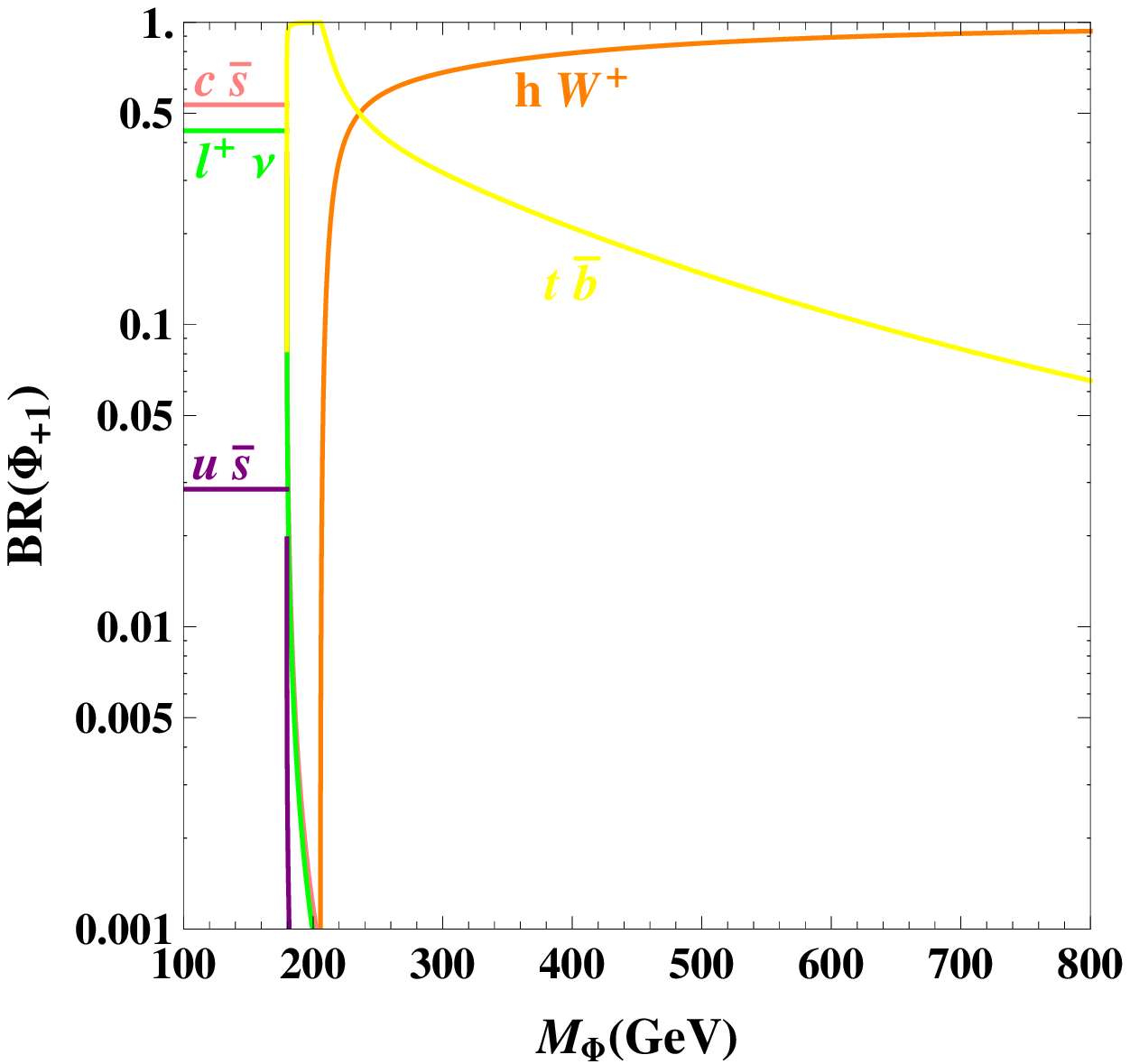}
\includegraphics[width=0.45\linewidth]{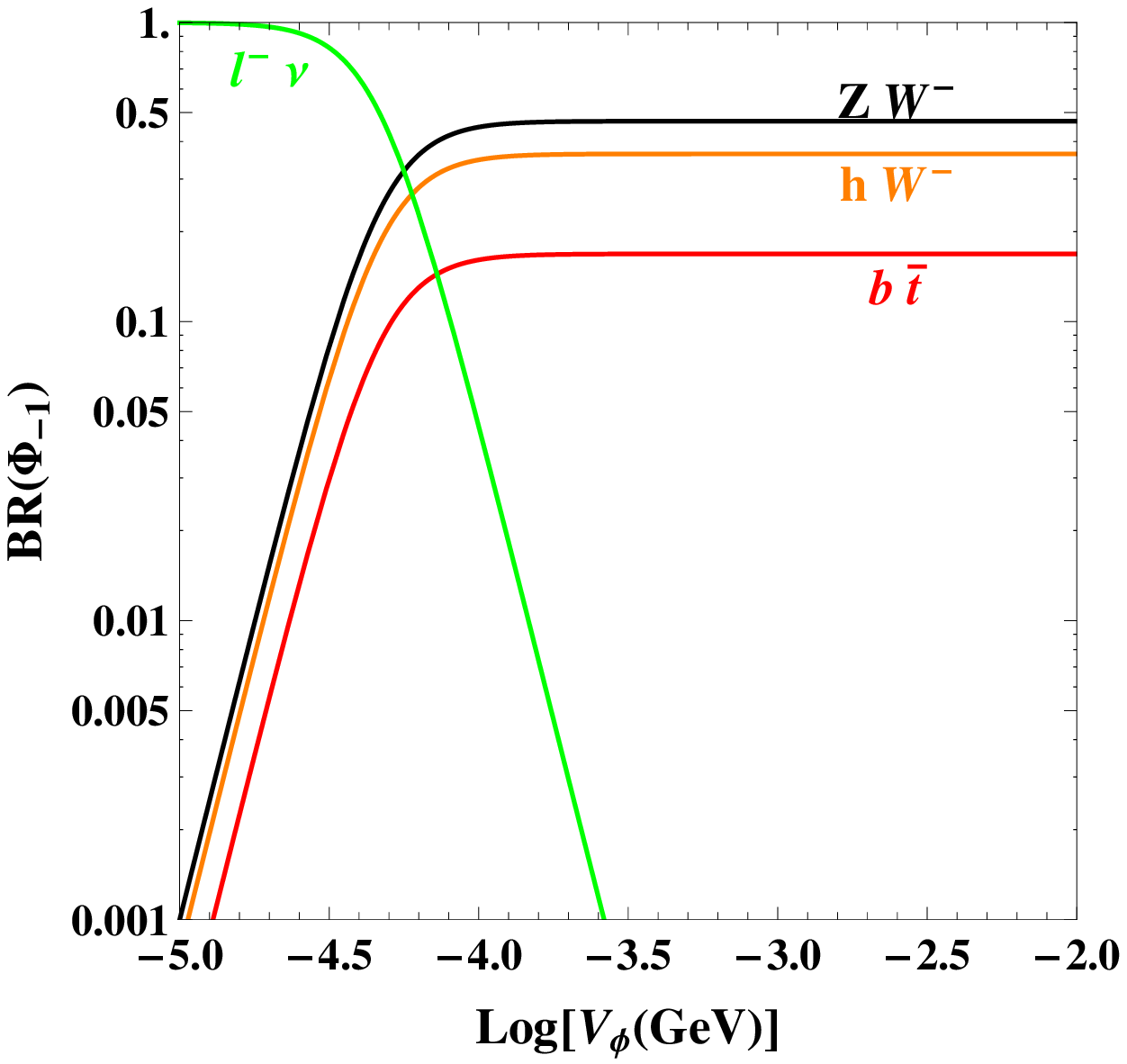}
\includegraphics[width=0.45\linewidth]{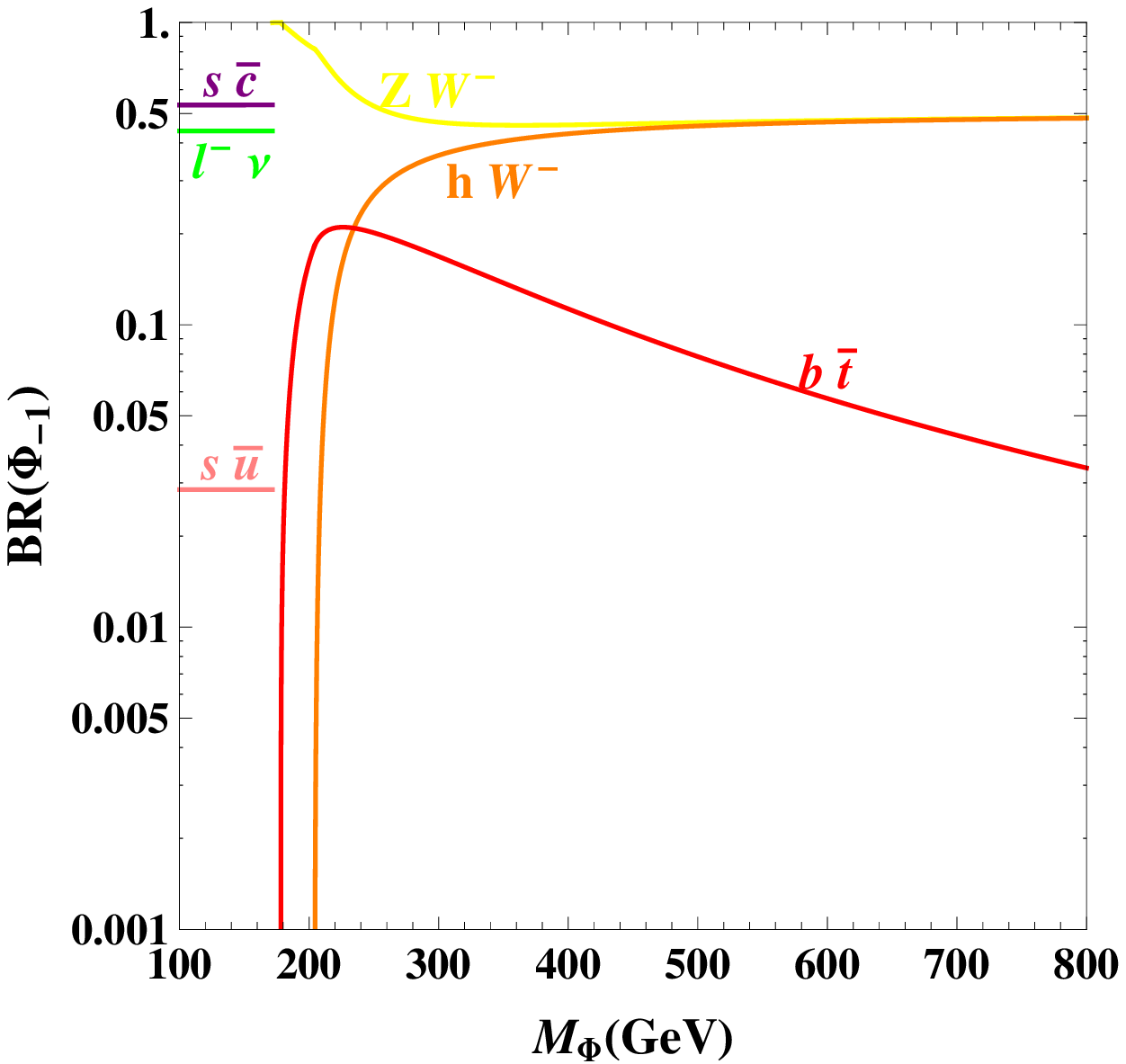}
\end{center}
\caption{Branching ratios of $\Phi_{+1}$ (upper panel) and $\Phi_{-1}$ (lower) as a
function of $v_{\Phi}$ for $M_{\Phi}=300~\GeV$ (left panel) and of $M_{\Phi}$ for
$v_{\Phi}=10^{-2}~\GeV$ (right).
\label{fig:decayphi2}}
\end{figure}

\subsubsection{Singly charged scalar $\Phi_{+1}$ and $\Phi_{-1}$ decays}
\label{subsubsec:phidecay2}

The decay modes of the singly charged scalars $\Phi_{+1}$ and $\Phi_{-1}$ are similar except that
$\Phi_{+1}\to ZW^+$ is absent. This difference arises from the mixing between the doublet and
quadruplet scalars due to vacuum expectation values, so that $\Phi_{+1}$ does not couple to $W^-Z$.
The decay amplitudes for $\Phi_{\pm 1}\to tb,~hW^\pm$ and $\Phi_{-1}\to ZW^-$ are proportional to
$v_\Phi$, while those for $\Phi_{\pm 1}\to\ell_i^\pm\nu_j$ are proportional to the Yukawa coupling of neutrinos. In the left panel of Fig. \ref{fig:decayphi2}, the relevant branching ratios are shown
as a function of $v_\Phi$ at $M_{\Phi_{\pm 1}}=300~\GeV$, and in the right panel as a function of
of $M_{\Phi_{\pm 1}}$ at $v_\Phi=10^{-2}~\GeV$.
For large values of $v_\Phi$, the important channels are $\Phi_{\pm 1}\to tb,~hW^\pm$ and
$\Phi_{-1}\to ZW^-$, while $\Phi_{\pm 1}\to\ell_i^\pm\nu_j$ dominate for small $v_\Phi$ and low $M_{\Phi}$.
Moreover, the decays $\Phi_{\pm 1}\to hW^\pm$ quickly dominate over
$\Phi_{\pm 1}\to tb$ once $M_\Phi$ is slightly above the threshold for $hW^\pm$, while the decay
width for $\Phi_{-1}\to ZW^-$ is always larger than for $\Phi_{-1}\to\bar{t}b$.

\begin{figure}
\begin{center}
\includegraphics[width=0.45\linewidth]{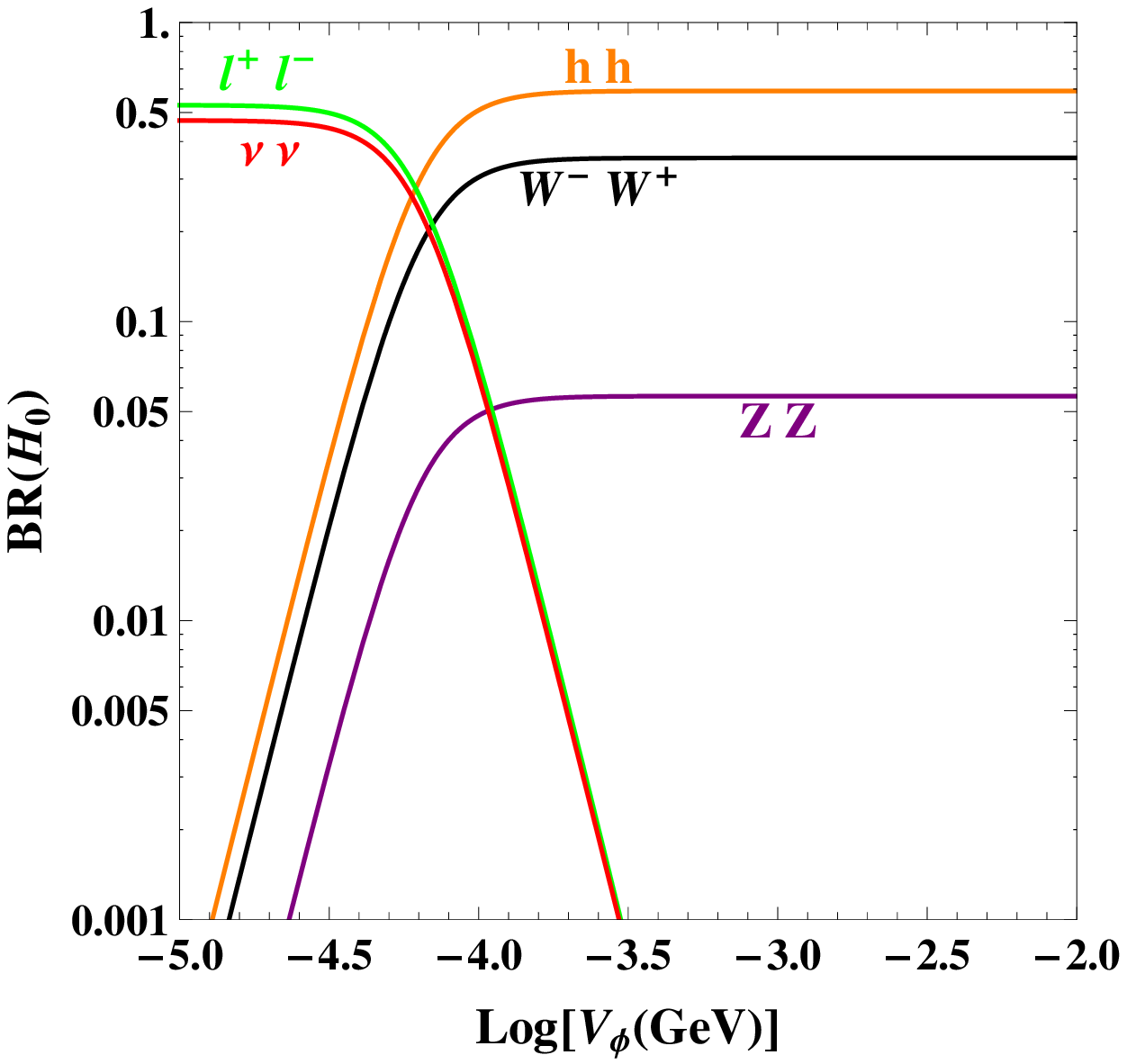}
\includegraphics[width=0.45\linewidth]{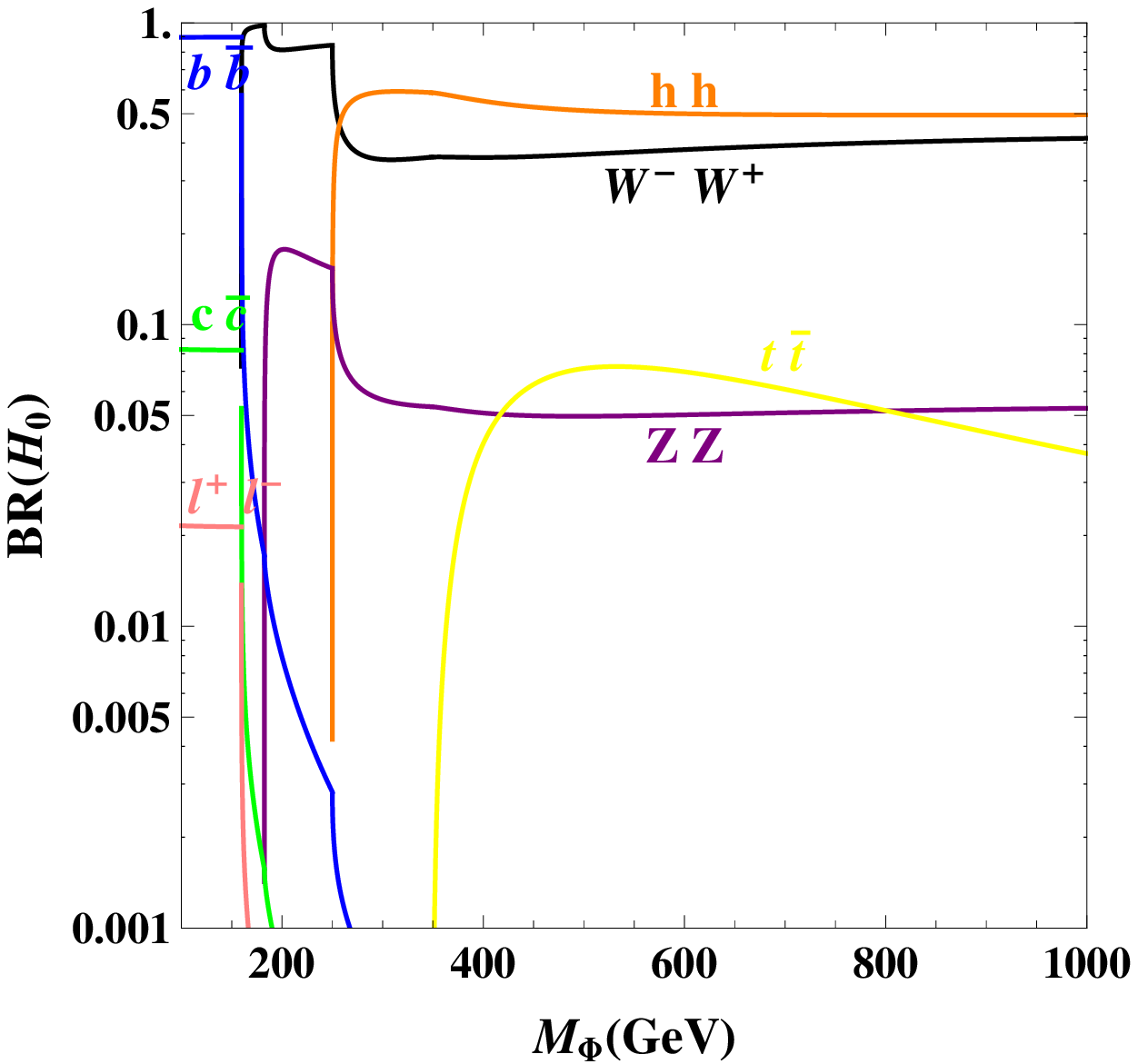}
\includegraphics[width=0.45\linewidth]{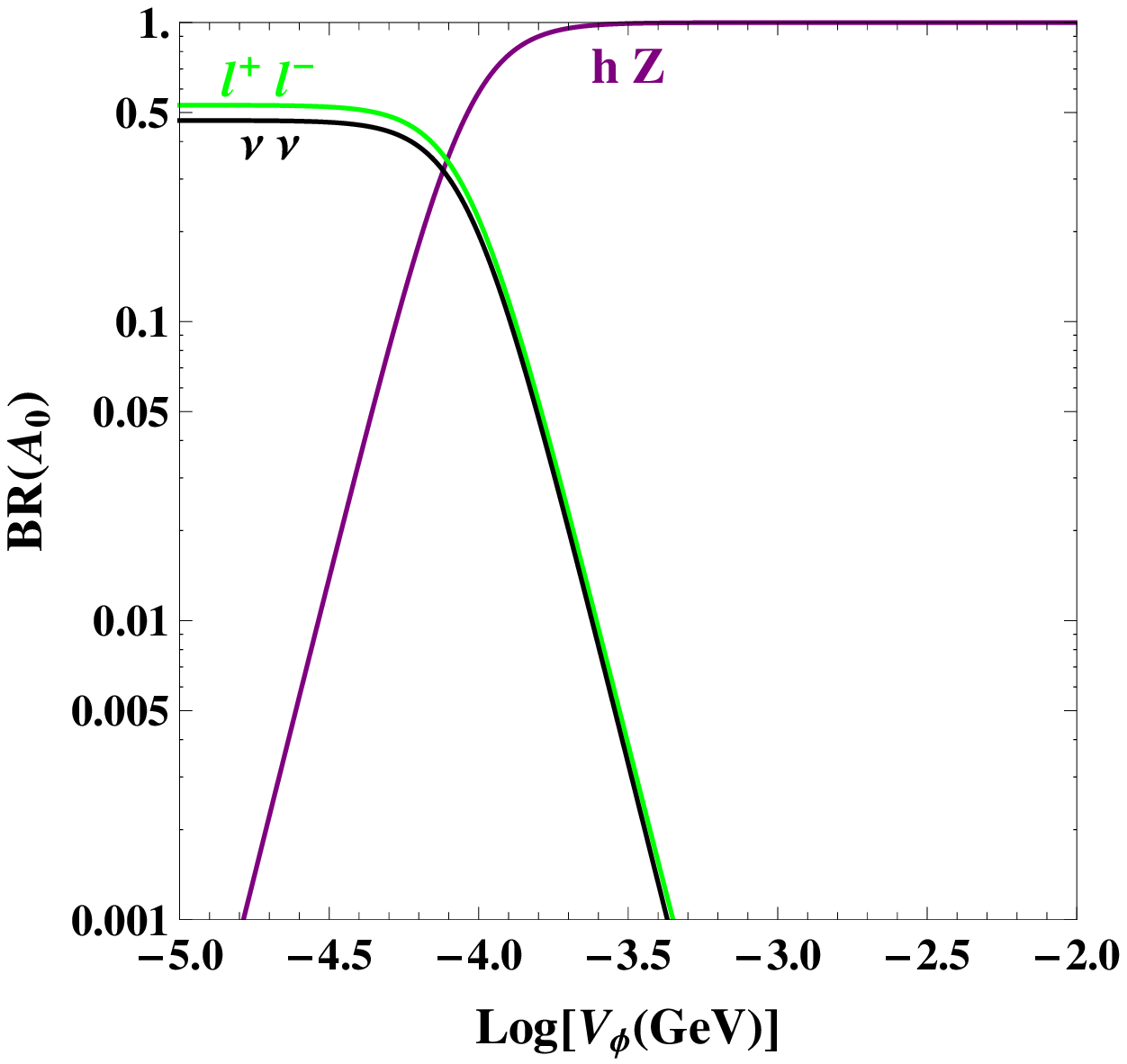}
\includegraphics[width=0.45\linewidth]{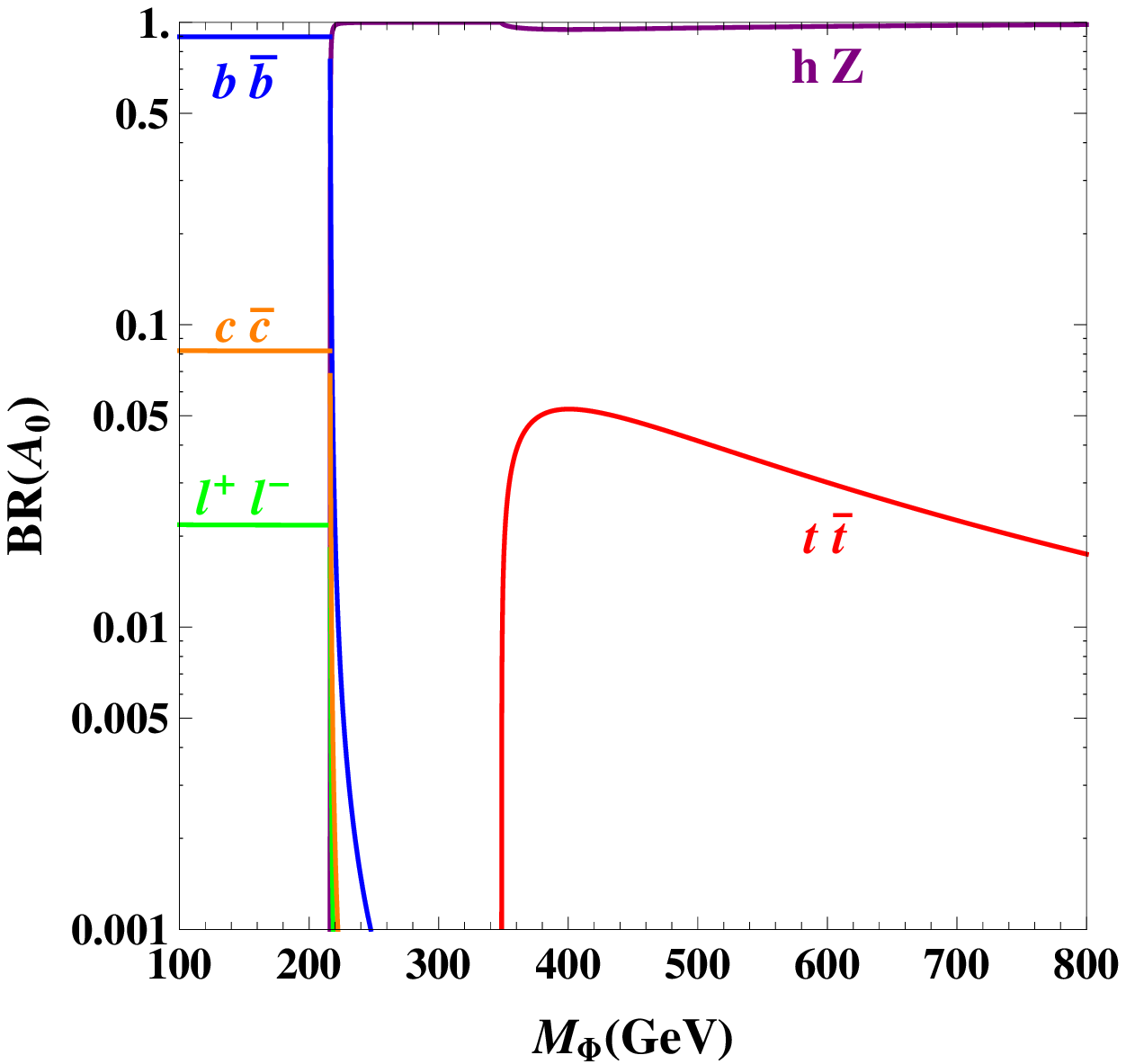}
\end{center}
\caption{Branching ratios of $H_0$ (upper panel) and $A_0$ (lower) as a function of $v_{\Phi}$
for $M_{\Phi}=300~\GeV$ (left panel) and of $M_{\Phi}$ for $v_{\Phi}=10^{-2}~\GeV$ (right).
\label{fig:decayphi3}}
\end{figure}

\subsubsection{CP-even scalar $H_0$ and CP-odd scalar $A_0$ decays}
\label{subsubsec:phidecay3}

The branching ratios for the neutral scalar decays are shown in Fig. \ref{fig:decayphi3}.
The relevant decay modes are, $H_0\to W^+W^-,~hh,~ZZ,~b\bar{b},~t\bar{t}$ and
$A_0\to hZ,~b\bar{b},~t\bar{t}$, which are proportional to $v_\Phi$,
and $H_0,~A_0\to \ell_i^+\ell_j^-,~\nu_i\nu_j$ which are proportional to the Yukawa coupling of
neutrinos.
For large $v_\Phi$, $H_0\to hh,~W^+W^-$ and $A_0\to hZ$ are the dominant channels while
$H_0\to ZZ$ is relatively suppressed. The latter is in contrast to the usual type II seesaw model
where the neutral member of the scalar triplet decays dominantly to a $Z$ pair \cite{Perez:2008ha}. This again arises from different scalar mixing patterns in the two seesaw models.
In the small $v_\Phi$ region, the channels $H_0,~A_0\to \ell_i^+ \ell_j^-,~\nu_i\nu_j$ become important.
Due to the constraints from the low energy LFV processes, we will work at the two benchmark
points in our signal analysis to demonstrate different features of the parameter space:
$v_\Phi=10^{-4}~\GeV$ for the signal channels involving LNV dilepton decays, and $v_\Phi=10^{-2}~\GeV$ for other channels.

\begin{figure}
\begin{center}
\includegraphics[width=0.45\linewidth]{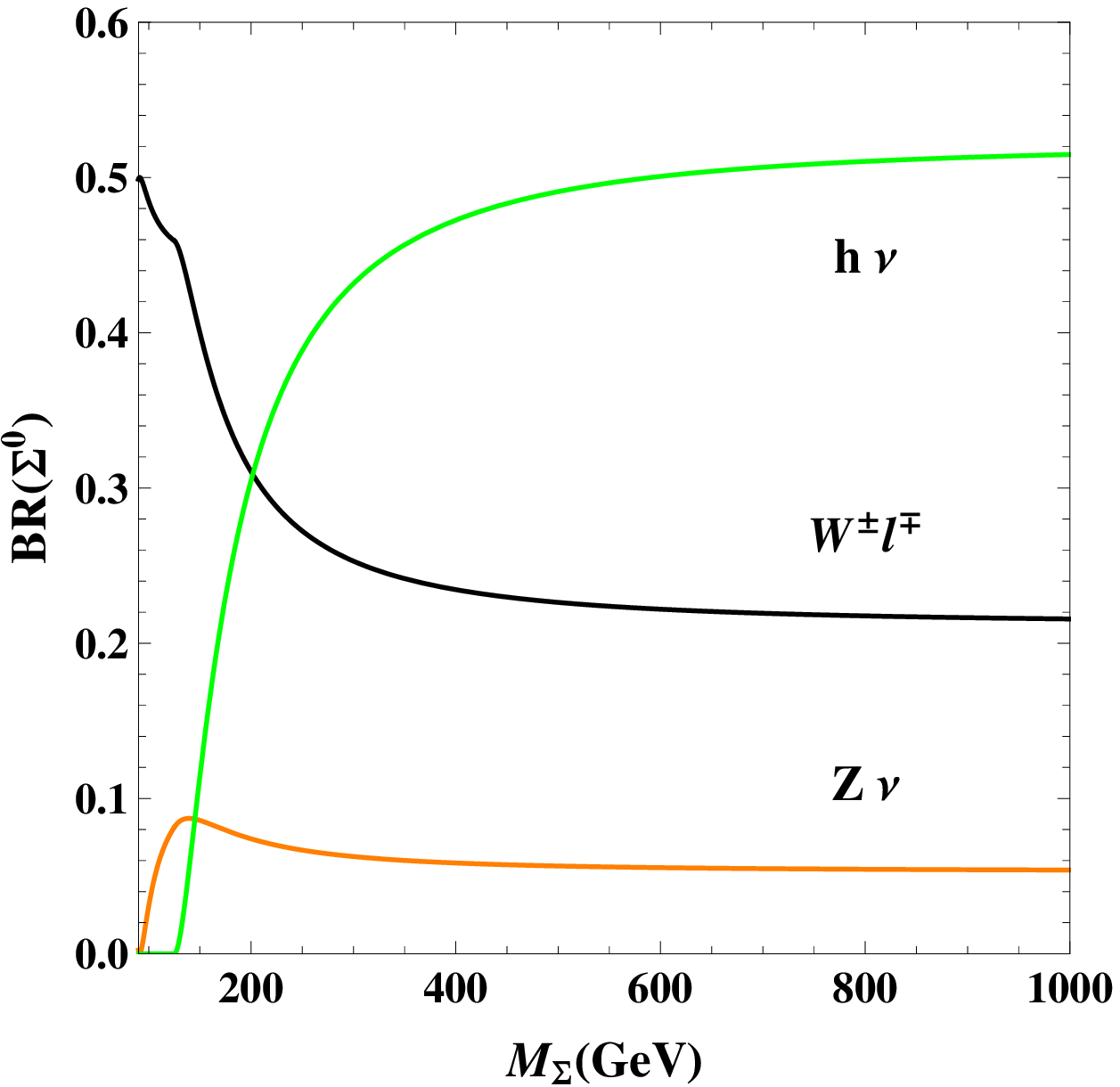}
\includegraphics[width=0.45\linewidth]{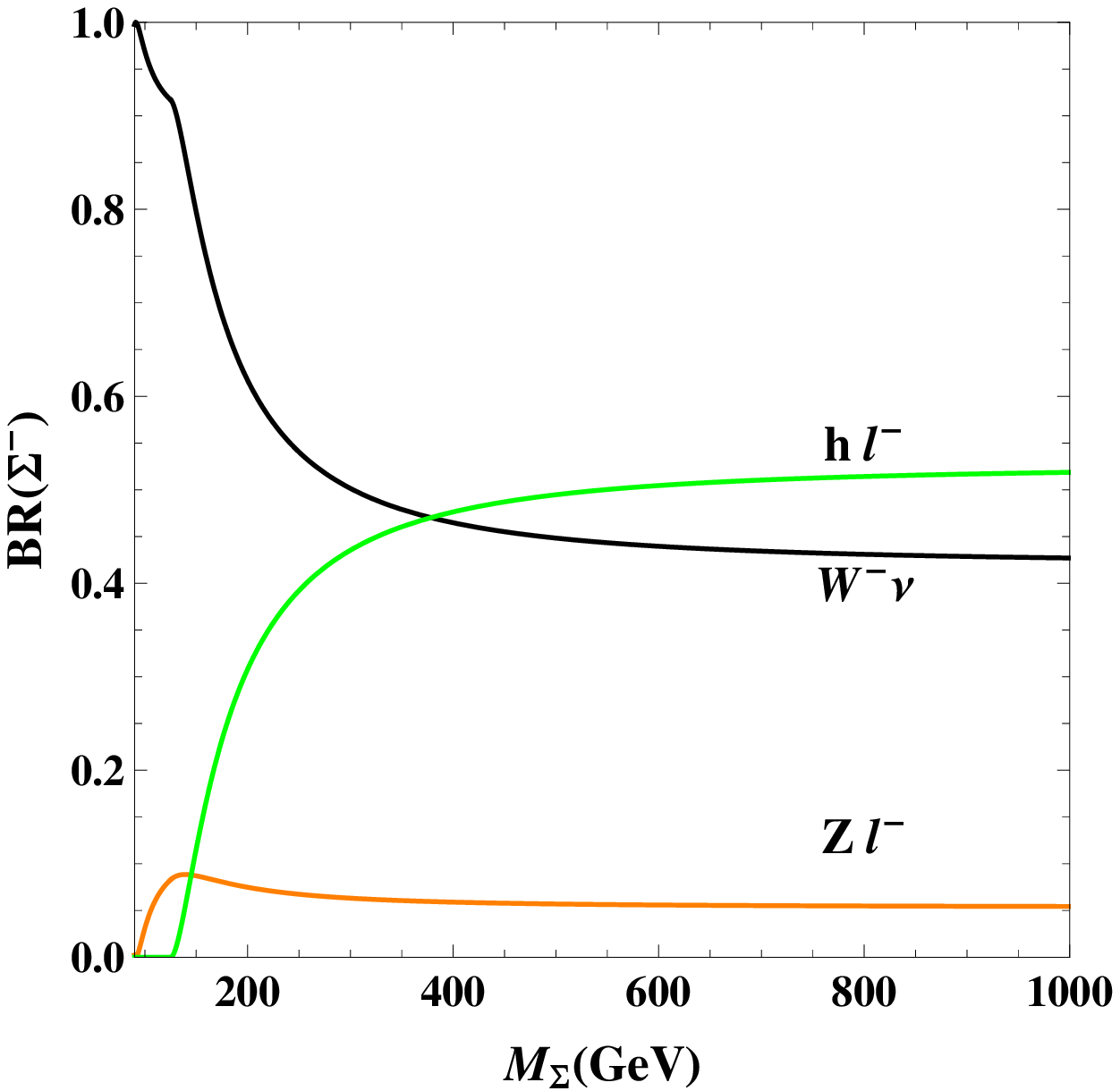}
\end{center}
\caption{Branching ratios of heavy fermions as a function of $M_\Sigma$.
\label{fig:decaysigma}}
\end{figure}

\subsubsection{Heavy quintuplet fermion decays}
\label{subsubsec:sigmadecay}

The most relevant decay channels of the quintuplet fermions are,
$\Sigma^0\to W^\pm\ell^\mp,~h\nu,~Z\nu$, and $\Sigma^-\to Z\ell^-,~h\ell^-,~W^-\nu$.
In Fig. \ref{fig:decaysigma}, we show the branching ratios for these channels versus the heavy fermion mass upon summing over the lepton flavors in the final states.
The results for $\Sigma^{--}$ are not presented since it has only one important decay,
$\Sigma^{--}\to W^-\ell^-$. For lepton-flavor specific final states, $\ell=e,~\mu,~\tau$, we observe the following relations,
\begin{eqnarray}
&&\Br(\Sigma^0\to W^\pm e^\mp) > \Br(\Sigma^0\to W^\pm \mu^\mp) \approx
\Br(\Sigma^0\to W^\pm \tau^\mp),
\nonumber\\
&&\Br(\Sigma^-\to he^-,~Z e^-) > \Br(\Sigma^-\to h\mu^-,~Z \mu^-) \approx
\Br(\Sigma^-\to h\tau^-,~Z \tau^-),
\nonumber\\
&&\Br(\Sigma^{--}\to W^- e^-) > \Br(\Sigma^{--}\to W^- \mu^-) \approx
\Br(\Sigma^{--}\to W^- \tau^-),
\label{relation_IH}%
\end{eqnarray}
for inverted neutrino mass hierarchy (IH), and
\begin{eqnarray}
&&\Br(\Sigma^0\to W^\pm \mu^\mp) \approx \Br(\Sigma^0\to W^\pm \tau^\mp)\gg
\Br(\Sigma^0\to W^\pm e^\mp),
\nonumber\\
&&\Br(\Sigma^-\to h\mu^-,~Z \mu^-) \approx \Br(\Sigma^-\to h\tau^-,~Z \tau^-)
\gg \Br(\Sigma^-\to he^-,~Z e^-),
\nonumber\\
&&\Br(\Sigma^{--}\to W^- \mu^-) \approx \Br(\Sigma^{--}\to W^- \tau^-)\gg
\Br(\Sigma^{--}\to W^- e^-),
\label{relation_NH}%
\end{eqnarray}
for normal hierarchy (NH). Similar relations are also found in the usual type III
seesaw model and can be understood as a consequence of the neutrino masses and mixing \cite{Li:2009mw}. The well-separated neutrino mass squared differences $\Delta m_{21}^2$ and $|\Delta m_{23}^2|$
indicate that the branching ratios to specific final states
can differ by a few times in the IH case and by an order of magnitude in the NH case.
This sensitivity to the mass hierarchy is considerably smeared out when summing over the lepton
flavors in the final states. We therefore do not distinguish between the IH and NH cases in Fig. \ref{fig:decaysigma}.


\subsection{Signals of new particles at the LHC}
\label{subsec:signal}

In this subsection we study the experimental signatures of new particles at the LHC.
We notice first that particles of equal charges appear also in the type II and III seesaw models.
In applying the LHC search results one must be careful since those particles have generally different
production and decay properties in different theoretical settings.
For instance, both CMS and ATLAS experiments set a lower bound on the doubly charged scalars ranging
from $204~\GeV$ to $459~\GeV$ \cite{Chatrchyan:2012ya} or from $375~\GeV$ to $409~\GeV$
\cite{ATLAS:2012hi}, assuming that they decay exclusively into like-sign dileptons in the
setting of type II seesaw.
These bounds obviously do not apply to our case under consideration since the branching ratio of $\Phi^{\pm\pm}\to\ell^\pm\ell^\pm$ can never get close to $100\%$ in the majority of the
parameter space due to the constraints from low-energy LFV transitions.
Similarly, both CMS \cite{CMS:2012ra} and ATLAS \cite{ATLAS:2013hma} have searched for pair
production of heavy leptons in type III seesaw, and set a lower bound on their mass to be in the range $180~\GeV$ to $210~\GeV$  or $245~\GeV$ respectively, assuming various patterns for the
heavy-light lepton mixing.

There are many possible final states resulting from $\Phi$ and $\Sigma$ production, given by the decay channels which we have discussed in subsection \ref{subsec:decay}.
In Tables \ref{tab:Phichannel} and \ref{tab:sigmachannel}, we collect the most relevant decay modes before including the sequential decays of SM particles.
These channels lead to various signatures which are conventionally classified according to the
multiplicity of the charged leptons.
We consider the following seven signal channels:
\begin{itemize}
   \item {$2\ell^{\pm}2\ell^{\mp}$, $2\ell^{\pm}4j$ and $2\ell^{\pm}4j+\cancel{E_T}$ channels from $\Phi$ production,}
   \item {$2\ell^{\pm}2\ell^{\mp}2j$, $3\ell^{\pm}\ell^{\mp}2j$, $3\ell^{\pm}2\ell^{\mp}\cancel{E_T}$ and $3\ell^\pm 3\ell^\mp$ channels from $\Sigma$ production.}
\end{itemize}
For clarity, we list all relevant final states and corresponding processes in Table \ref{tab:signal}.
They will be analyzed in detail in subsections \ref{subsubsec:phi 1}-\ref{subsubsec:sigma_4}.

\begin{table}[!h]
\begin{center}
\begin{tabular}{|c|c|c|c|c|}
\hline
& $\Phi_{+1}^{\ast} \to b\bar{t}\;(0.32)$ & $\Phi_{+1}^{\ast} \to hW^{-}\;(0.68)$ & $\Phi_{-1}^{\ast} \to hW^{+}\;(0.36)$ & $\Phi_{-1}^{\ast} \to ZW^{+}\;(0.47)$
\\
\hline
$\Phi_{+1} \to t\bar{b}\;(0.32)$ & $b\bar{b}t\bar{t}\;(0.10)$ & $t\bar{b}hW^{-}\;(0.22)$ & $-$ & $-$
\\
$\Phi_{+1} \to hW^{+}\;(0.68)$ & $b\bar{t}hW^{+}\;(0.22)$ & $hhW^{+}W^{-}\;(0.47)$ & $-$ & $-$
\\
$\Phi_{-1} \to hW^{-}\;(0.36)$ & $-$ & $-$ & $hhW^{+}W^{-}\;(0.13)$ & $hZW^{+}W^{-}\;(0.17)$
\\
$\Phi_{-1} \to ZW^{-}\;(0.47)$ & $-$ & $-$ & $hZW^{+}W^{-}\;(0.17)$ & $hhW^{+}W^{-}\;(0.22)$
\\
$A_{0} \to hZ\;(1.0)$ & $-$ & $-$ & $hhZW^{+}\;(0.36)$ & $hZZW^{+}\;(0.47)$
\\
$H_{0} \to W^{+}W^{-}\;(0.35)$ & $-$ & $-$ & $hW^{-}W^{+}W^{+}\;(0.13)$ & $ZW^{-}W^{+}W^{+}\;(0.17)$
\\
$H_{0} \to hh\;(0.60)$ & $-$ & $-$ & $hhhW^{+}\;(0.22)$ & $hhZW^{+}\;(0.28)$
\\
$\Phi_{+2} \to W^{+}W^{+}\;(1.0)$ & $b\bar{t}W^{+}W^{+}\;(0.32)$ & $hW^{-}W^{+}W^{+}\;(0.68)$ & $-$ & $-$\\
\hline
& $A_{0} \to hZ\;(1.0)$ & $H_{0} \to W^{+}W^{-}\;(0.35)$ & $H_{0} \to hh\;(0.60)$ & $\Phi_{+2}^{\ast} \to W^{-}W^{-}\;(1.0)$ \\
\hline
$\Phi_{+1} \to t\bar{b}\;(0.32)$ & $t\bar{b}hZ\;(0.32)$ & $t\bar{b}W^{+}W^{-}\;(0.10)$ & $t\bar{b}hh\;(0.20)$ & $-$\\
$\Phi_{+1} \to hW^{+}\;(0.68)$ & $hhZW^{+}\;(0.68)$ & $hW^{-}W^{+}W^{+}\;(0.24)$ & $hhhW^{+}\;(0.40)$
& $-$\\
$A_{0} \to hZ\;(1.0)$ & $-$ & $hZW^{+}W^{-}\;(0.35)$ & $hhhZ\;(0.60)$ & $-$
\\
$H_{0} \to W^{+}W^{-}\;(0.35)$ & $hZW^{+}W^{-}\;(0.35)$ & $-$ & $-$ & $-$
\\
$H_{0} \to hh\;(0.60)$ & $hhhZ\;(0.60)$ & $-$ & $-$ & $-$
\\
$\Phi_{+2} \to W^{+}W^{+}\;(1.0)$ & $-$ & $-$ & $-$ & $W^{+}W^{+}W^{-}W^{-}\;(1.0)$
\\
\hline
\end{tabular}
\end{center}
\caption{Final states from $\Phi$ production are shown with their branching ratios
given in the parentheses at $M_\Phi=300~\GeV$ and $v_\Phi=10^{-2}~\GeV$.
Only the modes with a branching ratio no less than $0.1$ are included.
}
\label{tab:Phichannel}
\end{table}

\begin{table}[!h]
\begin{center}
\begin{tabular}{|c|c|c|c|c|}
\hline
   & $\Sigma^{+}\to W^{+}\nu\;(0.5)$ & $\Sigma^{+}\to h\ell^{+}\;(0.44)$ & $\Sigma^{+}\to Z\ell^{+}\;(0.06)$ & $\Sigma^{++}\to W^{+}\ell^{+}\;(1.0)$
\\
\hline
   $\Sigma^{0}\to W^{\pm}\ell^{\mp}\;(0.5)$ & $W^{\pm}W^{+}\ell^{\mp}\nu\;(0.25)$   & $W^{\pm}h\ell^{\mp}\ell^{+}\;(0.22)$ & $W^{\pm}Z\ell^{\mp}\ell^{+}\;(0.03)$ & $-$
\\
   $\Sigma^{0}\to h\nu\;(0.44)$   & $W^{+}h\nu\nu\;(0.22)$ & $hh\ell^{+}\nu\;(0.19)$  & $Zh\ell^{+}\nu\;(0.026)$ & $-$
\\
   $\Sigma^{0}\to Z\nu\;(0.06)$   & $W^{+}Z\nu\nu\;(0.03)$ & $Zh\ell^{+}\nu\;(0.026)$  & $ZZ\nu\nu\;(0.0036)$ & $-$
\\
\hline
   $\Sigma^{-}\to W^{-}\nu\;(0.5)$       & $W^{+}W^{-}\nu\nu\;(0.25)$  & $hW^{-}\ell^{+}\nu\;(0.22)$ & $W^{-}Z\ell^{+}\nu\;(0.03)$ & $W^{+}W^{-}\ell^{+}\nu\;(0.5)$
\\
   $\Sigma^{-}\to h\ell^{-}\;(0.44)$         & $W^{+}h\ell^{-}\nu\;(0.22)$    & $hh\ell^{+}\ell^{-}\;(0.19)$   & $Zh\ell^{+}\ell^{-}\;(0.026)$ &  $W^{+}h\ell^{+}\ell^{-}\;(0.44)$
\\
   $\Sigma^{-}\to Z\ell^{-}\;(0.06)$         & $W^{+}Z\ell^{-}\nu\;(0.03)$    &
   $Zh\ell^{+}\ell^{-}\;(0.026)$   & $ZZ\ell^{+}\ell^{-}\;(0.0036)$ &
   $W^{+}Z\ell^{+}\ell^{-}\;(0.06)$
\\
\hline
   $\Sigma^{--}\to W^{-}\ell^{-}\;(1.0)$    & $W^{+}W^{-}\ell^{-}\nu\;(0.5)$ & $W^{-}h\ell^{+}\ell^{-}\;(0.44)$  & $W^{-}Z\ell^{+}\ell^{-}\;(0.06)$ &
   $W^{+}W^{-}\ell^{+}\ell^{-}\;(1.0)$
\\
\hline
\end{tabular}
\end{center}
\caption{Final states from $\Sigma$ production are shown with their branching ratios given in the parentheses at $M_\Sigma=300~\GeV$ and $v_\Phi=10^{-2}~\GeV$.}
\label{tab:sigmachannel}
\end{table}

\begin{table}[!h]
\begin{center}
\begin{tabular}{|l|l|}
\hline
final states & $\Phi$ production process in $pp$ collision
\\
\hline
$2\ell^{\pm}2\ell^{\mp}$ & $\Phi_{+2}\Phi_{+2}^{*}/A_0H_0 \to 2\ell^\pm 2\ell^\mp$
\\
\hline
$4j2\ell^{\pm}+\cancel{E_T}$ & $\Phi_{+2}\Phi_{+2}^{*} \to W^\pm W^\pm W^\mp W^\mp
\to jjjj\ell^\pm \ell^\pm \nu\nu$,
\\
& $\Phi_{+2}\Phi_{+1}^{*}(\Phi_{+2}^{*}\Phi_{+1}) \to W^\pm W^\pm + h W^\mp/\bar{t}b(t\bar{b})
\to jjb\bar{b}\ell^\pm \ell^\pm \nu\nu$
\\
\hline
$4j2\ell^{\pm}$ & $\Phi_{+2}\Phi_{+2}^{*} \to \ell^\pm \ell^\pm W^\mp W^\mp
\to jjjj\ell^\pm \ell^\pm$ ,
\\
& $\Phi_{+2}\Phi_{+1}^{*}(\Phi_{+2}^{*}\Phi_{+1}) \to \ell^\pm \ell^\pm + h W^\mp/\bar{t}b(t\bar{b})
\to jjb\bar{b}\ell^\pm \ell^\pm$
\\
\hline
\hline
final states & $\Sigma$ production process in $pp$ collision
\\
\hline
$2\ell^{\pm}2\ell^{\mp}2j$ &
$\Sigma^{\pm}\Sigma^{\mp}/\Sigma^{0}\Sigma^{\pm}/\Sigma^{\pm}\Sigma^{\mp\mp}
\to hZ(ZZ)\ell^\pm \ell^\mp/W^\pm \ell^\mp Z \ell^\pm/Z \ell^\pm W^\mp \ell^\mp
\to jj 2\ell^\pm 2\ell^\mp$
\\
\hline
$3\ell^{\pm}\ell^{\mp}2j$ & $\Sigma^{\pm}\Sigma^{0} \to W^\mp \ell^\pm Z \ell^\pm
\to jj 3\ell^\pm \ell^\mp$
\\
\hline
$3\ell^{\pm}2\ell^{\mp}+\cancel{E_T}$ &
$\Sigma^{\pm} \Sigma^{0}/\Sigma^{\pm\pm}\Sigma^{\mp}\to Z \ell^\pm W^\pm \ell^\mp (Z \ell^\pm Z \nu) /W^\pm \ell^\pm Z \ell^\mp \to 3 \ell^\pm 2\ell^\mp \nu$
\\
\hline
$3\ell^{\pm}3\ell^{\mp}$ & $\Sigma^{\pm} \Sigma^{\mp} \to \ell^\pm Z \ell^\mp Z
\to 3 \ell^\pm 3\ell^\mp$
\\
\hline
\end{tabular}
\end{center}
\caption{Signal channels considered and corresponding processes for $\Phi$ and $\Sigma$ production.}
\label{tab:signal}
\end{table}

\begin{figure}[!htbp]
\begin{center}
\includegraphics[width=0.45\linewidth]{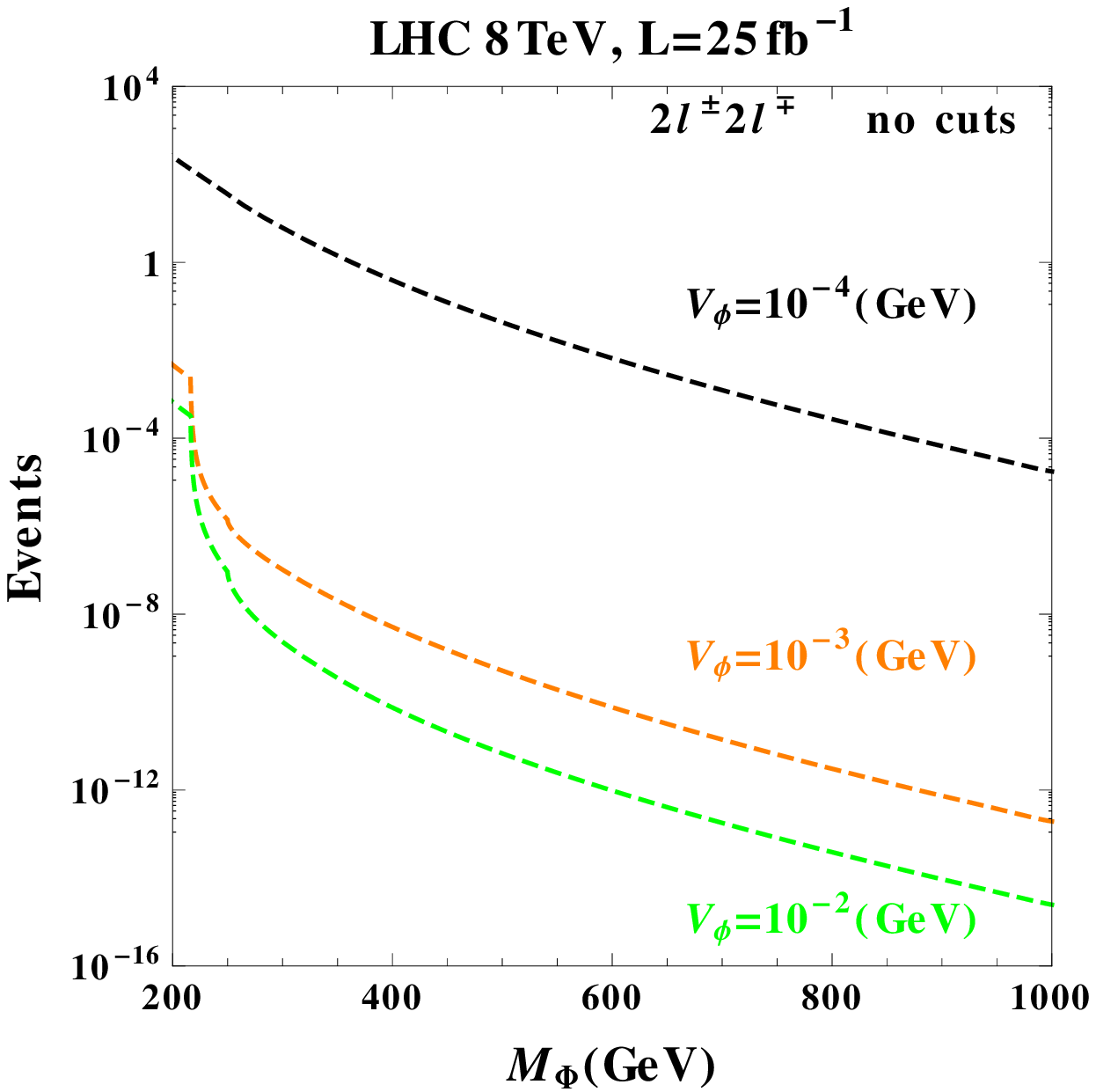}
\includegraphics[width=0.45\linewidth]{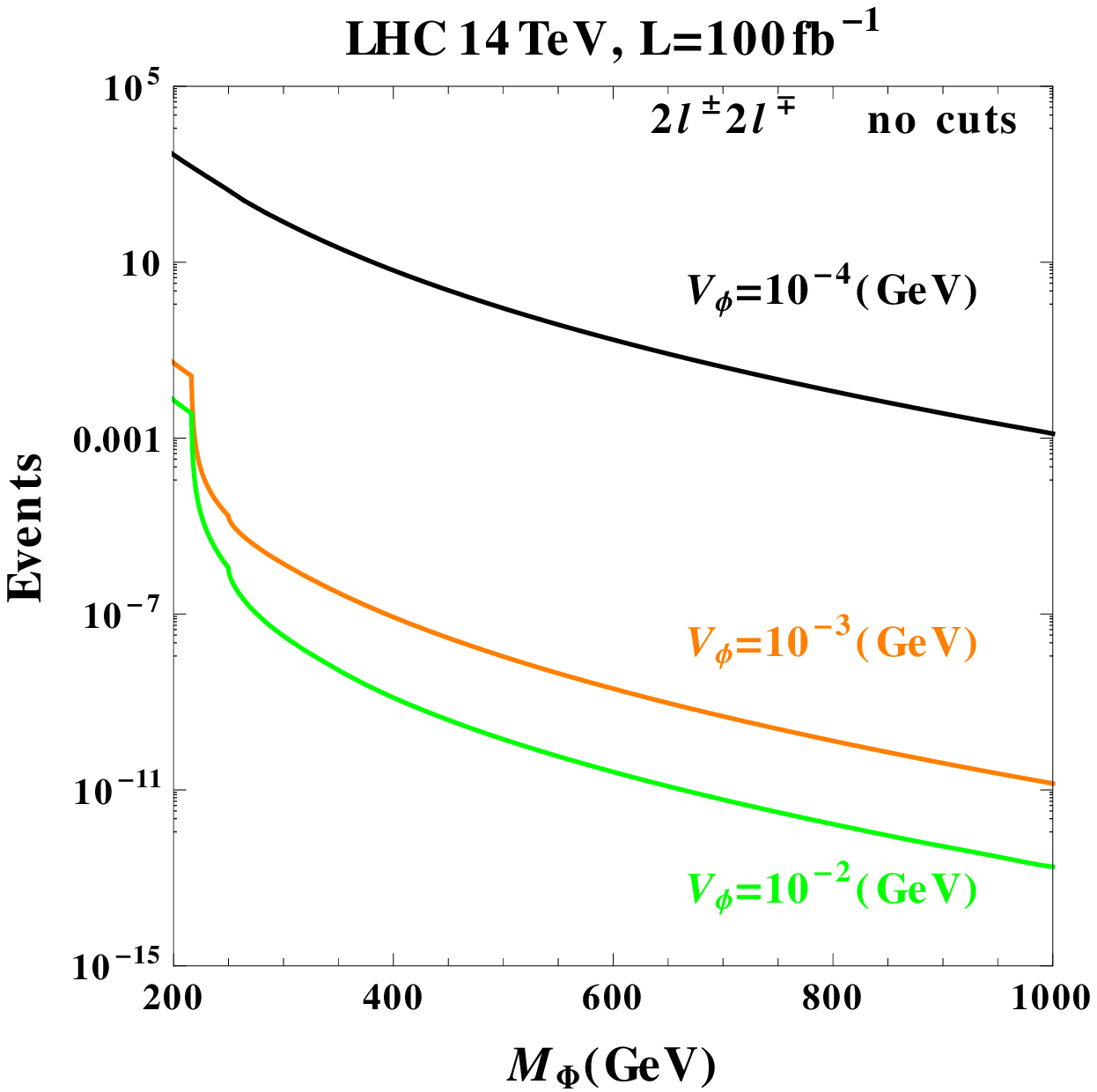}
\includegraphics[width=0.45\linewidth]{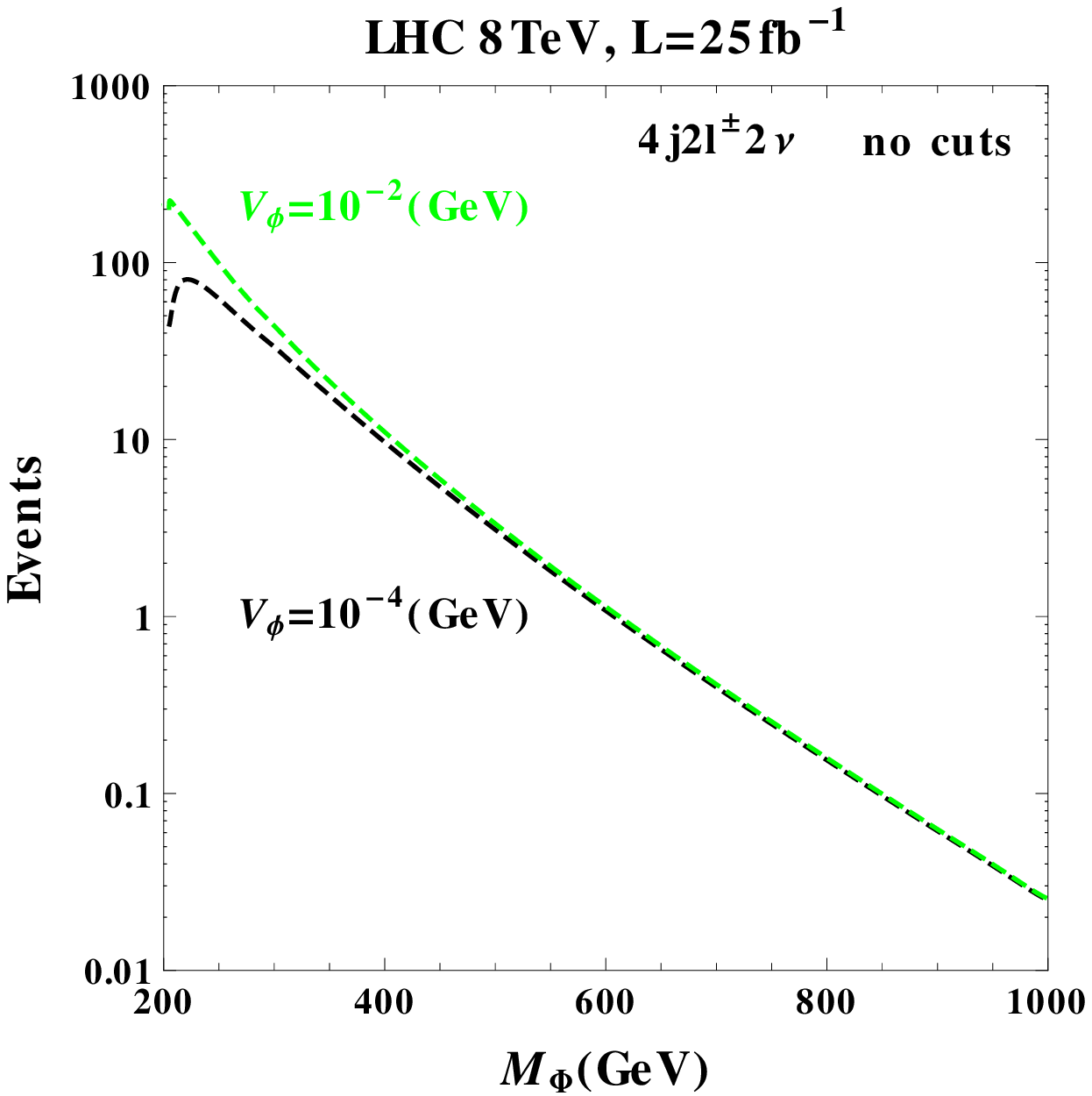}
\includegraphics[width=0.45\linewidth]{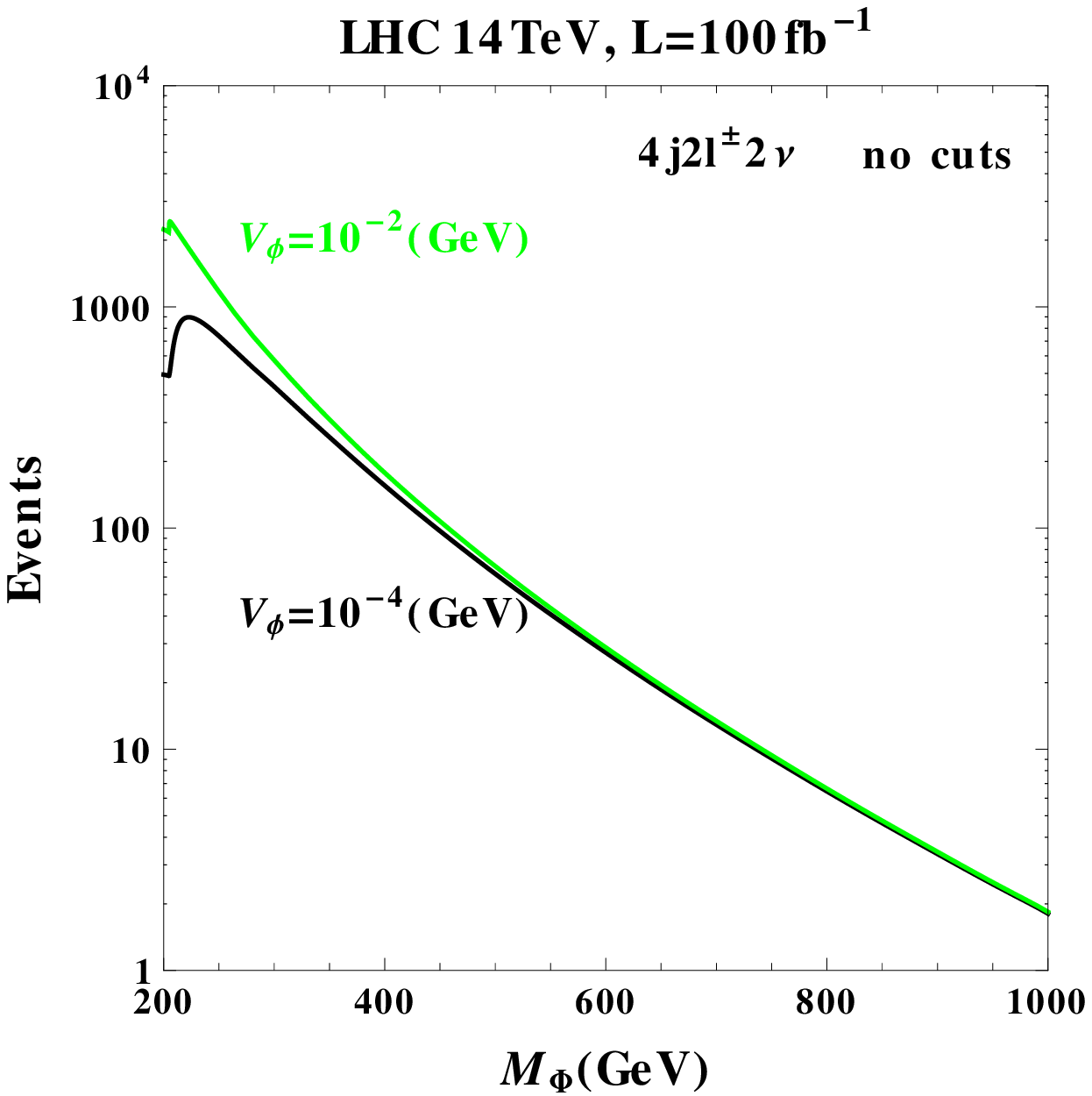}
\includegraphics[width=0.45\linewidth]{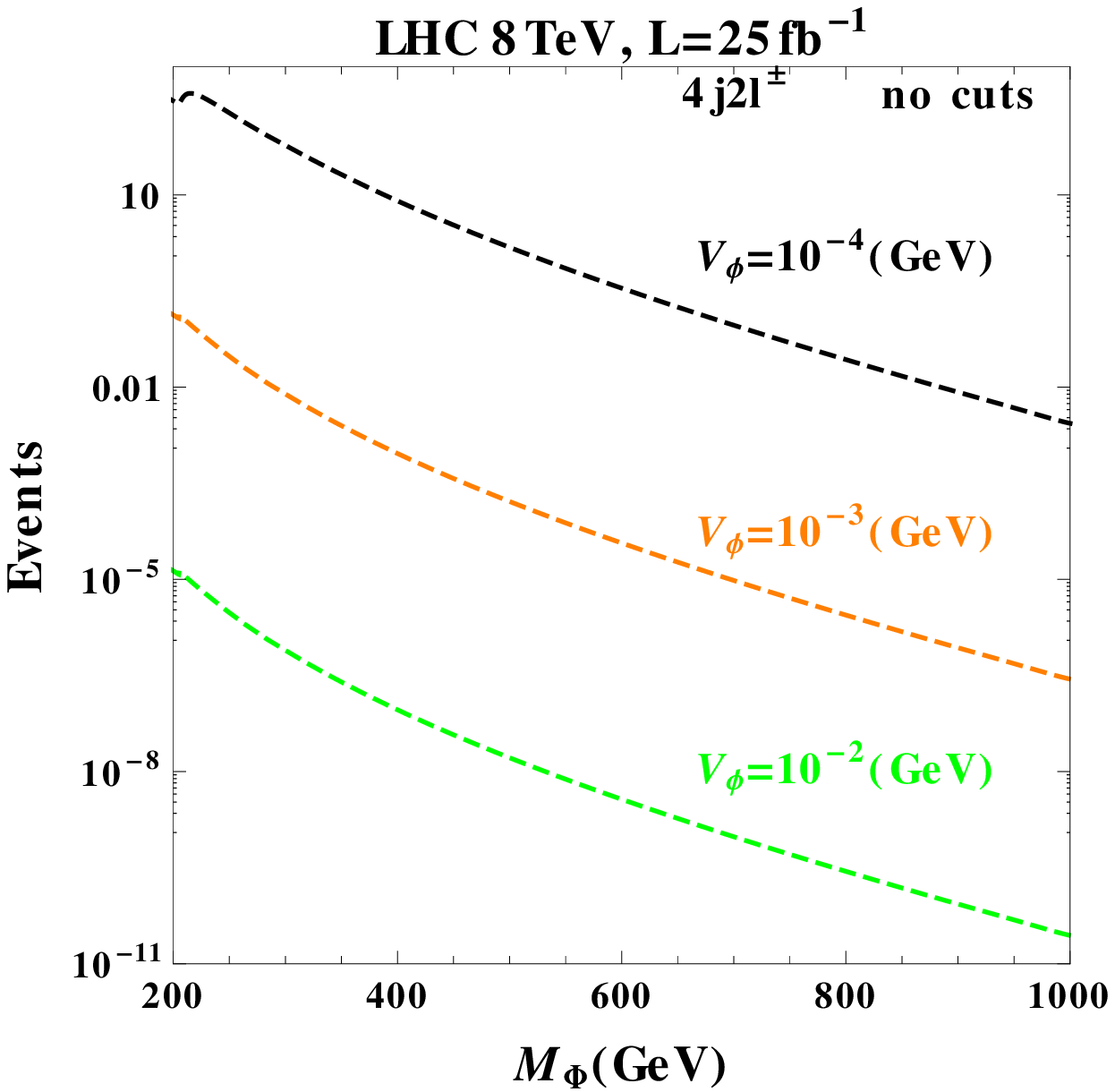}
\includegraphics[width=0.45\linewidth]{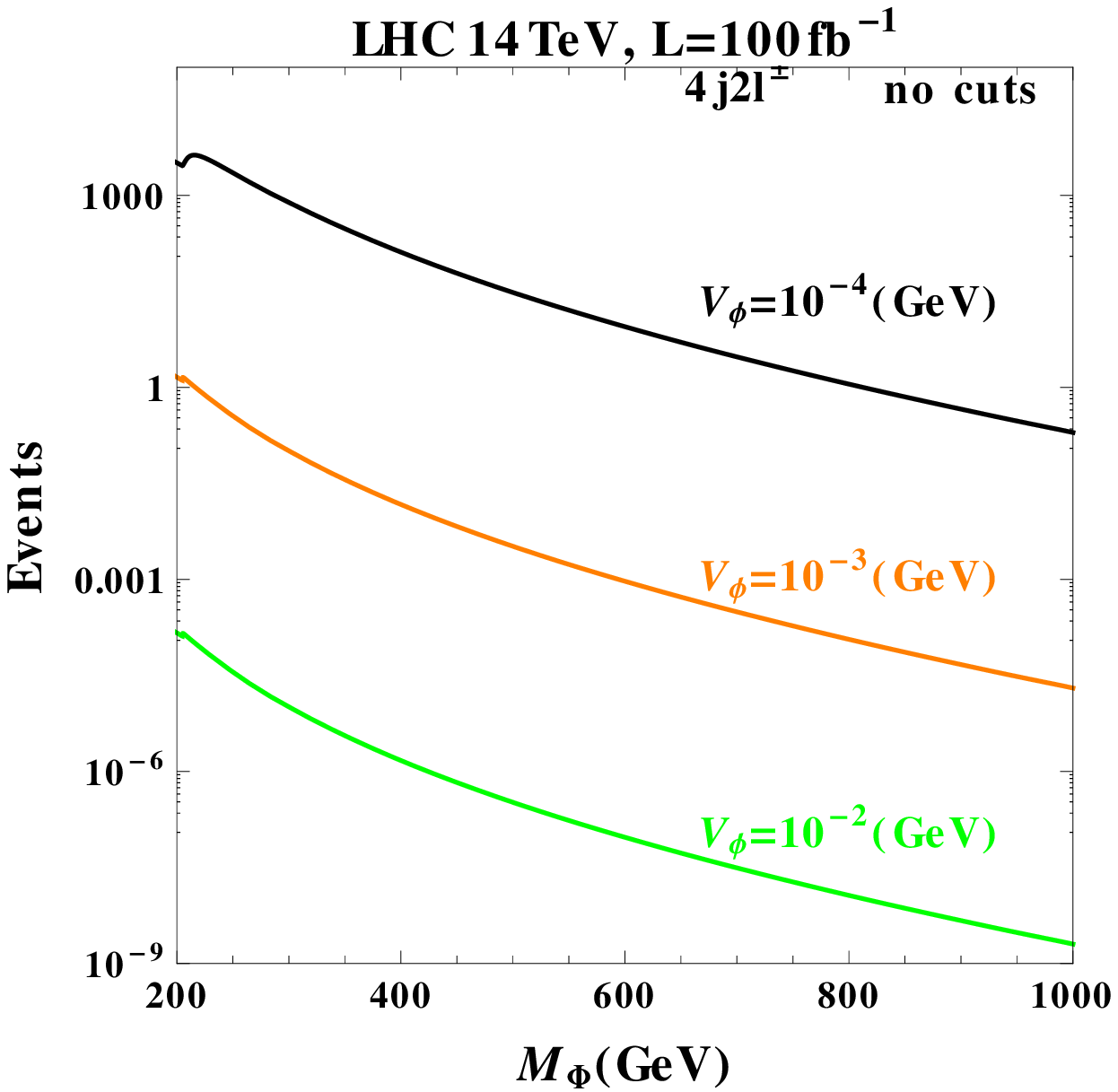}
\end{center}
\caption{Predicted number of signals in various channels of $\Phi$ production versus $M_\Phi$.
\label{fig:phievent1}}
\end{figure}

\begin{figure}[!htbp]
\begin{center}
\includegraphics[width=0.45\linewidth]{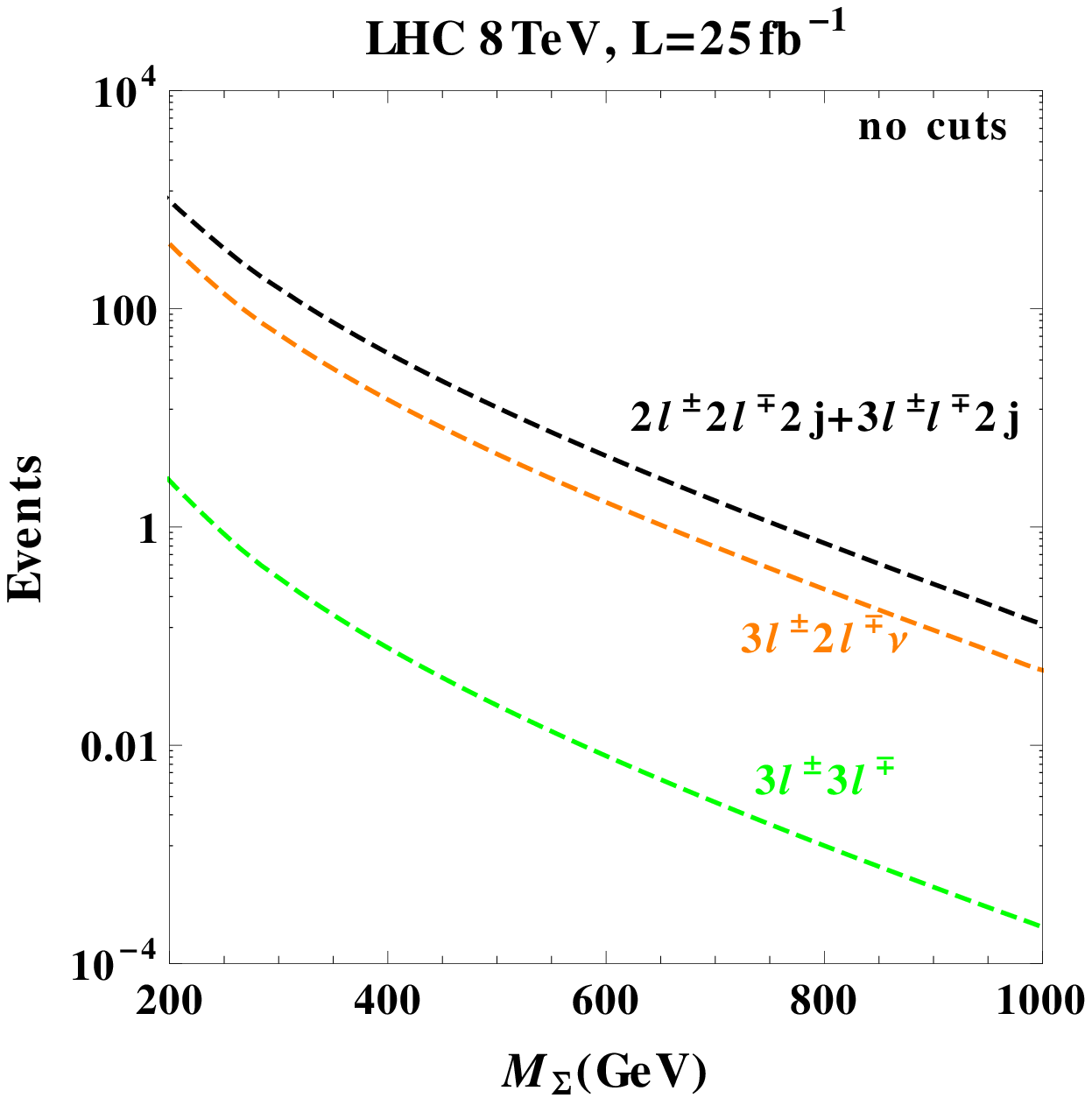}
\includegraphics[width=0.45\linewidth]{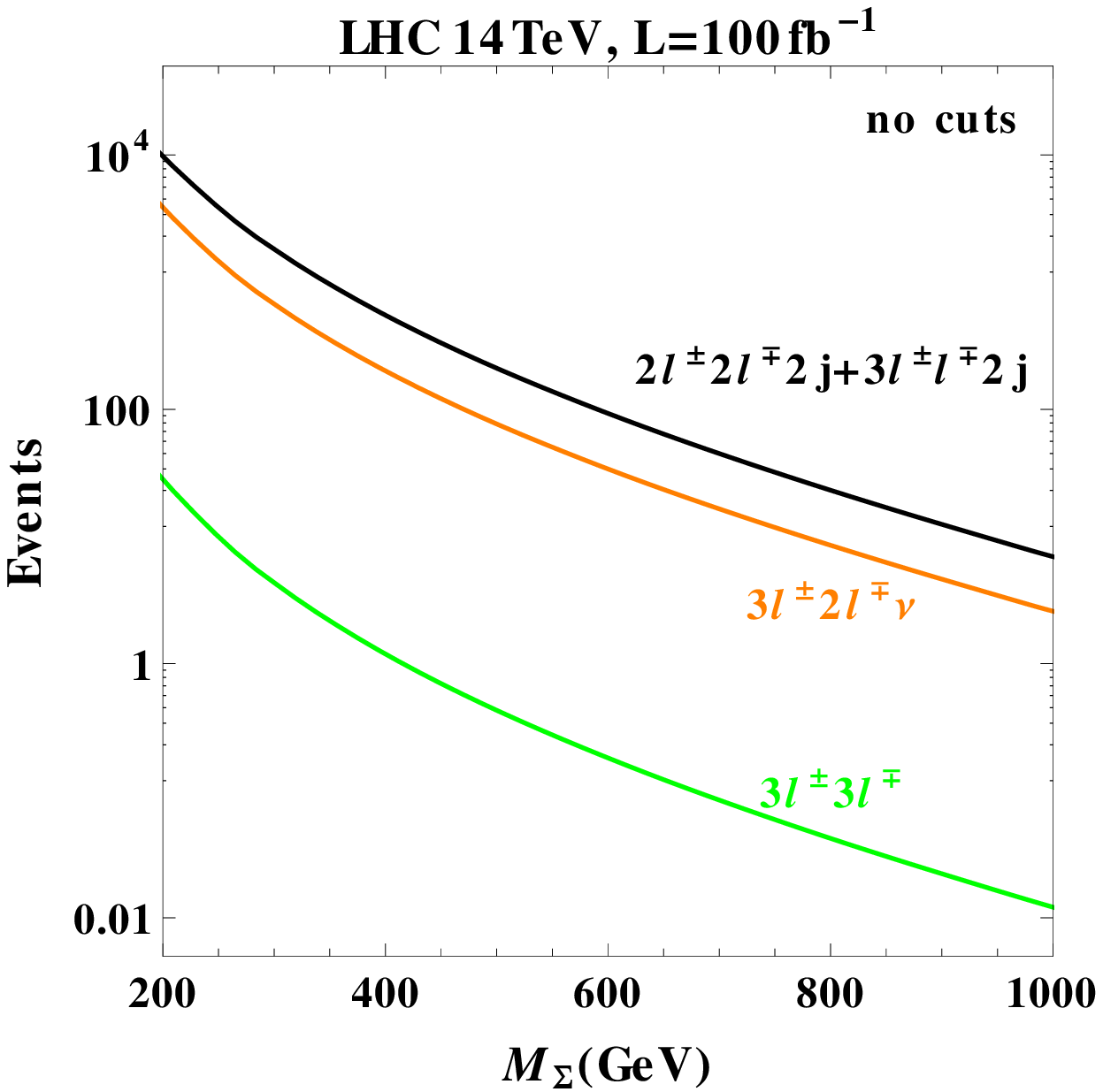}
\end{center}
\caption{Predicted number of signals in various channels of $\Sigma$ production versus $M_\Sigma$.
\label{fig:sigmaevent1}}
\end{figure}

Before studying the simulation and analysis of signal channels, we estimate the signal events
using the production cross sections and branching ratios discussed in the previous subsection.
The number of signal events can be formally written as
\begin{eqnarray}
N=L\times\rm{production\;cross\;section}\times\rm{decay\;branching\;ratios}
\end{eqnarray}
where $L$ is the integrated luminosity. Given a sufficient number of events $N$, the mass of a new particle is reconstructed by the invariant mass of combinations of particles in the final state.
This procedure can be applied to any signal channels. In Figs. \ref{fig:phievent1} and \ref{fig:sigmaevent1}, we present
the signal events for each channel versus the new particle masses $M_{\Phi,\Sigma}$ without imposing
any cuts. From Fig. \ref{fig:phievent1} one sees that the scalar signal channels are sensitive to
$v_\Phi$. In particular, as we discussed earlier,
the number of events of $2\ell^{\pm}2\ell^{\mp}$ and $2\ell^{\pm}4j$ channels drop rapidly with increasing $v_\Phi$,
while the $2\ell^{\pm}4j+\cancel{E_T}$ channels behave oppositely.
This is understandable since both $2\ell^{\pm}2\ell^{\mp}$ and $2\ell^{\pm}4j$ final states include the purely leptonic decay modes of the doubly charged scalars (see Table \ref{tab:signal}), which is significant only for $v_\Phi<10^{-4}~\GeV$.
For this reason, we choose two benchmarks in our simulation: $v_\Phi=10^{-4}~\GeV$ for the $2\ell^{\pm}2\ell^{\mp}$ and $2\ell^{\pm}4j$ channels, and $v_\Phi=10^{-2}~\GeV$ for the remaining
channels. Finally, we recall that the signal channels of $\Sigma$ do not depend on $v_\Phi$.

The SM backgrounds are also estimated by {\tt Madgraph5}. For simplicity, we only consider the irreducible backgrounds in our study. We do not include the following backgrounds:
(1) multijet final states like the $Wnj/Znj$ production where extra jets come from initial-state or/and final-state radiation and especially pile-up;
(2) isolated charged leptons from $b$ quark decays such as the $t\bar{t}/t\bar{t}nj/b\bar{b}nj$ backgrounds; and
(3) charged leptons missed by detectors or one jet misidentified as a lepton.
Some of them are analyzed and found to be relevant in multi-lepton signal
searches \cite{delAguila:2008cj}.
An accurate prediction of those backgrounds is difficult and can best be estimated from the
experimental data which is beyond the scope of our work. Fortunately, for the high $p_T$ leptons
which are most relevant to our signals, their effect is estimated to be small.

In the following subsections, we will present our analysis in each signal channel.
In signal simulation, we only consider electrons and and muons in our definition of a
lepton, i.e., $\ell=e,~\mu$.
For all the channels, we first impose the following basic cuts for the event selection,
\begin{eqnarray}
&&p_T(\ell)>15~\GeV,\;|\eta(\ell)|<2.5,
\nonumber\\
&&p_T(j)>20~\GeV,\;|\eta(j)|<2.5,
\nonumber\\
&&\Delta R_{\ell\ell}>0.4,\;\Delta R_{j\ell}>0.4,\;\Delta R_{jj}>0.4.
\end{eqnarray}
After that, specific cut selections are designed according to the properties of final states to
reduce the SM background in each channel.

\begin{figure}
\begin{center}
\includegraphics[width=0.45\linewidth]{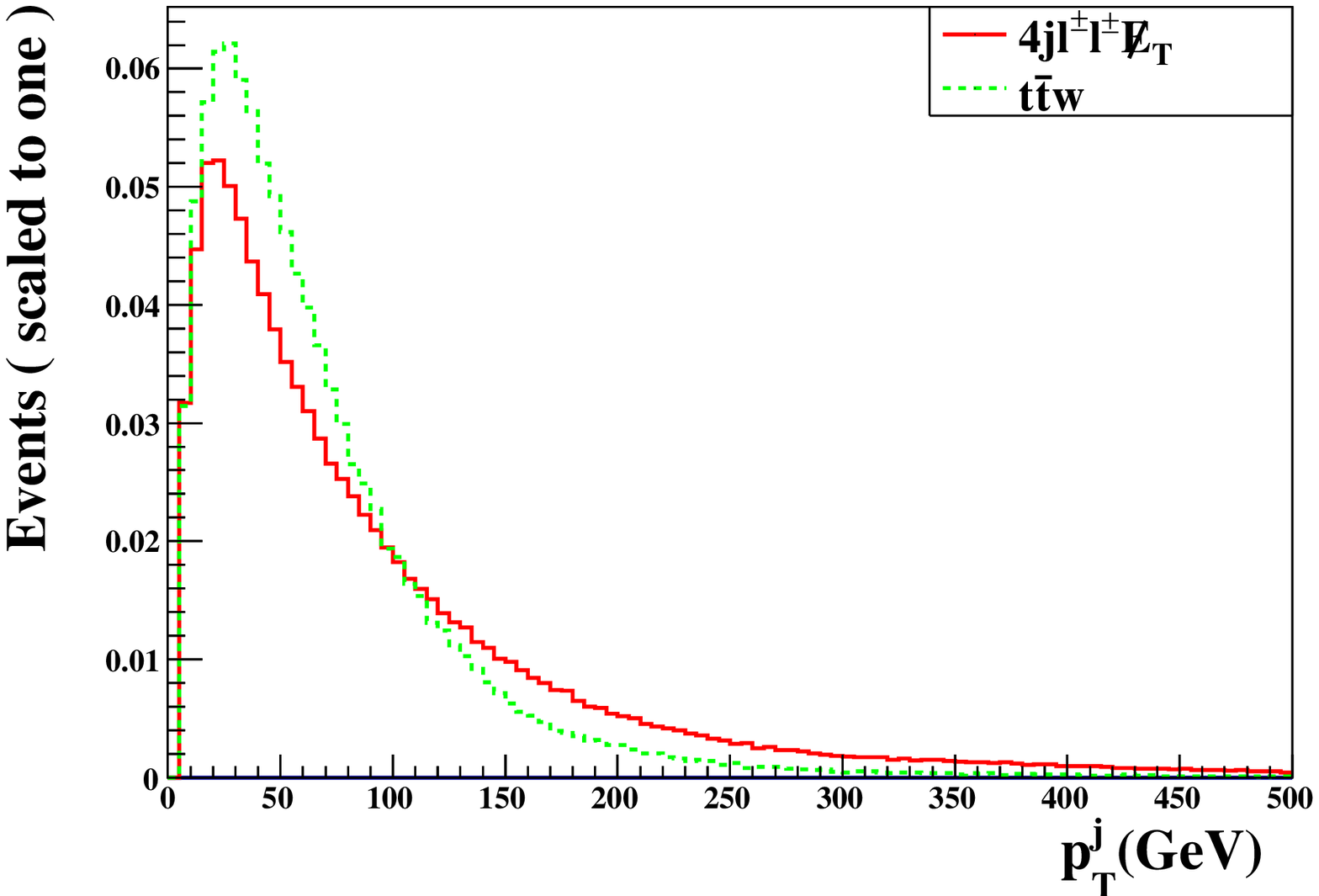}
\includegraphics[width=0.45\linewidth]{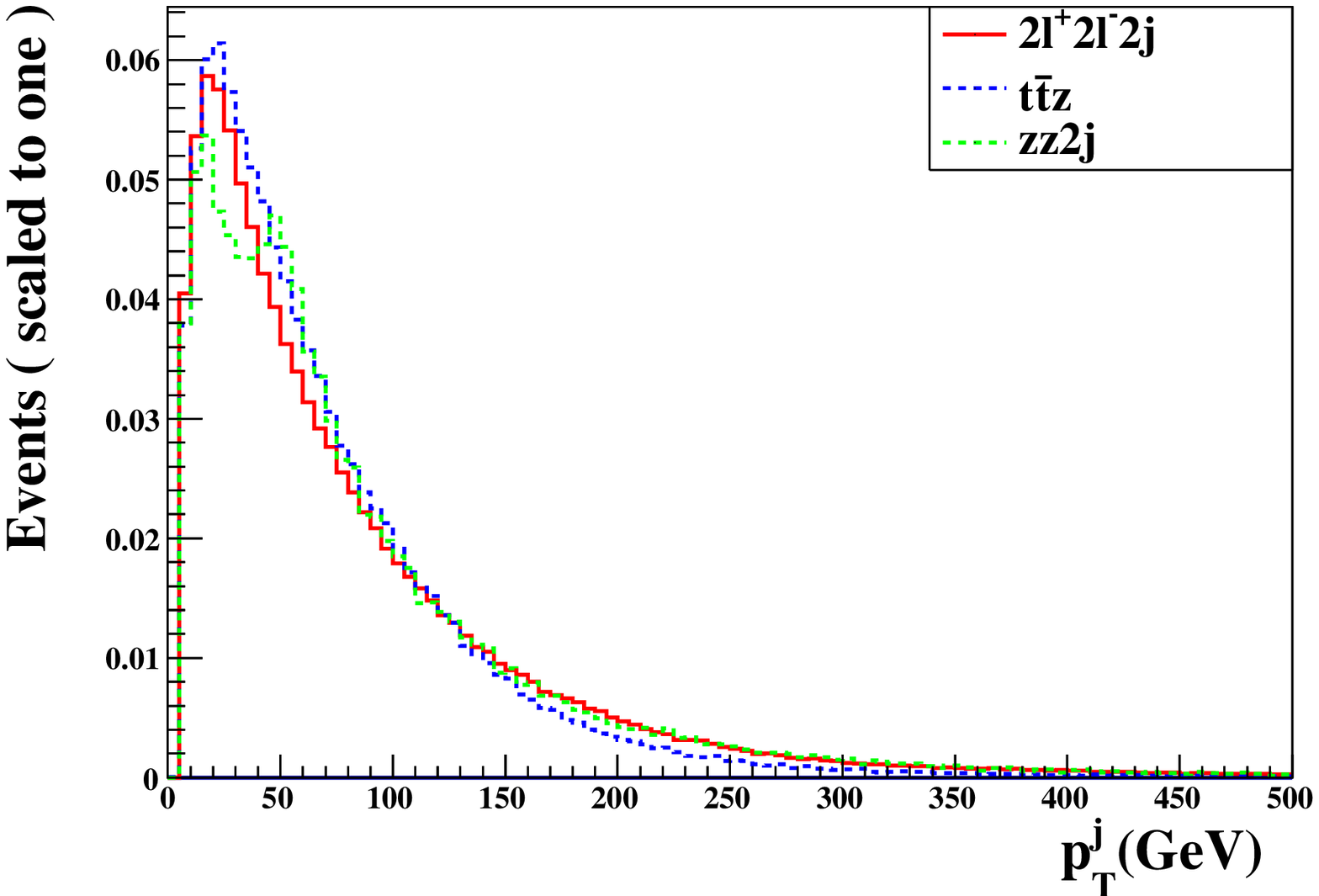}
\includegraphics[width=0.45\linewidth]{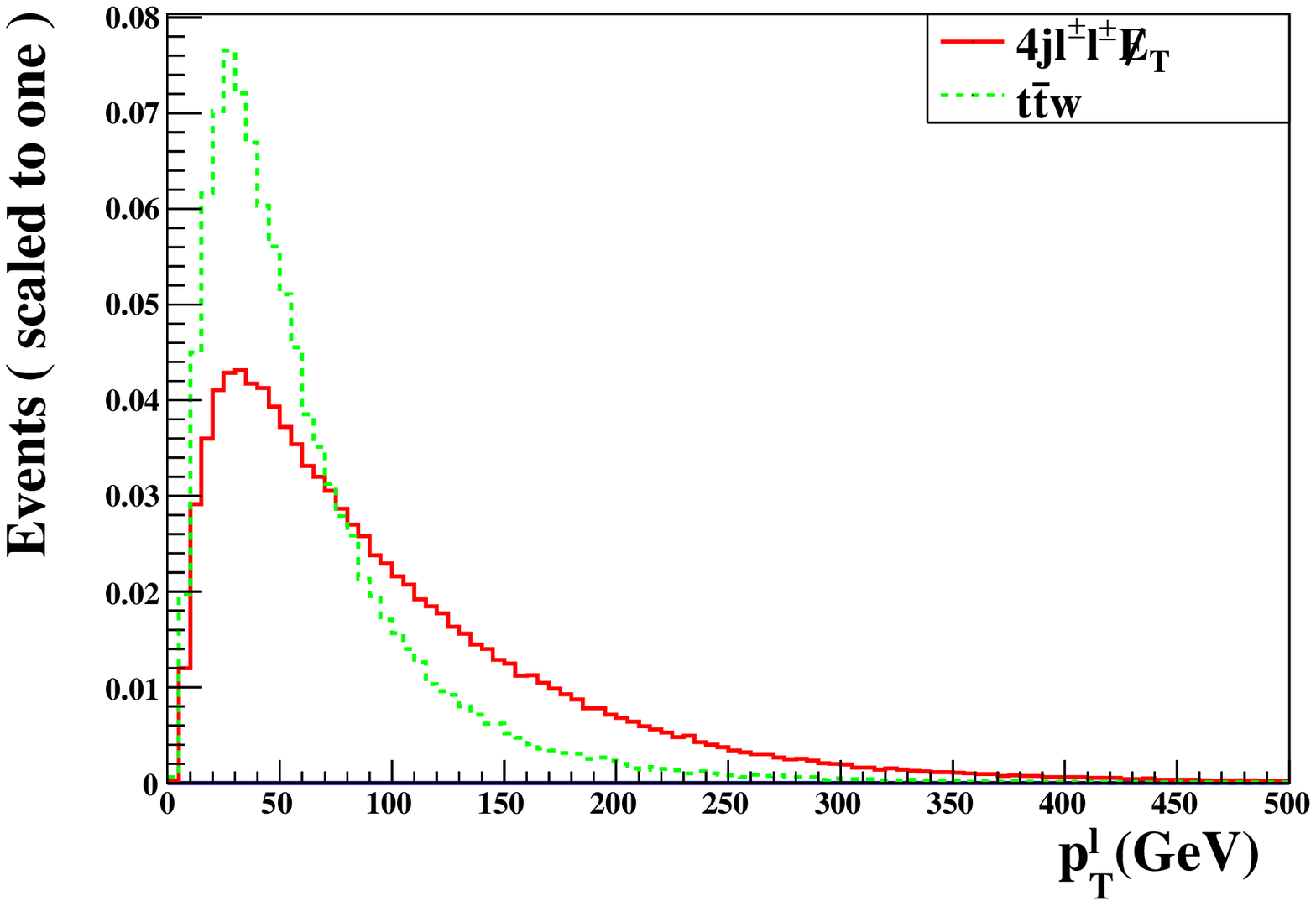}
\includegraphics[width=0.45\linewidth]{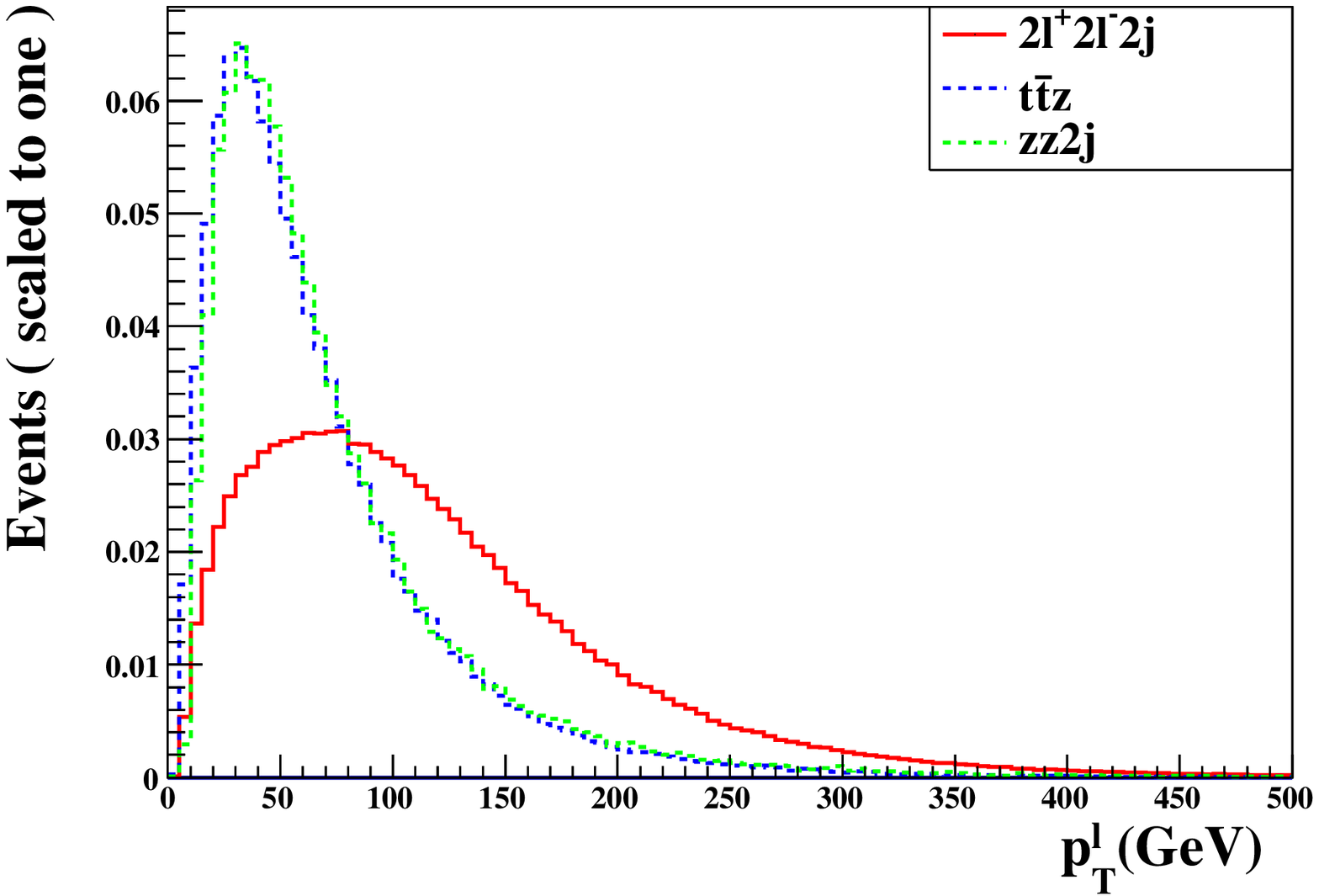}
\includegraphics[width=0.45\linewidth]{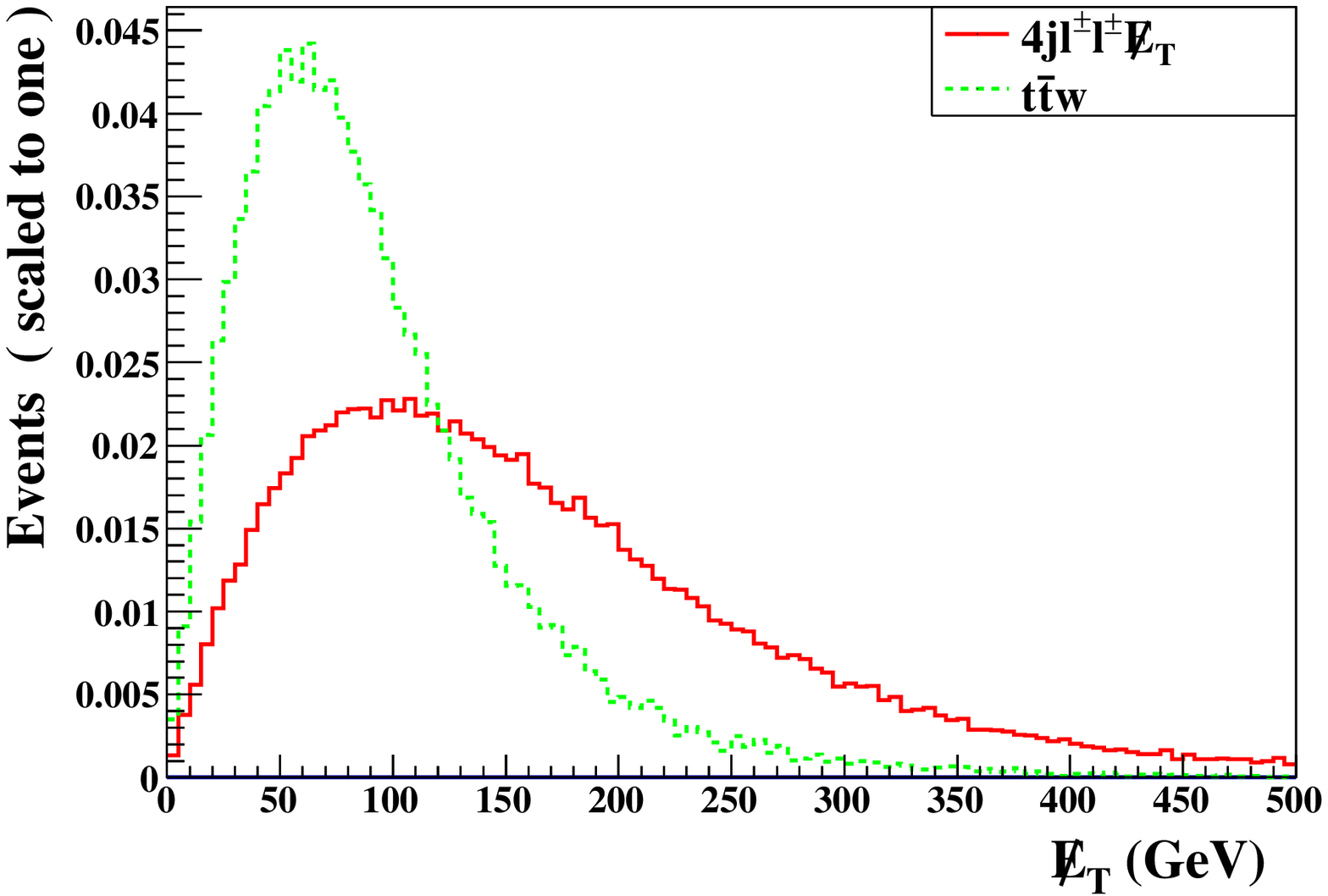}
\includegraphics[width=0.45\linewidth]{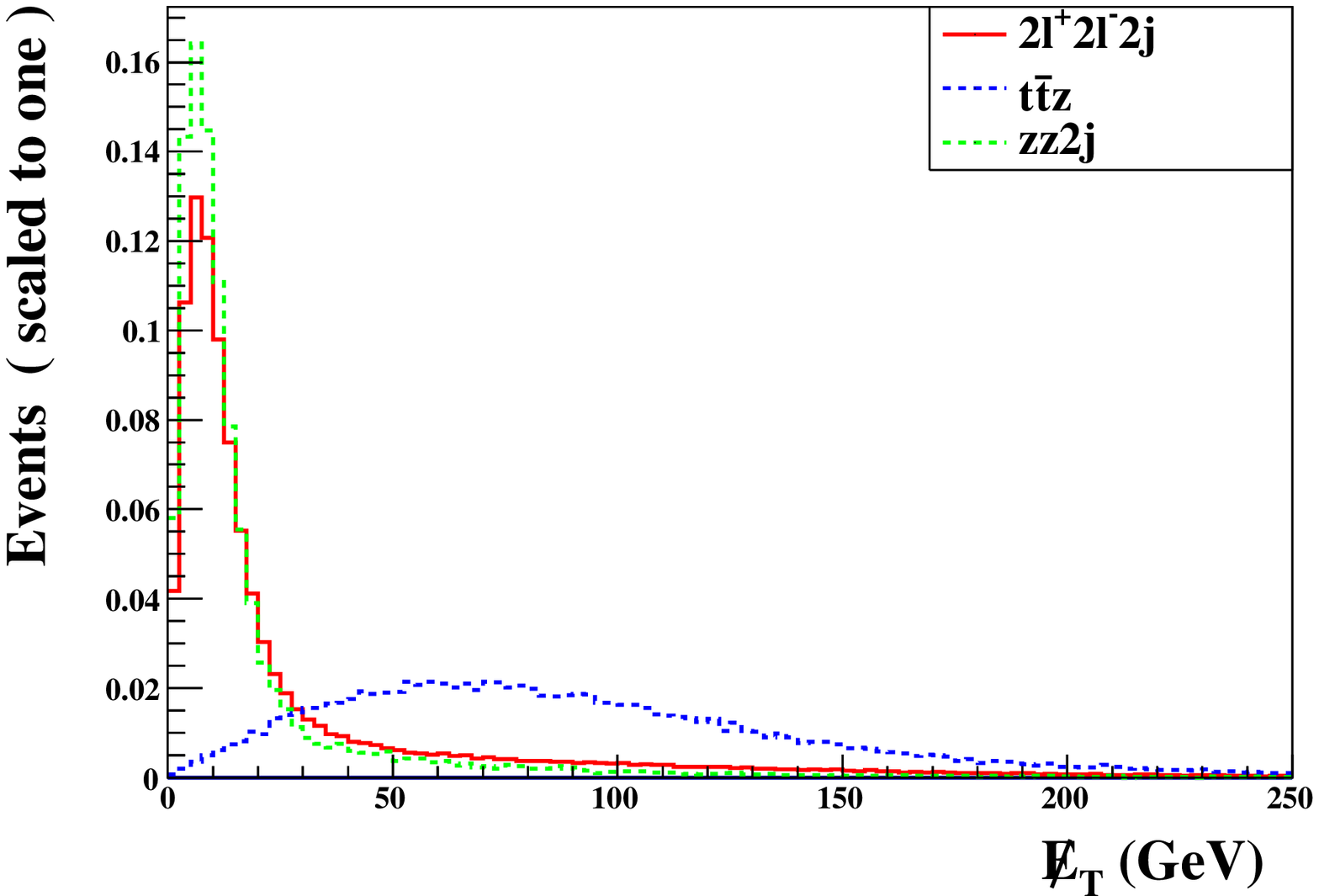}
\end{center}
\caption{Distributions of transverse momenta $p_T(\ell),~p_T(j)$ and missing energy $\cancel{E_T}$
after imposing basic cuts for the signals $4j2\ell^{\pm}+\cancel{E_T}$ (left panel) and
$2\ell^+2\ell^-2j$ (right) and the backgrounds.
\label{fig:phi_pt}}
\end{figure}

\begin{figure}
\begin{center}
\includegraphics[width=0.45\linewidth]{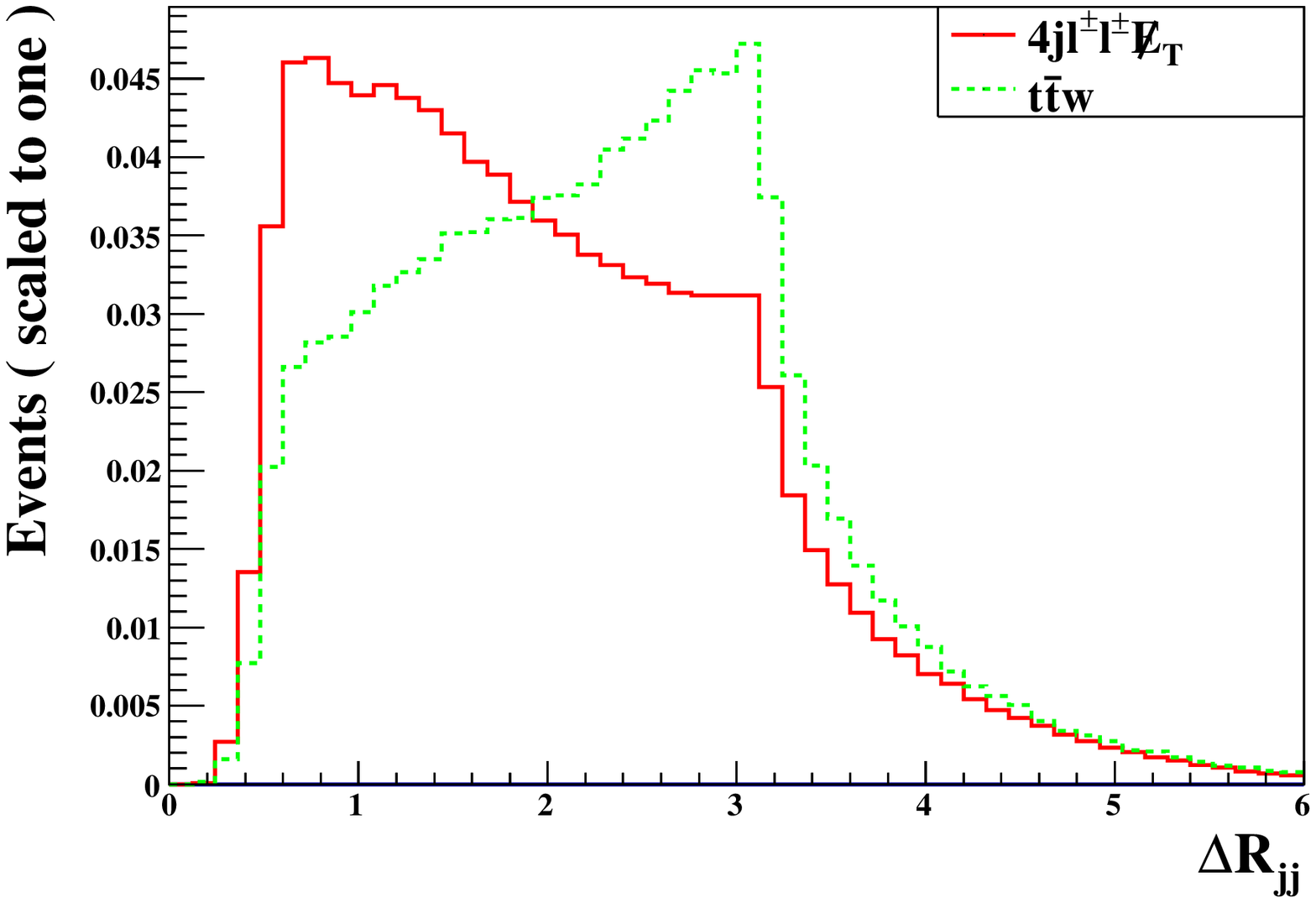}
\includegraphics[width=0.45\linewidth]{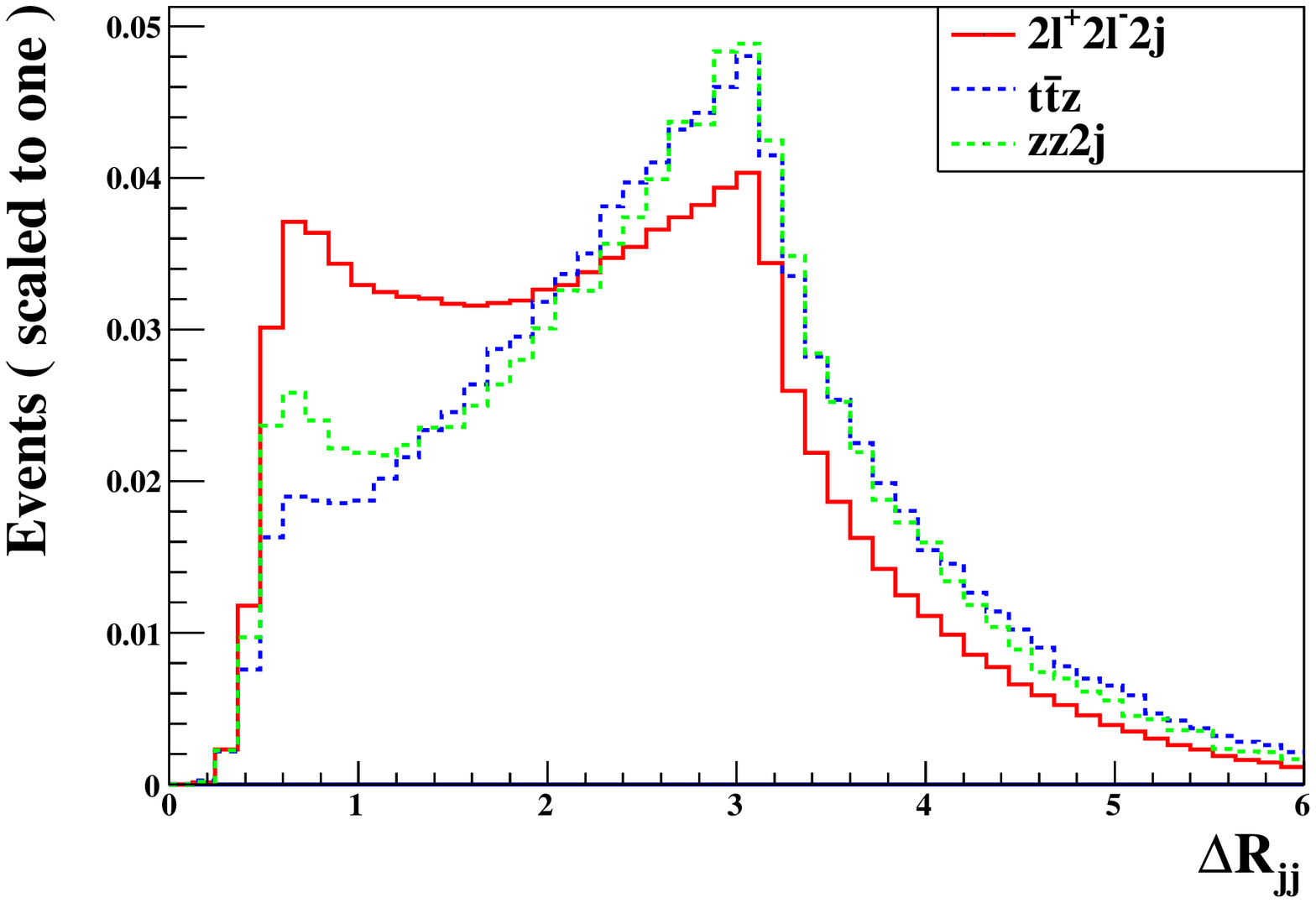}
\includegraphics[width=0.45\linewidth]{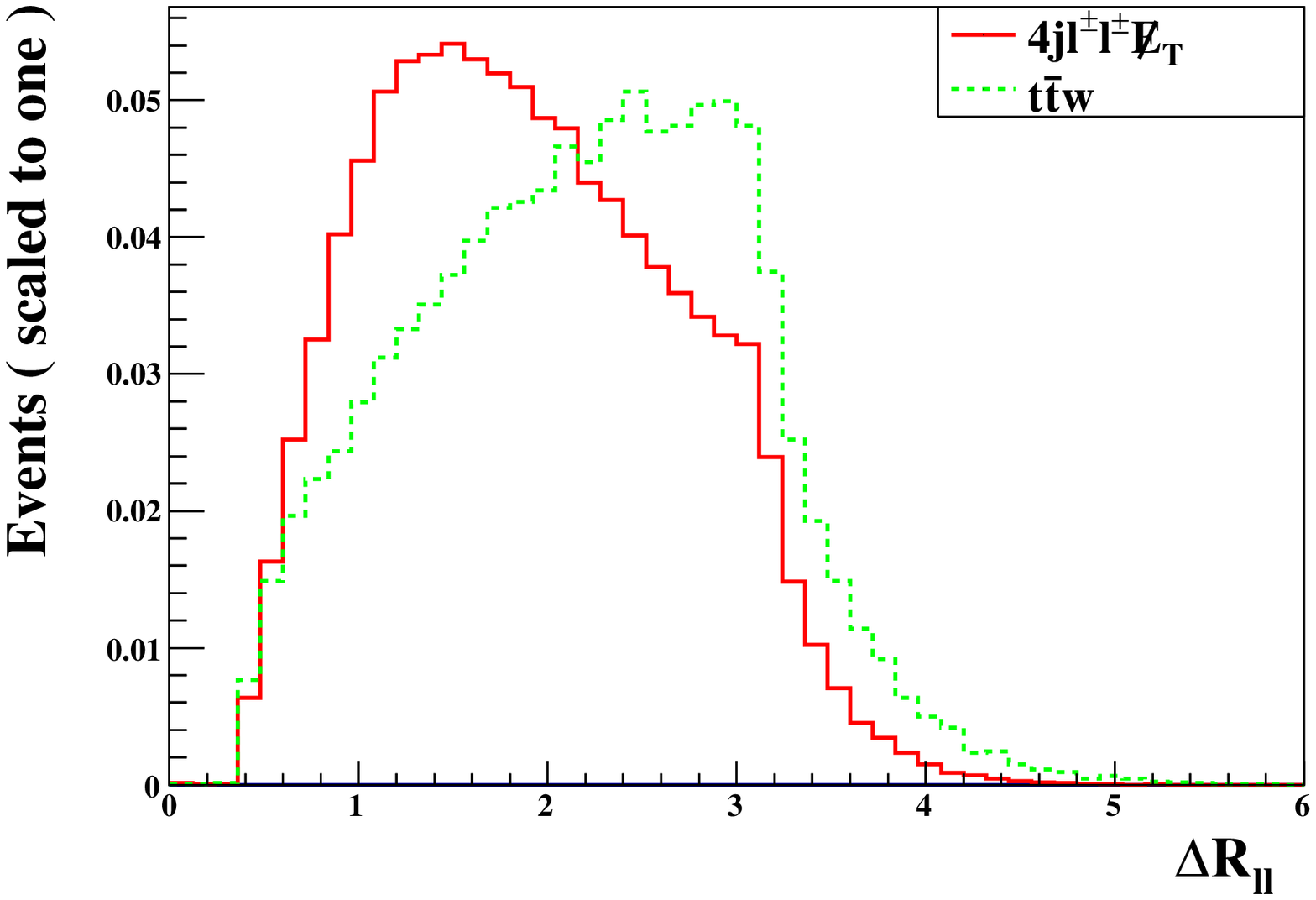}
\includegraphics[width=0.45\linewidth]{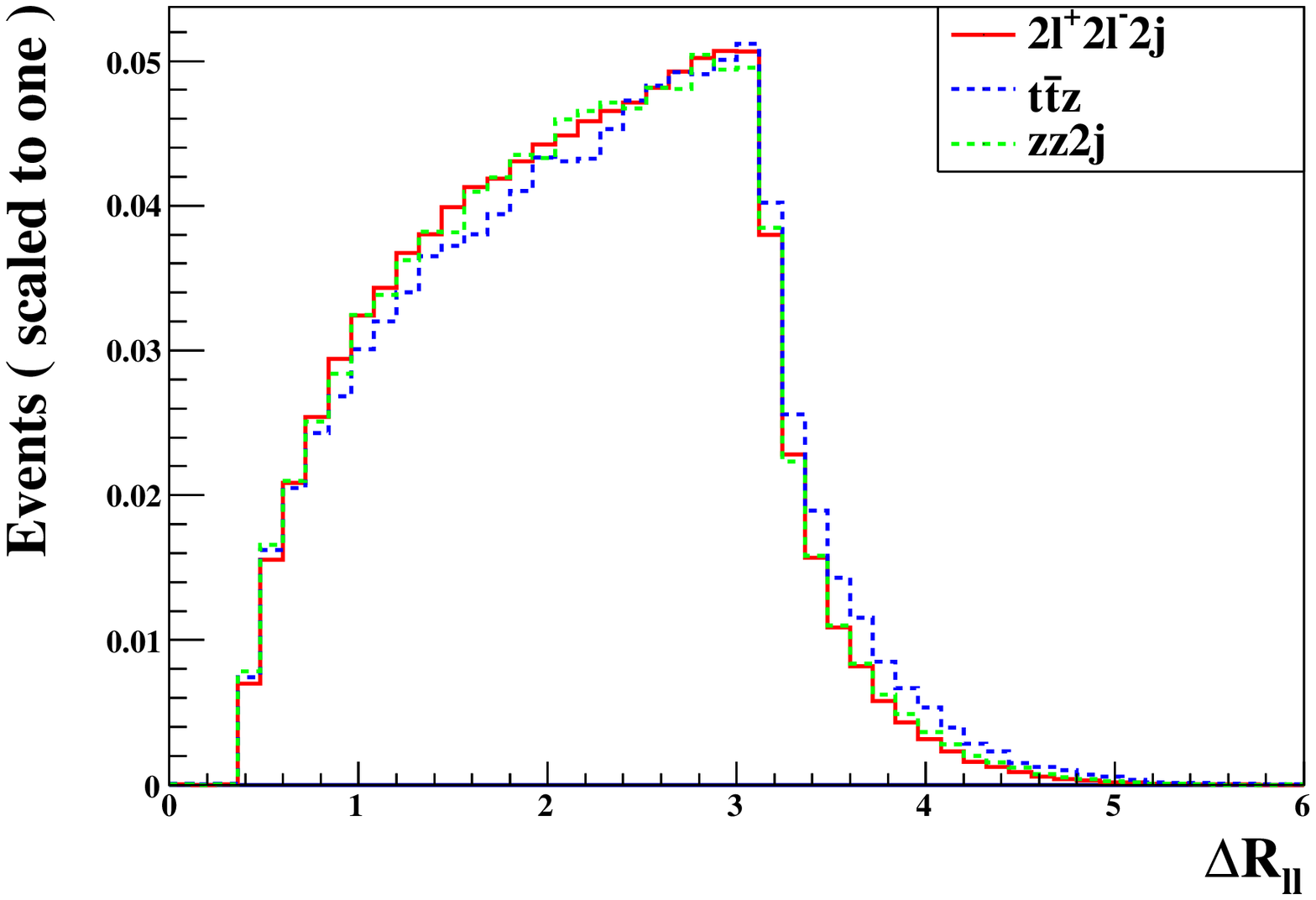}
\includegraphics[width=0.45\linewidth]{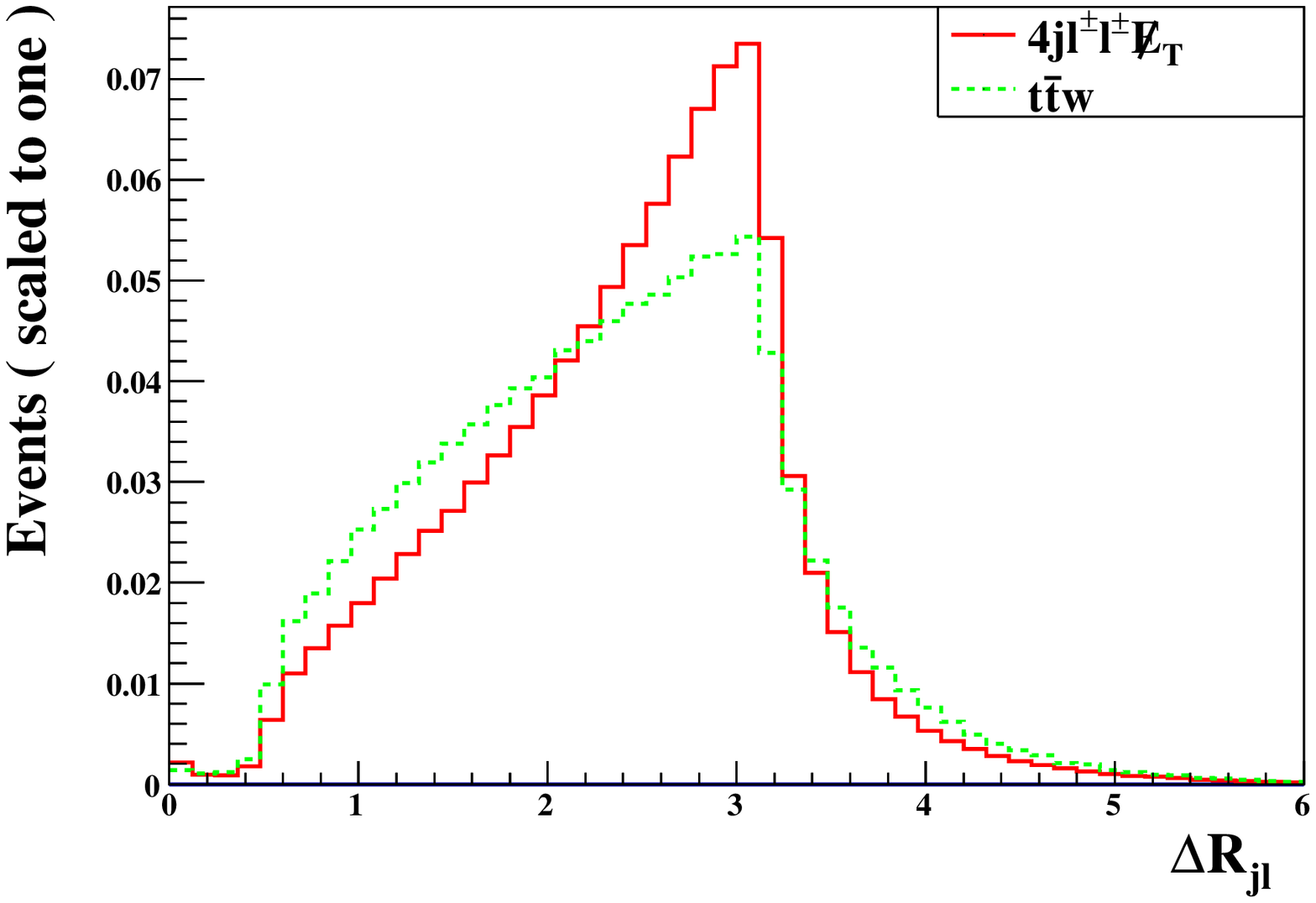}
\includegraphics[width=0.45\linewidth]{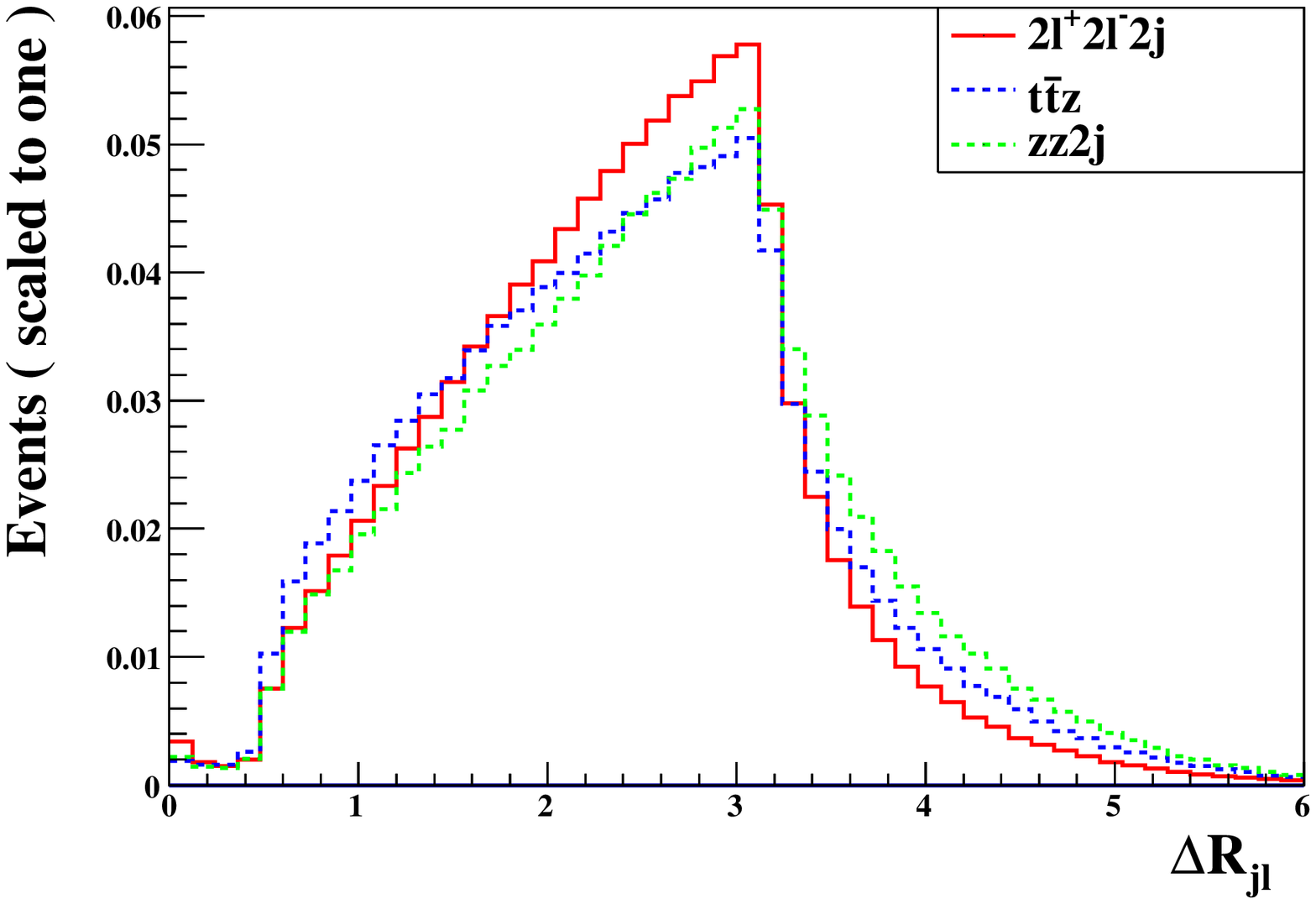}
\end{center}
\caption{Similar to Fig. \ref{fig:phi_pt}, but for distributions of particle separations
$\Delta R_{\ell\ell,jj,j\ell}$.
\label{fig:phi_deltaR}}
\end{figure}

\subsubsection{$\Phi$ production: $4j2\ell^{\pm}+\cancel{E_T}$ signal}
\label{subsubsec:phi 1}

As we discussed above, the pure gauge boson channel becomes dominant for
$v_\Phi > 10^{-4}~\GeV$, where the doubly charged scalars decay mainly into like-sign di-$W$'s.
Thus the channel $\Phi_{+2}\to W^+ W^+$ serves as the identifier for doubly charged scalars.
Although the absence of LNV decays prevents us extracting information on neutrino mass patterns
directly, the existence of mixing between new scalars and the SM Higgs would indicate that
some mechanism of neutrino mass generation is at work. It is helpful to search for channels
involving the mixing. These include the following decays whose amplitudes are proportional to
$v_\Phi$,
\begin{equation}
\Phi_{+1} \to  W^+ h/ t\bar{b},\quad
H_0  \to hh, \ W^+ W^- ,\quad
A_0  \to h Z.
\end{equation}
Both $\Phi_{+1}H_0$ and $\Phi_{+2}\Phi_{+1}^*$ production channels are useful to test gauge couplings
and confirm the nature of new scalars. However, it would be difficult to search the channel
$H_0\Phi_{+1}\to hhh W^+ $ which contains 6 $b$-jets in the final state. The reconstruction of three
SM Higgs bosons from multiple $b$ jets would suffer from large irreducible QCD backgrounds.
We thus focus on the $\Phi_{+2}\Phi_{+1}^*/\Phi_{+2}^*\Phi_{+1}$ channels. We reconstruct
the events by searching for hadronic decays of like-sign $W^\pm$ pairs from
$\Phi_{+2}/\Phi_{+2}^*$ decays and hadronic decays $W^\mp\to j j,~h\to b\bar{b}$ which in turn come from
$\Phi_{+1}/\Phi_{+1}^*$ decays,
\begin{eqnarray}
pp \to \Phi_{+2}\Phi_{+1}^{*}(\Phi_{+2}^{*}\Phi_{+1}) \to W^\pm W^\pm + hW^\mp/\bar{t}b(t\bar{b})\to jjb\bar{b}\ell^\pm \ell^\pm \nu\nu.
\end{eqnarray}
The decay branching ratios were given in Fig. \ref{fig:decayphi2} and in addition
$\Br(h \to b\bar{b})\approx 67.7\%$.

\begin{figure}
\begin{center}
\includegraphics[width=0.8\linewidth]{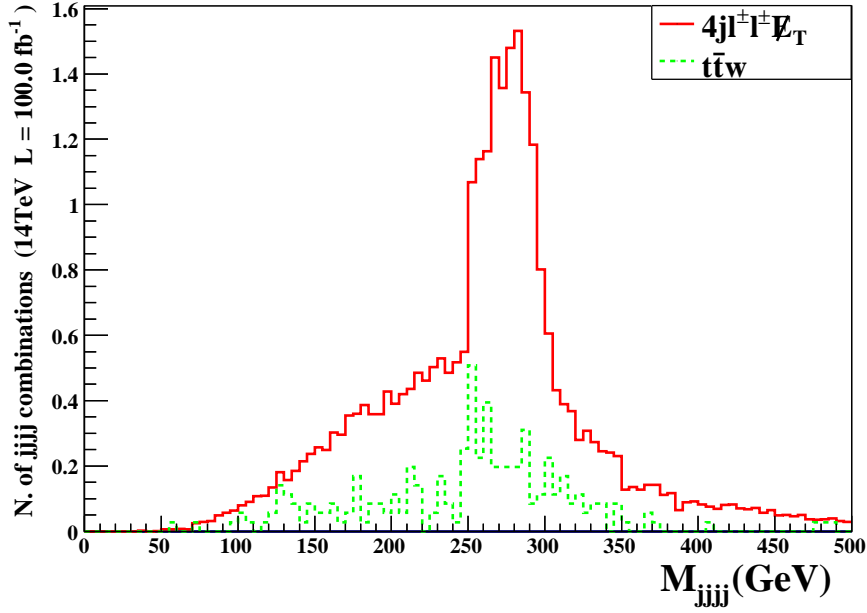}
\end{center}
\caption{Reconstruction of doubly and singly charged scalars via 4-jet invariant mass $M_{jjjj}$ for $M_{\Phi_{+2},\Phi_{+1}}=300~\GeV$ at LHC $14~\TeV,~L=100\;\fb^{-1}$.
The vertical axis displays the number of four-jet combinations.
\label{fig:phi_signal1}}
\end{figure}

The leading irreducible background to this signal is, $t \bar{t} W^\pm \to jjb\bar b W^\pm W^\pm$. Another irreducible background $jjjjW^\pm W^\pm$ is much smaller.
\footnote{This result is based on the following estimate.
{\tt Madgraph} gives that $jjjW^\pm W^\pm\to jjj\ell^\pm \ell^\pm\cancel{E}_T$ is about
$1.1~(0.31)~\fb$ at
$14~(8)~\TeV$. We chose the MLM scheme \cite{Hoche:2006ph} to perform a matching between the soft jets generated by {\tt Pythia} and the hard jets generated by {\tt Madgraph} to avoid double counting with
a matching scale ${\tt xqcut}\sim 60~\GeV$. Considering additional $\alpha_s$ and phase space
suppression, the cross section for $jjjjW^\pm W^\pm$ is much smaller than $t\bar{t}W^\pm$.}
The distributions of transverse momenta $p_T(\ell),~p_T(j)$, missing transverse energy $\cancel{E_T}$
and the particle separations $\Delta R _{\ell\ell,jj,j\ell}$ after imposing the basic cuts for both
signal and background are displayed in the left panel of Figs. \ref{fig:phi_pt} and \ref{fig:phi_deltaR}. There are several interesting features for the particle separation distributions. First, the peak of $\Delta R_{j\ell}$ is about 3.0, which indicates that the jets and leptons are isolated enough.
Second, the distributions of $\Delta R _{\ell\ell,jj}$ are distinct for signal and background -- The leptons and jets from $t\bar{t}W^\pm$ are more isolated than the signal channels.
We can thus distinguish between the two by this kinematical variable.
To be specific, we apply the following cuts,
\begin{eqnarray}
\Delta R_{jj}<2.5,\quad \Delta R_{\ell\ell}<2.5.
\end{eqnarray}
Additionally, instead of using $b$ tagging, we choose the following cuts on the transverse momentum
and missing energy to keep the maximal signal events.
\begin{eqnarray}
p_T(\ell) > 50~\GeV,\quad p_T(j) > 100~\GeV,\quad \cancel{E_T}>30~\GeV.
\end{eqnarray}
Since the dijets in the signal come from $W$ or Higgs decays, we require their invariant mass to be in
the $W/H$ mass window (with $M_W=80~\GeV$ and $M_h=125~\GeV$)
\begin{eqnarray}
M_W-20~\GeV < M_{jj} < M_h+25~\GeV.
\end{eqnarray}
For the $\Phi_{+2}\Phi_{+1}^*$ channel,
one branch of doubly and singly charged scalars gives like-sign dilepton pairs plus large missing energy while the other decays hadronically. We can thus fully
reconstruct them through the 4-jet invariant mass $M_{jjjj}$. At the benchmark point $M_{\Phi_{+2},\Phi_{+1}}=300~\GeV$, we require that $M_{jjjj}$ fall into the mass window
\begin{eqnarray}
250~\GeV < M_{jjjj} < 350~\GeV.
\label{eq_4jmass}%
\end{eqnarray}
The distribution of $M_{jjjj}$ for both the signal and the leading background $t\bar{t}W^\pm$ are
plotted in Fig. \ref{fig:phi_signal1} at LHC 14 TeV with $L=100\;\fb^{-1}$. We present in Table
\ref{tab:phi_signal1} the survival numbers of events and statistical significance $S/\sqrt{S+B}$
upon imposing the cuts step by step, for LHC 14 TeV and 8 TeV, respectively. We see that all the cuts chosen here are efficient enough in keeping the signal and suppressing the background. The signal to background ratio can reach $4:1$ and about 11 signal events survive at LHC 14 TeV.
However, the signal is too small to be observable at LHC 8 TeV.

\begin{table} [htbp]
\begin{center}
\begin{tabular}{|c|c|c|c|c|c|c|}
\hline
 cuts
& \multicolumn{2}{|c|}{signal $4j2\ell^{\pm}+\cancel{E_T}$}
& \multicolumn{2}{|c|}{bkg $t\bar{t}W^{\pm}$}
& \multicolumn{2}{|c|}{$S/\sqrt{S+B}$}\\
\hline
no cuts & \multicolumn{2}{|c|}{201 (14.7)} & \multicolumn{2}{|c|}{1409 (124)} & \multicolumn{2}{|c|}{5.02 (1.25)}  \\
basic cuts & \multicolumn{2}{|c|}{143 (11)} & \multicolumn{2}{|c|}{851 (82)} & \multicolumn{2}{|c|}{4.54 (1.14)}   \\
$(\cancel{E_T},p_T(\ell),p_T(j))>(30,50,100)~\GeV$
& \multicolumn{2}{|c|}{118.8 (8.8)} & \multicolumn{2}{|c|}{344.4 (30)} & \multicolumn{2}{|c|}{5.52 (1.41)} \\
$\Delta R_{jj}$, $\Delta R_{\ell\ell}<2.5$ & \multicolumn{2}{|c|}{33.04 (2.54)} & \multicolumn{2}{|c|}{31.7 (3.05)} & \multicolumn{2}{|c|}{4.1 (1.1)}\\
60$<M_{jj}/\GeV<150$ ($M_{W,h}$ reconst.) & \multicolumn{2}{|c|}{29.5 (2.3)} & \multicolumn{2}{|c|}{28.6 (2.7)} & \multicolumn{2}{|c|}{3.87 (1.03)}\\
250$<M_{jjjj}/\GeV<350$ & \multicolumn{2}{|c|}{11.3 (0.9)} & \multicolumn{2}{|c|}{2.5 (0.2)} & \multicolumn{2}{|c|}{3.04 (0.85)}\\
\hline
\end{tabular}
\end{center}
\caption{Survival numbers of events and statistical significance $S/\sqrt{S+B}$ after imposing
each cut sequentially at $M_{\Phi_{+2},\Phi_{+1}}=300~\GeV$ and for LHC $14~\TeV,~L=100\;\fb^{-1}$
($8~\TeV,~L=25\;\fb^{-1}$ in parentheses).}
\label{tab:phi_signal1}
\end{table}

\subsubsection{$\Phi$ production: $4j2\ell^{\pm}$ signal}
\label{subsubsec:phi_2}
\begin{figure} [htbp]
\begin{center}
\includegraphics[width=0.45\linewidth]{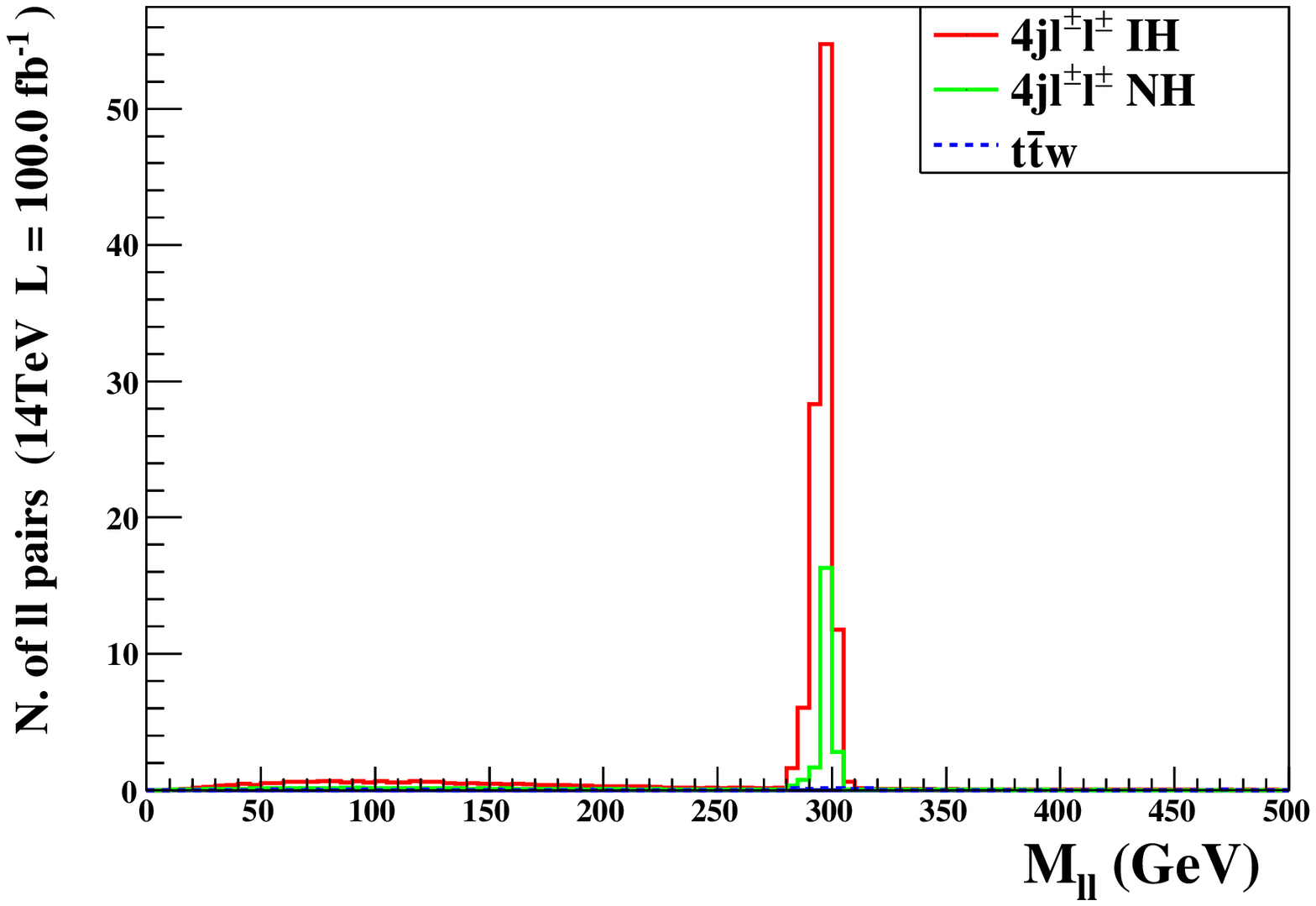}
\includegraphics[width=0.45\linewidth]{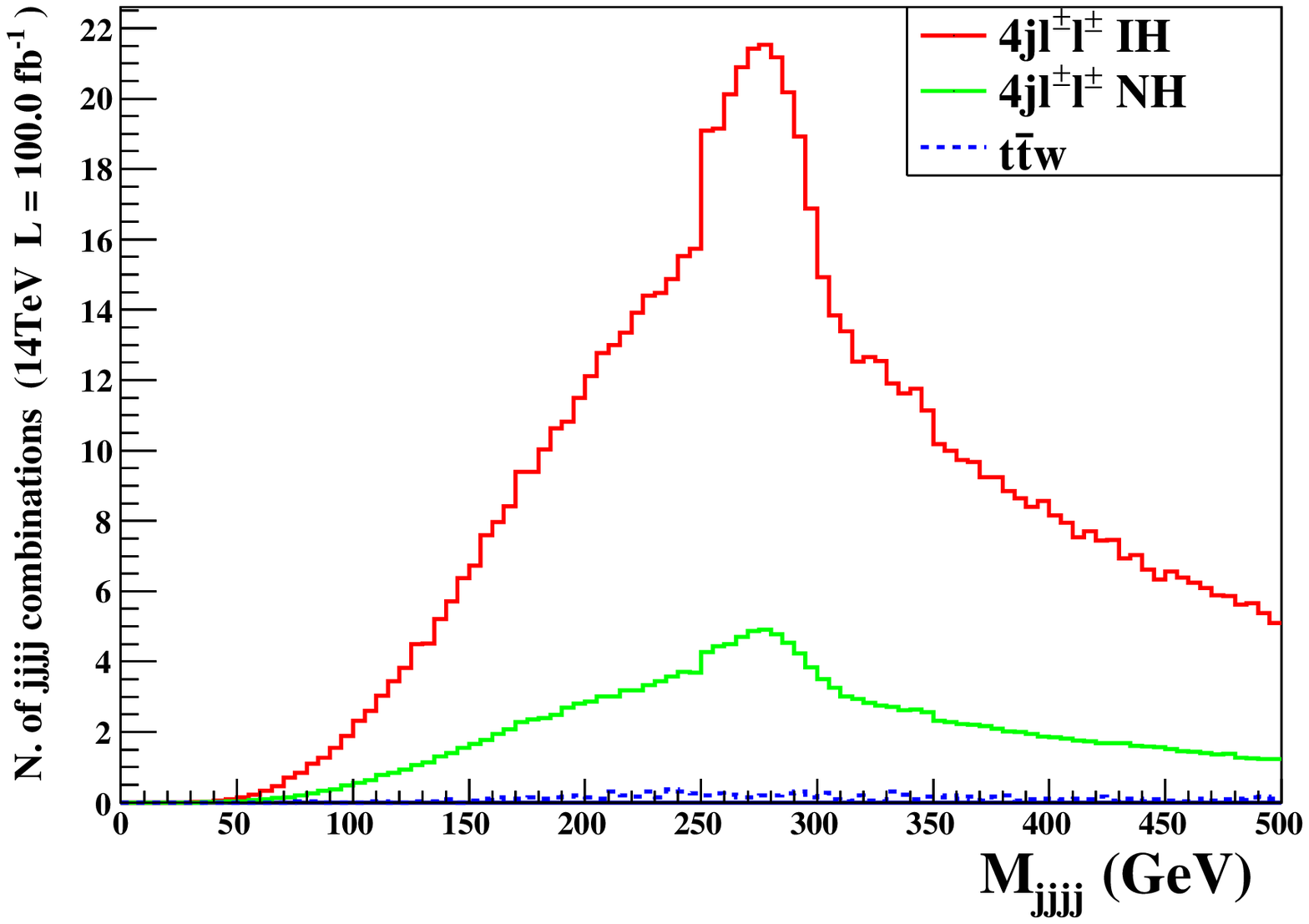}
\end{center}
\caption{Reconstruction of doubly and singly charged scalars via the dilepton
(left panel) and 4-jet (right) invariant mass for $M_{\Phi_{+2},\Phi_{+1}}=300~\GeV$ at LHC $14~\TeV,~L=100\;\fb^{-1}$.
\label{fig:phi_signal2}}
\end{figure}

Since the two main decays of $\Phi_{+2}$ and $\Phi_{+2}^*$ are roughly comparable around
$v_\Phi=10^{-4}~\GeV$, we found it advantageous to employ both to select signals in their
pair production, with one of them into like-sign dileptons and the other into like-sign di-$W$'s.
In addition, the associated $\Phi_{+2}\Phi_{+1}^{*}(\Phi_{+2}^{*}\Phi_{+1})$ production
contributes also to the signal. The singly-charged scalar decays to $W^\pm h$ and $tb$ have
some features that can be utilized for our purpose.
To reduce invisible neutrinos without cutting cross sections too much, we require both the $W$
boson and the SM Higgs decay into hadrons.
We apply similar cut selections as in the
$4j2\ell^{\pm}+\cancel{E_T}$ channel except that we do not use cuts on $\Delta R$ since the
$\Delta R _{\ell\ell,jj}$ distributions for signal and background are not distinct
enough, and that the missing energy cut for neutrinos is
replaced by a veto cut,
\begin{eqnarray}
\cancel{E_T}<30~\GeV.
\label{eq_vetocut}%
\end{eqnarray}
Another difference is that we can now fully reconstruct both doubly and singly charged scalars
by forming the 4-jet and dilepton invariant masses. For the former, we adopt the mass window
shown in eq. (\ref{eq_4jmass}), and for the latter, again at the benchmark point
$M_{\Phi_{+2},\Phi_{+1}}=300~\GeV$, we assume
\begin{eqnarray}
280~\GeV<M_{\ell\ell}<320~\GeV.
\label{eq_2lmass}%
\end{eqnarray}
Their distributions in the IH and NH cases are displayed in Fig. \ref{fig:phi_signal2} for
LHC $14~\TeV$, and the number of events is collected in Table \ref{tab:phi_signal2}.
This channel has considerable signal events and statistical significance, which can reach more than 100 events for the IH case and about 20 events even for the NH with an
integrated luminosity of $L=100\;\fb^{-1}$.
The better sensitivity to the IH case is common to both $\Phi$ and $\Sigma$ production
signals. It arises as a joint consequence of lepton-flavor dependence
in the decay branching ratios of heavy particles,
eqs. (\ref{relation_IH},\ref{relation_NH}) for heavy fermions,
and of the fact that only the electrons and muons are counted as leptons in signal simulation.
Actually, lepton flavor relations similar to eqs. (\ref{relation_IH},\ref{relation_NH}) also appear for heavy scalars.

\begin{table} [htbp]
\begin{center}
\begin{tabular}{|c|c|c|c|c|c|c|}
\hline
 cuts
& \multicolumn{2}{|c|}{signal $4j2\ell^{\pm}$}
& \multicolumn{2}{|c|}{bkg $t\bar{t}W^{\pm}$}
& \multicolumn{2}{|c|}{$S/\sqrt{S+B}$}\\
\hline
& IH  & NH & \multicolumn{2}{|c|}{} & IH  & NH \\
\hline
no cuts & 406 (29.7) & 81.6 (6) & \multicolumn{2}{|c|}{1409 (124)} & 9.53 (2.39) & 2.11 (0.52)\\
basic cuts & 296.6 (22.5) & 60.2 (4.7) & \multicolumn{2}{|c|}{851.3 (81.9)} & 8.75 (2.2) & 1.99 (0.5)
\\
$\cancel{E_T}<30~\GeV$,    & & & \multicolumn{2}{|c|}{} & &\\
$(p_T(\ell),p_T(j))>(50,100)~\GeV$ & 212.4 (16.2) & 42.7 (3.4) & \multicolumn{2}{|c|}{36.1 (3.2)} & 13.47 (3.68) & 4.81 (1.31)
\\
$60<M_{jj}/\GeV<150$ ($M_{W,h}$ reconst.),& & & \multicolumn{2}{|c|}{} & &\\
$280<M_{ll}/\GeV<320$ & 183.1 (13.9) & 37.1 (2.9) & \multicolumn{2}{|c|}{1.8 (0.1)} & 13.47 (3.72) & 5.94 (1.67)
\\
$250<M_{jjjj}/\GeV<350$ & 102.6 (7.7) & 21.8 (1.7) & \multicolumn{2}{|c|}{0.8 (0.04)} & 10.09 (2.76) & 4.59 (1.27)
\\
\hline
\end{tabular}
\end{center}
\caption{Similar to Table \ref{tab:phi_signal1}, but for the $4j2\ell^{\pm}$ signal and for both NH
and IH.
\label{tab:phi_signal2}}
\end{table}

\subsubsection{$\Phi$ production: $2\ell^{\pm}2\ell^{\mp}$ signal}
\label{subsubsec:phi_3}
This is a clean channel for the observation of pair production of doubly charged scalars with
practically little contamination from the SM background.
However, the signal events are also small compared to other channels. Only the
$\Phi_{+2}\Phi_{+2}^{*}$ and $A_0H_0$ production contributes, and the cross section for the latter
is smaller by about an order of magnitude than other channels.
Requiring the presence of four charged leptons further significantly reduces the signal especially
for the NH case, because the charged dilepton decays of $\Phi_{+2}/H_0/A_0$ are highly constrained
by the low energy LFV processes.
Although there is no intrinsic SM background for the LNV processes, there are some fake ones
which can lead to similar final states as our signal.
The main irreducible background comes from $ZZ\to \ell^+\ell^-\ell^+\ell^-$, and the
reducible background includes $ZW^+W^-\to \ell^+\ell^-\ell^+\nu \ell^-\nu$.

For the signal selection, we require the presence of four isolated charged leptons, two
positively charged and two negatively charged, whose individual transverse momentum $p_T(\ell)$
must be larger than $50~\GeV$.
The veto cut for the missing transverse energy in
eq. (\ref{eq_vetocut}) is applied to reduce the $ZW^+W^-$ background.
And the events containing a pair of oppositely charged leptons with an invariant mass within
$10~\GeV$ around $M_Z\approx 90~\GeV$ are vetoed.
This effectively cuts the $ZZ$ background almost without affecting the signal.
Finally, to reconstruct the new scalars, both like-sign dilepton pairs must pass the invariant
mass cut in eq. (\ref{eq_2lmass}).

\begin{figure} [htbp]
\begin{center}
\includegraphics[width=0.8\linewidth]{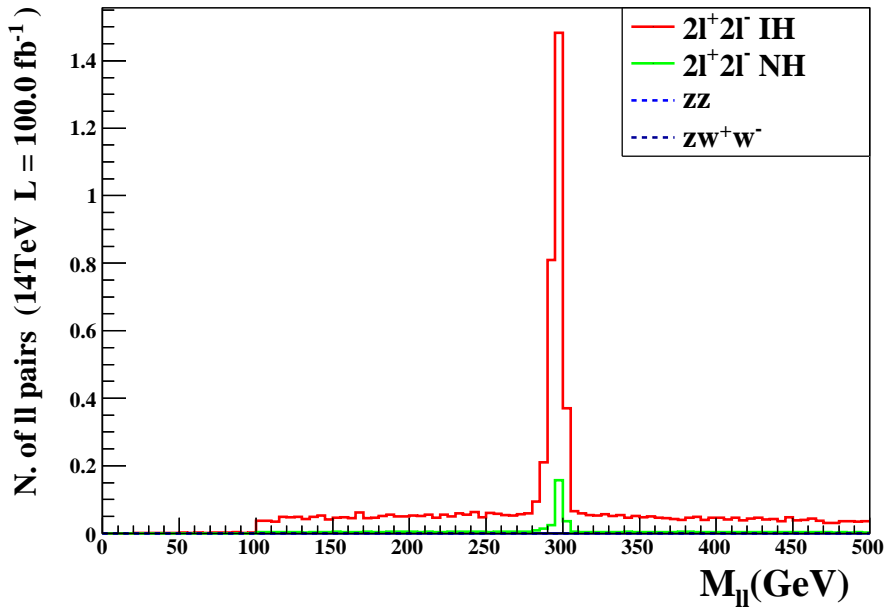}
\end{center}
\caption{Reconstruction of doubly charged and neutral scalars via dilepton invariant mass
$M_{\ell\ell}$ for $M_{\Phi_{+2},H_0,A_0}=300~\GeV$ at LHC $14~\TeV,~L=100\;\fb^{-1}$.
\label{fig:phi_signal3}}
\end{figure}

The invariant mass distribution of the dilepton pairs after all above cuts is displayed in
Fig. \ref{fig:phi_signal3}. In Table \ref{tab:phi_signal3}, we collect the event numbers for
signal and background upon imposing the cuts step by step.
No SM background survives these selections, while only $1.4$ ($0.11$) signal events can be reached
for the IH (NH) case.
It is worth recalling that the discovery of doubly charged scalars does not require to observe both dilepton pairs with an invariant mass around $M_{\Phi_{+2}}$, but it is sufficient to identify
a clear peak in the $M_{\ell\ell}$ distribution \cite{delAguila:2008cj}.
As can be seen in Fig. \ref{fig:phi_signal3}, the peaks are indeed clearly visible for both
IH and NH cases, though for NH the number of events at the peak is small even
with an integrated luminosity of $100~\fb^{-1}$.

\begin{table} [htbp]
\begin{center}
\begin{tabular}{|c|c|c|c|c|c|c|}
\hline
cuts & \multicolumn{2}{|c|}{signal $2\ell^{\pm}2\ell^{\mp}$} & bkg $ZZ$ & bkg $ZW^+W^-$ & 
\multicolumn{2}{|c|}{$S/\sqrt{S+B}$}
\\
\hline
& IH & NH & & & IH & NH
\\
\hline
no cuts & 31.6 (2.3) & 2.1 (0.15) & 4765 (555) & 31 (2) & 0.45 (0.096) & 0.03 (0.0064)\\
basic cuts & 9.7 (0.6) & 0.7 (0.04) & 610.4 (63.3) & 6.4 (0.5) & 0.39 (0.07) & 0.027 (0.005)\\
$\cancel{E_T}<30~\GeV$, $p_T^\ell>50~\GeV$ &
8 (0.5) & 0.5 (0.03) & 404.2 (43.5) & 0.7 (0.06) & 0.39 (0.072) & 0.026 (0.005)
\\
$80<M_{\ell^+\ell^-}/\GeV<100$ ($Z$ veto)&
7 (0.4) & 0.4 (0.03) & 81.3 (8.7) & 0.2 (0.02) & 0.74 (0.14) & 0.05 (0.009)
\\
$280<M_{\ell^\pm\ell^\pm}/\GeV<320$ &
1.4 (0.08) & 0.11 (0.006) & 0.0 (0.0) & 0.0 (0.0) & 1.16 (0.28) & 0.33 (0.08)
\\
\hline
\end{tabular}
\end{center}
\caption{Similar to Table \ref{tab:phi_signal1}, but for the $2\ell^{\pm}2\ell^{\mp}$ signal
and for both IH and NH.
\label{tab:phi_signal3}}
\end{table}

\subsubsection{$\Sigma$ production: $2\ell^{\pm}2\ell^{\mp}2j$ signal}
\label{subsubsec:sigma_1}

\begin{figure} [htbp]
\begin{center}
\includegraphics[width=0.45\linewidth]{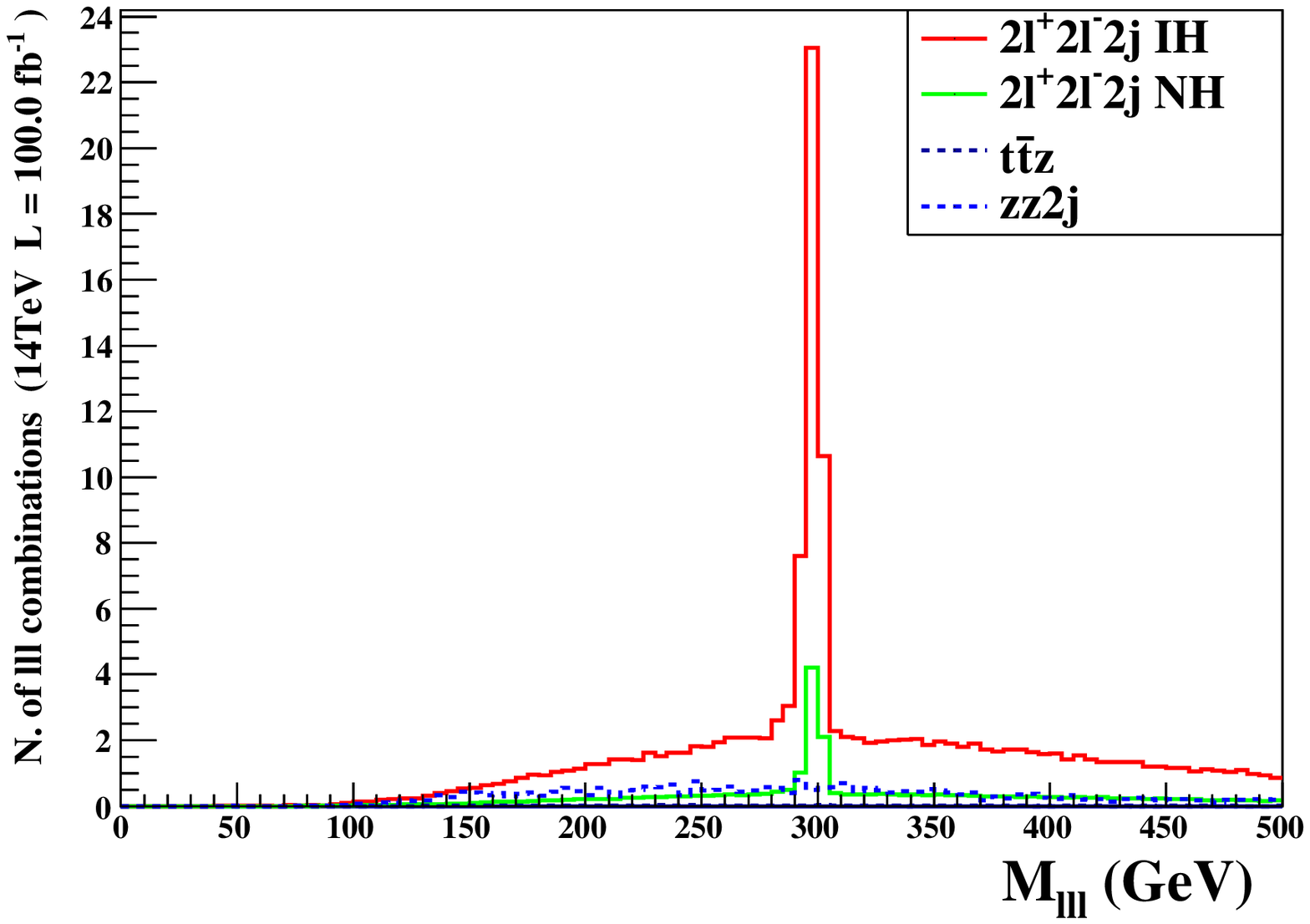}
\includegraphics[width=0.45\linewidth]{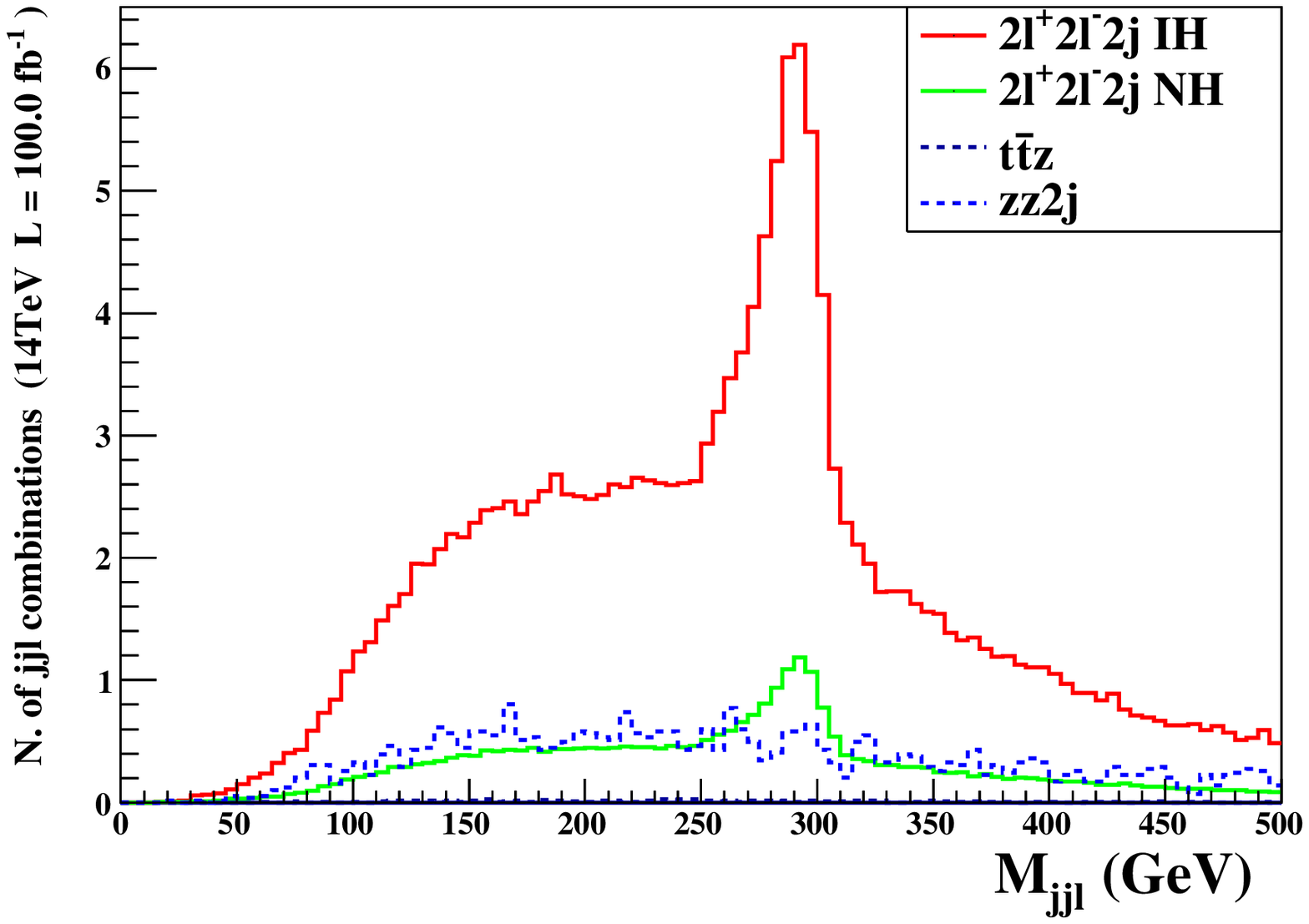}
\end{center}
\caption{Reconstruction of $\Sigma^{\pm}$, $\Sigma^{0}$ and $\Sigma^{\mp\mp}$ via a trilepton
(left panel) and a dijet-plus-one-lepton (right) invariant mass for $M_{\Sigma^{\pm},\Sigma^{0}}=300~\GeV$
at LHC $14~\TeV,~L=100\;\fb^{-1}$.
\label{fig:sigma_signal1}}
\end{figure}

In contrast to the previous LNV four-lepton final states, this signal is common to production of
new scalars and fermions. In the latter case,
the signal can result from many decay channels of the pair or associated production of fermions,
\begin{eqnarray}
&&\Sigma^{\pm}\Sigma^{\mp} \to \ell^\pm Z \ell^\mp Z \,,
\text{ with }ZZ \to \ell^+ \ell^-q \bar q \,,
   \nonumber \\
&&\Sigma^{\pm}\Sigma^{\mp} \to \ell^\pm Z \ell^\mp h \,,
\text{ with }Z \to \ell^+ \ell^- , h \to q \bar q \,,
   \nonumber \\
&& \Sigma^{0}\Sigma^{\pm} \to W^\pm \ell^\mp Z \ell^\pm \,,
\text{ with }Z\to \ell^+ \ell^- , W \to q \bar q' \,,
   \nonumber \\
&& \Sigma^{\pm}\Sigma^{\mp\mp} \to Z \ell^\pm W^\mp \ell^\mp \,,
\text{ with }Z \to \ell^+ \ell^- , W \to q \bar q'\,.
\label{eq:sigma_1}
\end{eqnarray}
The main backgrounds are $t \bar{t} Z \to b\bar b\ell^+\ell^-\ell^+\ell^-\nu\nu$ and
$ZZ2j\to \ell^+\ell^-\ell^+\ell^-jj$. Both of them are estimated using {\tt Madgraph}.
For $ZZ2j$, we use the {\tt MLM} matching scheme assuming ${\tt xqcut}=35~\GeV$.
The kinematical distributions upon imposing the basic cuts were displayed in the right panel of Figs. \ref{fig:phi_pt} and \ref{fig:phi_deltaR}.
After this, each of the four isolated charged leptons is required to have a transverse momentum
no smaller than $50~\GeV$, and a veto cut $\cancel{E_T}<30~\GeV$ facilitates reducing the
$t\bar{t}Z$ background. Analogous to the $4j2\ell^{\pm}+\cancel{E_T}$ final state,
the jet separation (see the right panel of Fig. \ref{fig:phi_deltaR})
is demanded to be smaller than $2.5$ to suppress further the background.
Since the signal dijet comes from $W,~Z,~h$ decays, it helps to separate it from the
background by concentrating on the invariant mass window, $60~\GeV<M_{jj}<150~\GeV$.
Considering that all channels in eq. (\ref{eq:sigma_1}) involve the decay chain
$\Sigma^{\pm}\to \ell^{\pm}Z \to \ell^{\pm}\ell^+\ell^-$, we do not apply $Z$ veto on the
dilepton invariant mass $M_{\ell^+\ell^-}$.
The heavy mass of $\Sigma^{\pm},~\Sigma^{0}$ and $\Sigma^{\mp\mp}$
can be fully reconstructed by forming a trilepton invariant mass $M_{\ell\ell\ell}$ and a
dijet-plus-one-lepton invariant mass $M_{jj\ell}$, by focusing on the windows respectively,
\begin{eqnarray}
280~\GeV<M_{\ell\ell\ell}<320~\GeV,~250~\GeV<M_{jj\ell}<350~\GeV.
\end{eqnarray}
The resulting distributions are shown in Fig. \ref{fig:sigma_signal1} for both NH and IH cases,
and the numbers of events after sequential cuts are collected in Table \ref{tab:sigma_signal1}.
The final number of signal events can reach $53$ ($9$) at LHC $14~\TeV,~L=100\;\fb^{-1}$ in the IH
(NH) case, which looks considerable.

\begin{table} [htbp]
\begin{center}
\begin{tabular}{|c|c|c|c|c|c|c|}
\hline
cuts& \multicolumn{2}{|c|}{signal $2\ell^{\pm}2\ell^{\mp}2j$} &
bkg $ZZ2j$ & bkg $t\bar{t}Z$ & \multicolumn{2}{|c|}{$S/\sqrt{S+B}$}
\\
\hline
& IH & NH & & & IH & NH \\
\hline
no cuts & 369 (30) & 64.5 (5.3) & 402 (34) & 198 (10) & 11.9 (3.46) & 2.5 (0.75)
\\
basic cuts & 315.9 (25.8) & 55.3 (4.6) & 378.8 (32.3) &170.4 (9.3) & 10.74 (3.15) & 2.25 (0.67)
\\
$\cancel{E_T}<30~\GeV,~p_T^\ell>50~\GeV$,& & & & & &
\\
$\Delta R_{jj}<2.5$ & 91.9 (9.1) & 15.3 (1.6) & 67.6 (6.7) & 2.8 (0.2) & 7.22 (2.29) & 1.66 (0.54)
\\
$60\!<\!M_{jj}/\GeV\!<\!150$ ($M_{W,Z,h}$ reconst.) & 74.7 (7.4) & 12.3 (1.3) & 44 (4.3) & 1.7 (0.13) & 6.81 (2.15) & 1.62 (0.53)
\\
$280<M_{\ell\ell\ell}/\GeV<320$,  & & & & & &
\\
$250<M_{jj\ell}/\GeV<350$& 52.9 (5.3) & 9.04 (0.94) & 12.5 (1.1) & 0.3 (0.016) & 6.52 (2.11) & 1.94 (0.66)
\\
\hline
\end{tabular}
\end{center}
\caption{Survival numbers of events and statistical significance after imposing each cut
sequentially at $M_{\Sigma^{\pm},\Sigma^0}=300~\GeV$ and for LHC $14~\TeV,~L=100~\fb^{-1}$
($8~\TeV,~L=25~\fb^{-1}$).
\label{tab:sigma_signal1}}
\end{table}

\subsubsection{$\Sigma$ production: $3\ell^{\pm}\ell^{\mp}2j$ signal}
\label{subsubsec:sigma_2}

\begin{figure} [htbp]
\begin{center}
\includegraphics[width=0.45\linewidth]{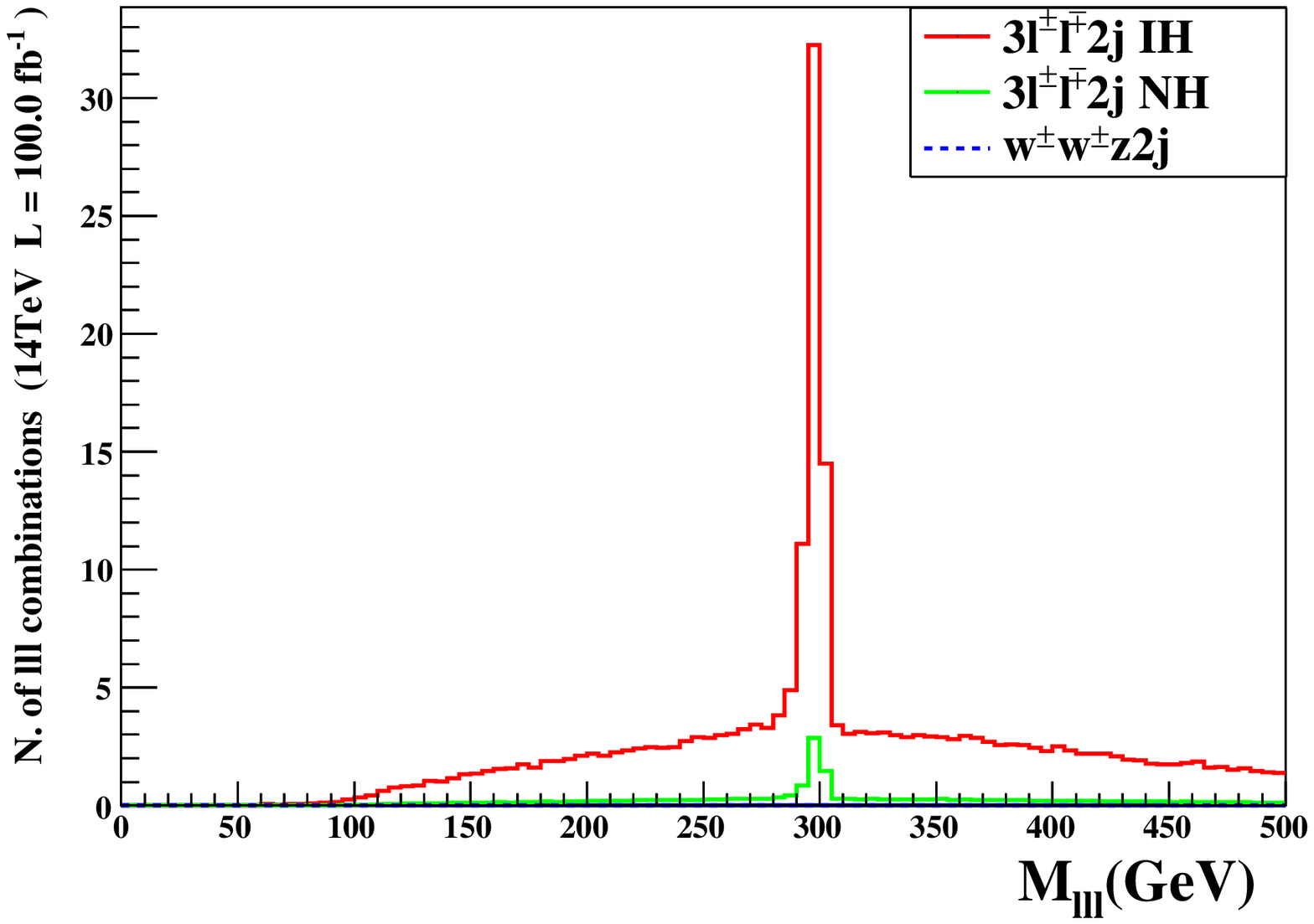}
\includegraphics[width=0.45\linewidth]{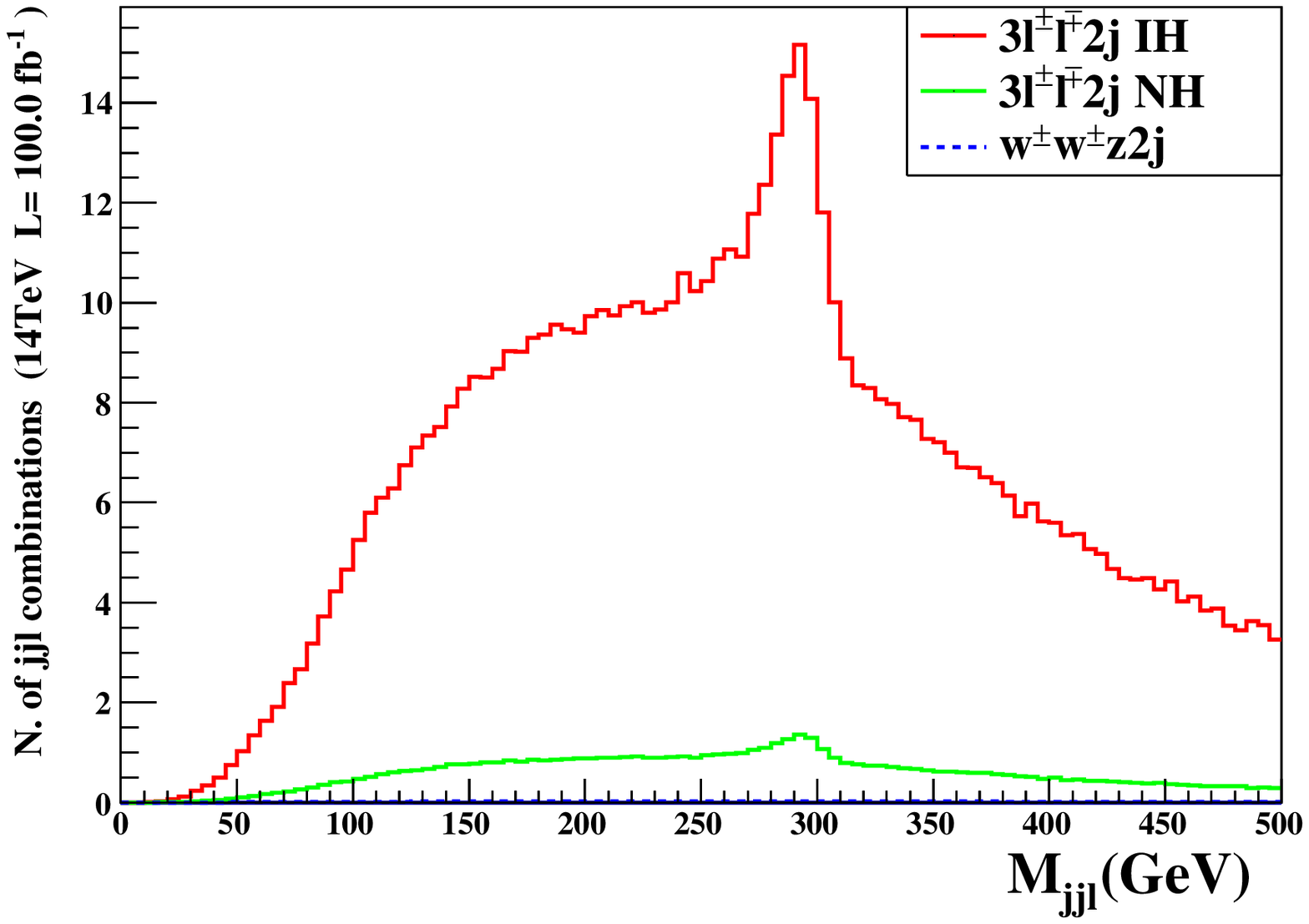}
\end{center}
\caption{Reconstruction of $\Sigma^{\pm},\Sigma^{0}$ via a trilepton (left panel) and a dijet-plus-one-lepton (right) invariant mass for $M_{\Sigma^{\pm},\Sigma^{0}}=300~\GeV$
at LHC $14~\GeV,~L=100~\fb^{-1}$.
\label{fig:sigma_signal2}}
\end{figure}

The associated production $\Sigma^{\pm}\Sigma^{0}$ with decays
\begin{align}
&&\Sigma^{\pm}\Sigma^{0} \to \ell^\pm Z\ell^\pm W^\mp;
~Z \to\ell^+\ell^-,~W \to q \bar q'
\label{ec:ch4q2}%
\end{align}
can produce a final state containing three leptons of same charge plus one lepton
of opposite charge. The irreducible SM background
$W^{\pm}W^{\pm}Z2j$ is small enough compared to the signal, thus the basic cuts are sufficient.
The $\Sigma^{\pm}$ and $\Sigma^{0}$ masses can be reconstructed in a manner similar to that for
the preceding $2\ell^{\pm}2\ell^{\mp}2j$ final state.
The resulting two invariant masses $M_{\ell\ell\ell}$
and $M_{jj\ell}$ are plotted in Fig.
\ref{fig:sigma_signal2} for $M_{\Sigma^{\pm},\Sigma^{0}}=300~\GeV$.
These plots display a clear peak from which $M_{\Sigma^{\pm},\Sigma^{0}}$ can be measured.
From Table \ref{tab:sigma_signal2}, we see that one can reach statistical significance $S/\sqrt{S+B}\simeq10$ and expect about 100 signal events in the IH case at LHC
$14~\TeV,~L=100~\fb^{-1}$. It looks also optimistic to discover a signal in this channel
for the NH case.

\begin{table} [htbp]
\begin{center}
\begin{tabular}{|c|c|c|c|c|c|c|}
\hline
cuts & \multicolumn{2}{|c|}{signal $3\ell^{\pm}\ell^{\mp}2j$} &
\multicolumn{2}{|c|}{bkg $W^{\pm}W^{\pm}Z2j$} &
\multicolumn{2}{|c|}{$S/\sqrt{S+B}$}
\\
\hline
& IH & NH & \multicolumn{2}{|c|}{} & IH & NH
\\
\hline
no cuts & 121 (9.9) & 10.6 (0.8) & \multicolumn{2}{|c|}{0.23 (0.02)} & 11 (3.15) & 3.23 (0.9)
\\
basic cuts & 103.3 (8.5) & 9.1 (0.7) & \multicolumn{2}{|c|}{0.2 (0.019)} & 10.15 (2.91) & 3 (0.83)
\\
\hline
\end{tabular}
\end{center}
\caption{Similar to Table \ref{tab:sigma_signal1}, but for the $3\ell^{\pm}\ell^{\mp}2j$ signal.}
\label{tab:sigma_signal2}
\end{table}

\subsubsection{$\Sigma$ production: $3\ell^{\pm}2\ell^{\mp}+\cancel{E_T}$ signal}
\label{subsubsec:sigma_3}

The five leptons in the final state of this channel can be produced via the decays,
\begin{eqnarray}
&& \Sigma^{\pm} \Sigma^{0} \to \ell^\pm Z \, \ell^\mp W^\pm \,,
  \text{ with } Z \to \ell^+ \ell^-, W \to \ell \nu \,,
   \nonumber\\
&& \Sigma^{\pm} \Sigma^{0} \to \ell^\pm Z \, Z \nu \,,
  \text{ with both } Z \to \ell^+ \ell^- \,,
   \nonumber\\
&& \Sigma^{\pm\pm}\Sigma^{\mp} \to \ell^\pm W^\pm \, \ell^\mp Z \,,
  \text{ with } Z \to \ell^+ \ell^-, W \to \ell \nu \,.
\label{ec:ch5l}
\end{eqnarray}
This signal has a much larger branching ratio than the six-lepton signal
(see Fig. \ref{fig:sigmaevent1}), but still a tiny background, and is thus expected to be more
significant.
In the event selection, we do not apply any additional criteria beyond the basic cuts.
The numbers of events are shown in Table~\ref{tab:sigma_signal3}.
In the IH (NH) scenario, a signal of $50$ ($15$) events is achievable at LHC $14~\TeV$ while it is not
quite observable at $8~\TeV$.
Since none of scalar production produces a five-lepton final state, this channel would signal the
occurrence of heavy fermion production, albeit only at a relatively large luminosity.

\begin{figure} [htbp]
\begin{center}
\includegraphics[width=0.8\linewidth]{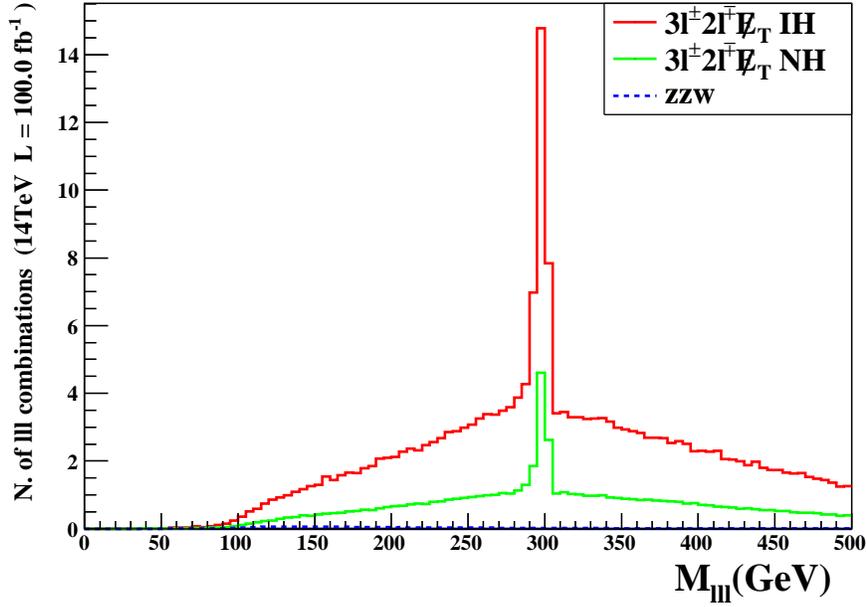}
\end{center}
\caption{Reconstruction of $\Sigma^{\pm}, \Sigma^{0}$ via a trilepton invariant mass for $M_{\Sigma^{\pm}}=300~\GeV$ at LHC $14~\TeV,~L=100\;\fb^{-1}$.
\label{fig:sigma_signal3}}
\end{figure}

\begin{table} [htbp]
\begin{center}
\begin{tabular}{|c|c|c|c|c|c|c|}
\hline
cuts & \multicolumn{2}{|c|}{signal $3\ell^{\pm}2\ell^{\mp}+\cancel{E_T}$} &
\multicolumn{2}{|c|}{bkg $ZZW^{\pm}$} & \multicolumn{2}{|c|}{$S/\sqrt{S+B}$}
\\
\hline
& IH & NH & \multicolumn{2}{|c|}{} & IH & NH
\\
\hline
no cuts & 157 (12.9) & 46.5 (3.8) & \multicolumn{2}{|c|}{3 (0.3)} & 12.4 (3.55) & 6.6 (1.87)\\
basic cuts & 51.2 (3.4) & 15 (1) & \multicolumn{2}{|c|}{0.7 (0.06)} & 7.11 (1.84) & 3.78 (0.97)\\
\hline
\end{tabular}
\end{center}
\caption{Similar to Table \ref{tab:sigma_signal1}, but for the $3\ell^{\pm}2\ell^{\mp}+\cancel{E_T}$
signal.}
\label{tab:sigma_signal3}
\end{table}

\subsubsection{$\Sigma$ production: $3\ell^{\pm}3\ell^{\mp}$ signal}
\label{subsubsec:sigma_4}

This final channel is the cleanest one but has a tiny cross section. It proceeds exclusively through
the following chain,
\begin{eqnarray}
\Sigma^{\pm} \Sigma^{\mp} \to \ell^\pm Z \ell^\mp Z,\text{with both }Z \to \ell^+ \ell^-.
\label{ec:ch6l}
\end{eqnarray}
Upon imposing the basic cuts, we seek six isolated charged leptons, each with a transverse momentum
$p_T>50~\GeV$. We further require that both invariant masses $M_{\ell\ell\ell}$ fall
in the window $280-320~\GeV$. The surviving background events after these cuts become practically
negligible. However, the signals are also tiny: only $1.6$ events are found in the IH scenario and
$0.3$ events in the NH scenario, even with a very high integrated luminosity of $3000~\fb^{-1}$.
Therefore, this signal channel seems irrelevant for the current and near future LHC run.

\begin{figure} [htbp]
\begin{center}
\includegraphics[width=0.8\linewidth]{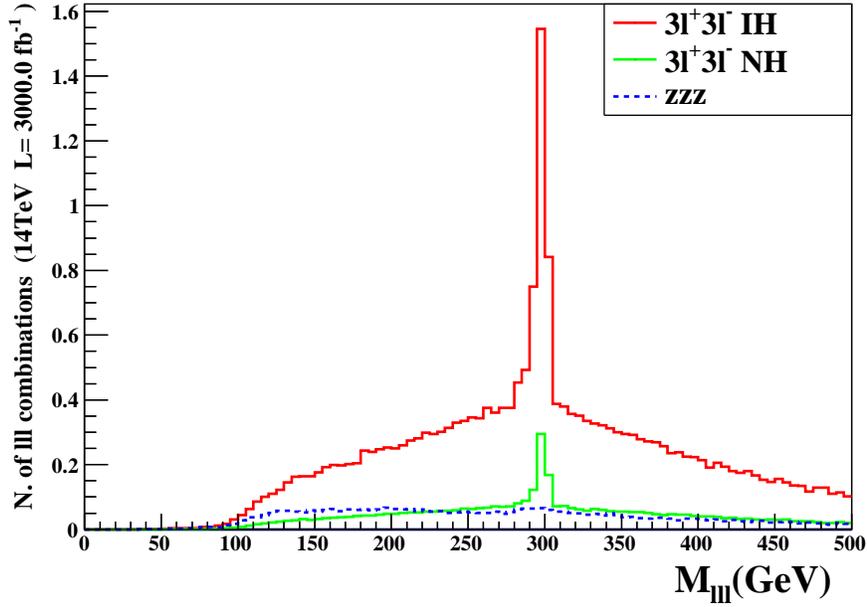}
\end{center}
\caption{Reconstruction of $\Sigma^{\pm}$ via trilepton invariant mass for $M_{\Sigma^{\pm}}=300~\GeV$
at LHC $14~\TeV,~L=3000\;\fb^{-1}$.
\label{fig:sigma_signal4}}
\end{figure}

\begin{table} [htbp]
\begin{center}
\begin{tabular}{|c|c|c|c|c|c|c|}
\hline
cuts & \multicolumn{2}{|c|}{signal $3\ell^{\pm}3\ell^{\mp}$} & \multicolumn{2}{|c|}{bkg $ZZZ$} &
\multicolumn{2}{|c|}{$S/\sqrt{S+B}$}
\\
\hline
& IH & NH & \multicolumn{2}{|c|}{} & IH & NH
\\
\hline
no cuts & 13.3 (0.035) & 2.4 (0.006) & \multicolumn{2}{|c|}{9.6 (0.031)} & 2.78 (0.14) & 0.68 (0.032)
\\
basic cuts & 4.6 (0.01) & 0.83 (0.0018) & \multicolumn{2}{|c|}{2.5 (0.007)} & 1.74 (0.079) & 0.45 (0.019)\\
$p_T^\ell>50~\GeV$ & 4.6 (0.01) & 0.82 (0.0018) & \multicolumn{2}{|c|}{2.4 (0.0065)} & 1.75 (0.08) &
0.46 (0.02)
\\
$M_{\ell^+\ell^-}>90~\GeV$ & 4.4 (0.01) & 0.8 (0.0017) & \multicolumn{2}{|c|}{2.1 (0.0057)} &
1.74 (0.08) & 0.47 (0.02)
\\
$280<M_{\ell\ell\ell}/\GeV<320$ & 1.6 (0.0038) & 0.3 (0.0007) &
\multicolumn{2}{|c|}{0.26 (0.0007)} & 1.15 (0.056) & 0.39 (0.019)
\\
\hline
\end{tabular}
\end{center}
\caption{Similar to Table \ref{tab:sigma_signal1}, but for the $3\ell^{\pm}3\ell^{\mp}$ signal at
LHC $14~\TeV,~L=3000\;\fb^{-1}$ ($8~\TeV,~L=25\;\fb^{-1}$ in parentheses).
\label{tab:sigma_signal4}}
\end{table}

\section{Conclusions}

We have carried out a careful study of the minimal version of the cascade seesaw
\cite{Liao:2010cc} in both theoretical and phenomenological aspects.
We have made a comprehensive analysis on low-energy LFV constraints and LHC signatures.
For this, we have developed a {\tt UFO} model by means of {\tt FeynRules} package,
which can also be applied to phenomenological studies for other seesaw mechanisms.

The main features and results are as follows:
\begin{itemize}
  \item {We introduced a convenient framework to handle Yukawa couplings.
  Based on a parametrization trick in Ref. \cite{Liao:2009fm}, all mixing matrices are expressed in terms of the quadruplet scalar VEV $v_\Phi$, a complex parameter $t$
  and known neutrino parameters. Together with heavy particle masses this fixes all
  production rates and decay branching ratios of heavy particles. This facilitates our
  phenomenological analysis significantly.}

  \item {We considered systematically the contributions of new interactions to the
  stringently constrained LFV transitions, including the decays $\mu\to e\gamma,3e$ and
  $\mu$-$e$ conversion in nuclei. We found that the strictest constraint comes from the
  upper bound on the decay $\mu\to e\gamma$. For instance, for heavy masses of
  $200-300~\GeV$, the scalar VEV $v_\Phi$ must be at least of order $10^{-4}~\GeV$.
  This significantly affects the decays of heavy particles. Inclusion of low-energy
  constraints makes our collider study realistic.}

  \item {We examined all relevant decays of new particles at some benchmark points of
  free parameters, keeping an eye on their impact on the detection strategy at LHC.
  We explored LHC signatures by surveying potentially interesting signal channels.
  For the detection of quadruplet scalars, the $4j2\ell^{\pm}$ signal is most important,
  and it has significant signal events and statistical significance. And for the
  quintuplet fermions, the $2\ell^{\pm}2\ell^{\mp}2j$,
  $3\ell^{\pm}\ell^{\mp}2j$ and $3\ell^{\pm}2\ell^{\mp}+\cancel{E_T}$ signals are quite promising.}

\end{itemize}

\section*{Notes added}
During the finishing stage of this work, a new preprint \cite{Chen:2013xpa}
appeared that also studied the LHC signatures of the model. Here we discuss briefly some
of the differences between that work and ours.
(1) The authors in \cite{Chen:2013xpa} did not consider the mixing of quadruplet
scalars and quintuplet fermions, so that their decay modes of the new particles are much
less then ours.
For instance, they claimed that the $\Phi_{+2}$ coupling to dileptons is absent, thus
$\Phi_{+2}$ decays always dominantly into di-$W$'s. Our study indicates that the two
decay modes are actually comparable around $v_\Phi\sim 10^{-4}~\GeV$.
(2) They treated the Yukawa couplings and the VEV $v_\Phi$ as free parameters. Our
analysis tells that the two are correlated by constraints from low-energy LFV transitions.
These two differences affect the LHC analysis in a significant manner.
(3) The choice of signal channels and the corresponding cut selections are distinct
between the two papers.

\section*{Acknowledgement}
RD would like to thank Kai Wang for useful discussion at the beginning of this work and Liang-Liang Zhang for help on data analysis. This work was supported in part by the
grants NSFC-11025525 and NSFC-11205113.

\section*{APPENDIX A: Some details on the minimal cascade seesaw model}
\label{appendix_A}

We first list the gauge interactions of the $\Phi$ field. Together with the usual
interactions of the SM $\phi$ field and upon incorporating their mixing, one obtains
the gauge couplings of the physical scalars and the would-be Goldstone bosons.

The trilinear terms linear in $v_\Phi$ are
\begin{eqnarray}
  \calL&\supset&g_2^2v_\Phi\Bigg[\left(\sqrt{3}(W_\mu^+)^2\Phi_{+2}^*+\text{h.c.}\right)
  +\frac{7}{\sqrt{2}}W_\mu^+W^{\mu-}\rmRe\Phi_0+\frac{1}{2\sqrt{2}c_W^2}Z_\mu Z^\mu\rmRe\Phi_0
  \nonumber\\
  &&\hspace{2em}+c_W^{-1}Z^\mu\left(\frac{1}{2}\sqrt{6}(1+s_W^2)W_\mu^+\Phi_{-1}
  -2s_W^2W_\mu^+\Phi_{+1}^*+\text{h.c.}\right)
  \Bigg]\nonumber\\
  &&\hspace{2em}+\frac{1}{\sqrt{2}}eg_2v_\Phi A^\mu
  \left[W_\mu^+(2\Phi_{+1}^*-\sqrt{3}\Phi_{-1})+\text{h.c.}\right],
\end{eqnarray}
while the other trilinear terms are
\begin{eqnarray}
  \calL&\supset&\frac{ig_2}{\sqrt{2}}W^{\mu+}\Big[\sqrt{3}
  (\Phi_0^*\partial_\mu\Phi_{-1}-\Phi_{-1}\partial_\mu\Phi_0^*)
  +2(\Phi_{+1}^*\partial_\mu\Phi_0-\Phi_0\partial_\mu\Phi_{+1}^*)
  \nonumber\\
  &&\hspace{4em}
+\sqrt{3}(\Phi_{+2}^*\partial_\mu\Phi_{+1}-\Phi_{+1}\partial_\mu\Phi_{+2}^*)\Big]
  +\text{h.c.}\nonumber\\
  &&+\frac{ig_2}{2c_W}Z^{\mu}\Big[(3-4s_W^2)\Phi_{+2}^*\partial_\mu\Phi_{+2}
  +(1-2s_W^2)\Phi_{+1}^*\partial_\mu\Phi_{+1}\nonumber\\
  &&\hspace{4em}-\Phi_0^*\partial_\mu\Phi_0
  +(-3+2s_W^2)\Phi_{-1}^*\partial_\mu\Phi_{-1}\Big]\nonumber\\
  &&+ieA^{\mu}\Big[2\Phi_{+2}^*\partial_\mu\Phi_{+2}
  +\Phi_{+1}^*\partial_\mu\Phi_{+1}-\Phi_{-1}^*\partial_\mu\Phi_{-1}\Big].
\end{eqnarray}
The quartic gauge interaction terms of $\Phi$ are
\begin{eqnarray}
\calL&\supset&
+\sqrt{3}g_2^2\Big[W^+_\mu W^{+\mu}\big(\Phi_{-1}\Phi_{+1}^*+\Phi_0\Phi_{+2}^*\big)+\text{h.c.}\Big]
  \nonumber\\
&&+\frac{1}{2}g_2^2W^+_\mu W^{-\mu}\big(3|\Phi_{+2}|^2+7|\Phi_{+1}|^2
  +7|\Phi_0|^2+3|\Phi_{-1}|^2\big)
  \nonumber\\
&&+\frac{g_2^2}{\sqrt{2}c_W}Z^\mu\Big[W^+_\mu \Big(\sqrt{3}(2-3s_W^2)\Phi_{+1}\Phi_{+2}^*
  -2s_W^2\Phi_0\Phi_{+1}^*+\sqrt{3}(-2+s_W^2)\Phi_{-1}\Phi_0^*\Big)+\text{h.c.}\Big]
  \nonumber\\
&&+\frac{eg_2}{\sqrt{2}}A^\mu\Big[W^+_\mu \left(3\sqrt{3}\Phi_{+1}\Phi_{+2}^*
  +2\Phi_0\Phi_{+1}^*-\sqrt{3}\Phi_{-1}\Phi_0^*\right)+\text{h.c.}\Big]
  \nonumber\\
&&+\frac{g_2^2}{4c_W^2}Z_\mu Z^\mu\Big[(3-4s_W^2)^2|\Phi_{+2}|^2
  +(1-2s_W^2)^2|\Phi_{+1}|^2+|\Phi_0|^2+(3-2s_W^2)^2|\Phi_{-1}|^2\Big]
  \nonumber\\
&&+\frac{eg_2}{c_W}Z_\mu A^\mu\Big[2(3-4s_W^2)|\Phi_{+2}|^2
  +(1-2s_W^2)|\Phi_{+1}|^2-(3+2s_W^2)|\Phi_{-1}|^2\Big]
  \nonumber\\
&&+e^2A_\mu A^\mu \Big[4|\Phi_{+2}|^2+|\Phi_{+1}|^2+|\Phi_{-1}|^2\Big].
\end{eqnarray}

The explicit forms of the mixing coupling matrices appearing in
eq. (\ref{eq_currents}) are,
\begin{eqnarray}
  &&\hspace{-2em}\calW_L=U_N^\dag w_LU_L=\begin{pmatrix}
  U_\text{PMNS}^\dag\left(\mathbf{1}_3+\frac{7}{4M_\Sigma}ZZ^\dag\right)
   & -\sqrt{\frac{3}{2}}U_\text{PMNS}^\dag Z/\sqrt{M_\Sigma}\\
  -2Z^\dag/\sqrt{M_\Sigma} & \sqrt{6}\mathbf{1}_2
  \end{pmatrix},\nonumber\\[1em]
  &&\hspace{-2em}\calW_R=U_N^Tw_RU_R=\begin{pmatrix}
  0_3 & -\sqrt{6}U_\text{PMNS}^TZ^\ast/\sqrt{M_\Sigma}\\
  -3\eta^\dag &\sqrt{6}\mathbf{1}_2
  \end{pmatrix},\nonumber\\[1em]
  &&\hspace{-2em}\calW_L^D=U_L^\dag w_D=\begin{pmatrix}
  -\sqrt{\frac{3}{2M_\Sigma}}Z\\
  \mathbf{1}_2
  \end{pmatrix},~
  \calW_R^D=U_R^\dag w_D=\begin{pmatrix}
  -\sqrt{\frac{3}{2}}\eta\\
  \mathbf{1}_2
  \end{pmatrix},\nonumber\\[1em]
  &&\hspace{-2em}\calZ_L^\nu=U_N^\dag z_L^NU_N
  =\frac{1}{2}\begin{pmatrix}
  \mathbf{1}_3-U_\text{PMNS}^\dag ZZ^\dag U_\text{PMNS}/M_\Sigma
  & U_\text{PMNS}^\dag Z/\sqrt{M_\Sigma}\\
  Z^\dag U_\text{PMNS}/\sqrt{M_\Sigma} & Z^\dag Z/M_\Sigma
  \end{pmatrix},\nonumber\\[1em]
  &&\hspace{-2em}\calZ^\ell_L=U_L^\dagger z_L^EU_L=\left(-\frac{1}{2}+s_W^2\right)\mathbf{1}_5
  -\frac{1}{2}\begin{pmatrix}
  \frac{3}{2M_\Sigma}ZZ^\dag
  &-\sqrt{\frac{3}{2M_\Sigma}}Z\\
  -\sqrt{\frac{3}{2M_\Sigma}}Z^\dag
  &\mathbf{1}_2
  \end{pmatrix},\nonumber\\[1em]
  &&\hspace{-2em}\calZ^\ell_R=U^\dagger_Rz_R^EU_R=s_W^2\mathbf{1}_5
  -\begin{pmatrix}
  \frac{3}{2}\eta\eta^\dag  &-\sqrt{\frac{3}{2}}\eta\\
  -\sqrt{\frac{3}{2}}\eta^\dag &\mathbf{1}_2
  \end{pmatrix}.
\end{eqnarray}

\section*{APPENDIX B: Loop functions}
\label{appendix_B}

The functions appearing in the radiative transitions are
\begin{eqnarray}
  F^a(r)&=&\frac{1}{12(1-r)^4}\big[1-6r+3r^2+2r^3-6r^2\ln r\big],
  \nonumber\\
  F^b(r)&=&-\frac{1}{12(1-r)^4}\big[2+3r-6r^2+r^3+6r\ln r\big],
  \nonumber\\
  G^a(r)&=&\frac{1}{36(1-r)^4}\big[2-9r+18r^2-11r^3+6r^3\ln r\big],
  \nonumber\\
  G^b(r)&=&\frac{1}{36(1-r)^4}\big[-16+45r-36r^2+7r^3-12\ln r+18r\ln r\big],
\end{eqnarray}
and the function from the box diagram in Fig. \ref{box.diag} is
\begin{eqnarray}
  H(r)=\frac{1}{4(1-r)^3}r(1-r^2+2r\ln r).
\end{eqnarray}

\section*{APPENDIX C: Decay widths of heavy particles}
\label{appendix_C}

Listed below are the approximate expressions for the relevant decay widths in
the degenerate case studied in this work.

Doubly charged scalar $\Phi_{+2}$:
\begin{eqnarray}
\Gamma(\Phi_{+2} \to \ell^+_i \ell^+_j) & = &
\frac{3 M_{\Phi}|(ZZ^{\dag})_{ij}|^2}{16 \pi v_{\Phi}^2 (1+\delta_{ij})},
\\
\Gamma(\Phi_{+2} \to W^+ W^+) & = &
\frac{3 v_{\Phi}^2 M_{\Phi}^3}{2 \pi v_\phi^4}
\left(1-4\frac{M_W^2}{M_{\Phi}^2}\right)^{\frac{1}{2}}
\left(1-4\frac{M_W^2}{M_{\Phi}^2}+12\frac{M_W^4}{M_{\Phi}^4}\right).
\end{eqnarray}
Singly charged scalar $\Phi_{+1}$:
\begin{eqnarray}
\Gamma(\Phi_{+1} \to t \bar{b}) & = &
\frac{3 M_t^2 v_{\Phi}^2 M_{\Phi}}{\pi v_\phi^4}\left(1-\frac{M_t^2}{M_{\Phi}^2}\right)^2,
\\
\Gamma(\Phi_{+1} \to \ell^+_i \nu_{j} ) & = & \frac{M_{\Phi}|(ZZ^{\dag})_{ij}|^2}{64\pi v_{\Phi}^2},\\
\Gamma(\Phi_{+1} \to h W^+ ) & = &
\frac{2v_{\Phi}^2 M_{\Phi}^3}{\pi v_\phi^4}\left[\frac{M_h^4}{M_{\Phi}^4}
+\left(1-\frac{M_W^2}{M_{\Phi}^2}\right)^2-2\frac{M_h^2}{M_{\Phi}^2}
\left(1+\frac{M_W^2}{M_{\Phi}^2}\right)^2\right]^\frac{3}{2}.
\end{eqnarray}
CP-even neutral scalar $H_0$:
\begin{eqnarray}
\Gamma(H_0 \to b \bar{b}) & = &
\frac{27 M_b^2 v_{\Phi}^2 M_{\Phi}}{4 \pi v_\phi^4},
\\
\Gamma(H_0 \to t \bar{t}) & = &
\frac{27 M_t^2 v_{\Phi}^2 M_{\Phi}}{4 \pi v_\phi^4}\left(1-\frac{M_t^2}{M_{\Phi}^2}\right)^2,
\\
\Gamma(H_0 \to \ell_i^+ \ell_j^-) & = &
\frac{9 M_{\Phi}|(ZZ^{\dag})_{ij}|^2}{32 \pi v_{\Phi}^2(1+\delta_{ij})},
\\
\Gamma(H_0 \to \nu_i \nu_j) & = & \frac{M_{\Phi}|(ZZ^{\dag})_{ij}|^2}{4 \pi v_{\Phi}^2(1+\delta_{ij})},\\
\Gamma(H_0 \to hh) & \approx &
\frac{7 v_{\Phi}^2 M_{\Phi}^3}{2 \pi v_\phi^4}
\left(1-4\frac{M_h^2}{M_{\Phi}^2}\right)^{\frac{1}{2}},
\\
\Gamma(H_0 \to W^+ W^-) & = &
\frac{2 v_{\Phi}^2 M_{\Phi}^3}{\pi v_\phi^4}
\left(1-4\frac{M_W^2}{M_{\Phi}^2}\right)^{\frac{1}{2}}
\left(1-4\frac{M_W^2}{M_{\Phi}^2}+12\frac{M_W^4}{M_{\Phi}^4} \right),
\\
\Gamma(H_0 \to Z Z) & = &
\frac{v_{\Phi}^2 M_{\Phi}^3}{4 \pi v_\phi^4}
\left(1-4\frac{M_Z^2}{M_{\Phi}^2}\right)^{\frac{1}{2}}
\left(1-4\frac{M_Z^2}{M_{\Phi}^2}+12\frac{M_Z^4}{M_{\Phi}^4} \right).
\end{eqnarray}
CP-odd neutral scalar $A_0$:
\begin{eqnarray}
\Gamma(A_0 \to b \bar{b}) & = &
\frac{3M_b^2 v_{\Phi}^2 M_{\Phi}}{4 \pi v_\phi^4}, \\
\Gamma(A_0 \to t \bar{t}) & = &
\frac{3M_t^2 v_{\Phi}^2 M_{\Phi}}{4 \pi v_\phi^4}
\left(1-\frac{M_t^2}{M_{\Phi}^2}\right)^2, \\
\Gamma(A_0 \to \ell_i^+ \ell_j^-) & = &
\frac{9 M_{\Phi}|(ZZ^{\dag})_{ij}|^2}{32 \pi v_{\Phi}^2(1+\delta_{ij})},
\\
\Gamma(A_0 \to \nu_i \nu_j) & = & \frac{M_{\Phi}|(ZZ^{\dag})_{ij}|^2}{4 \pi v_{\Phi}^2(1+\delta_{ij})},\\
\Gamma(A_0 \to h Z) & = &
\frac{2 v_{\Phi}^2 M_{\Phi}^3}{\pi v_\phi^4}
\left[1+\left(\frac{M_h^2}{M_{\Phi}^2}-\frac{M_Z^2}{M_{\Phi}^2}\right)^2
-2\left(\frac{M_h^2}{M_{\Phi}^2}-\frac{M_Z^2}{M_{\Phi}^2}\right)\right]^{\frac{3}{2}}.
\end{eqnarray}
Singly charged scalar $\Phi_{-1}$:
\begin{eqnarray}
\Gamma(\Phi_{-1} \to \bar{t} b) & = &
\frac{9 M_t^2 v_{\Phi}^2 M_{\Phi}}{4 \pi v_\phi^4}
\left(1-\frac{M_t^2}{M_{\Phi}^2}\right)^2,
\\
\Gamma(\Phi_{-1} \to \ell^-_i \nu_j) & = &
\frac{3 M_{\Phi}|(ZZ^{\dag})_{ij}|^2}{64\pi v_{\Phi}^2},
\\
\Gamma(\Phi_{-1} \to h W^-) & = &
\frac{3 v_{\Phi}^2 M_{\Phi}^3}{4\pi v_\phi^4}\left[\frac{M_h^4}{M_{\Phi}^4}
+\left(1-\frac{M_W^2}{M_{\Phi}^2}\right)^2-2\frac{M_h^2}{M_{\Phi}^2}
\left(1+\frac{M_W^2}{M_{\Phi}^2}\right)^2\right]^\frac{3}{2},
\\
\Gamma(\Phi_{-1} \to Z W^- ) & = & \frac{3 v_{\Phi}^2 M_{\Phi}^3}{2 \pi v_\phi^4}
\bigg[1+\left(\frac{M_W^2}{M_{\Phi}^2}-\frac{M_Z^2}{M_{\Phi}^2}\right)^2
-2\left(\frac{M_W^2}{M_{\Phi}^2}+\frac{M_Z^2}{M_{\Phi}^2}\right)\bigg]^{1/2},
\nonumber
\\
&&
\times\left[1+\frac{M_W^4}{M_{\Phi}^4}+10\frac{M_W^2}{M_{\Phi}^2}\frac{M_Z^2}{M_{\Phi}^2}
 +\frac{M_Z^4}{M_{\Phi}^4}
 -2\left(\frac{M_W^2}{M_{\Phi}^2}+\frac{M_Z^2}{M_{\Phi}^2}\right)\right].
\end{eqnarray}
Neutral heavy fermion $\Sigma^0$:
\begin{eqnarray}
\Gamma(\Sigma^0_i \to W^{\pm}\ell^{\mp}_j)& =& \frac{g^2_2}{64 \pi}~4|Z_{ji}|^2
\frac{M_{\Sigma}^2}{M_W^2}\left(1-\frac{M_W^2}{M_{\Sigma}^2}\right)
\left(1+\frac{M_W^2}{M_{\Sigma}^2}-2\frac{M_W^4}{M_{\Sigma}^4}\right),
\\
\sum_{l=e}^{\tau}\Gamma(\Sigma^0_i \to Z \nu_l) & = & \frac{g^2_2}{64 \pi c_W^2}
\sum_{l=e}^{\tau}|Z_{li}|^2\frac{M_{\Sigma}^2}{M_Z^2}\left(1-\frac{M_Z^2}{M_{\Sigma}^2}\right)^2
\left(1+2\frac{M_Z^2}{M_{\Sigma}^2}\right),
\\
\sum_{l=e}^{\tau}\Gamma(\Sigma^0_i \to h \nu_l) & \approx & \frac{g_2^2}{64\pi}~9\sum_{l=e}^{\tau}|Z_{li}|^2\frac{M_{\Sigma}^2}{M_W^2}
\left(1-\frac{M_h^2}{M_{\Sigma}^2}\right)^2.
\end{eqnarray}
Singly charged heavy fermion $\Sigma^{-}$:
\begin{eqnarray}
\sum_{l=e}^{\tau}\Gamma(\Sigma^{-}_i \to W^- \nu_l)& = &
\frac{g^2_2}{64 \pi}\sum_{l=e}^{\tau}\frac{15}{2}|Z_{li}|^2
\frac{M_{\Sigma}^2}{M_W^2}\left(1-\frac{M_W^2}{M_{\Sigma}^2}\right)^2
\left(1+2\frac{M_W^2}{M_{\Sigma}}\right) ,
\\
\Gamma(\Sigma^{-}_i \to Z \ell^-_j) & = &
\frac{g^2_2}{64 \pi c_W^2}~\frac{3}{4}|Z_{ji}|^2
\frac{M_{\Sigma}^2}{M_Z^2}\left(1-\frac{M_Z^2}{M_{\Sigma}^2}\right)
\left(1+\frac{M_Z^2}{M_{\Sigma}^2}-2\frac{M_Z^4}{M_{\Sigma}^4}\right),
\\
\Gamma(\Sigma^{-}_i \to h l^-) & \approx & \frac{g_2^2}{64\pi}~\frac{27}{4}|Z_{li}|^2\frac{M_{\Sigma}^2}{M_W^2}
\left(1-\frac{M_h^2}{M_{\Sigma}^2}\right)^2.
\end{eqnarray}
Doubly charged heavy fermion $\Sigma^{--}$:
\begin{equation}
\Gamma(\Sigma^{--}_i \to W^- \ell^-_j ) =
\frac{g^2_2}{64 \pi}~6|Z_{ji}|^2\frac{M_{\Sigma}^3}{M_W^2}
\left(1-\frac{M_W^2}{M_{\Sigma}^2}\right)
\left(1+\frac{M_W^2}{M_{\Sigma}^2}-2\frac{M_W^4}{M_{\Sigma}^4}\right).
\end{equation}

\end{document}